\documentclass[twocolumn,aps,superscriptaddress,multicol,amsmath,amssymb]{revtex4-2}
\makeatletter
\newcommand*{\rom}[1]{\expandafter\@slowromancap\romannumeral #1@}
\makeatother
\usepackage{color}
\usepackage{xcolor}
\usepackage{graphicx}
\usepackage[colorlinks=true, linkcolor=blue, citecolor=blue, urlcolor=blue]{hyperref}
\usepackage{amsmath,amssymb}
\usepackage{epstopdf}

\usepackage[font=small,labelfont=bf,justification=justified]{caption}
\usepackage[font=small,labelfont=bf,justification=justified]{subcaption}
\usepackage{ragged2e} 
\usepackage{mathtools}
\newcommand{\overparent}[1]{\overbracket[0.8pt][0.8ex]{#1}}

\usepackage{tikz}
\usepackage{feynmp}
\usepackage{tikz-feynman}
\tikzfeynmanset{compat=1.1.0}
\usepackage{extarrows}

\usepackage{cancel}
\usepackage{float}
 
\usepackage{mathrsfs}
\graphicspath{{figs/}}
\def\beq{\begin{equation}}
\def\eeq{\end{equation}}
\def\bea{\begin{eqnarray}}
\def\eea{\end{eqnarray}}

\begin{document}

\title{Quantum Many-Body Theory for $kq$-Deformed Particles}
\author {Habib Esmaili}
\affiliation{Department of Physics, University of Mohaghegh Ardabili, P.O. Box 179, Ardabil, Iran}
\author {Hosein Mohammadzadeh}
\affiliation{Department of Physics, University of Mohaghegh Ardabili, P.O. Box 179, Ardabil, Iran}
\author {Mehdi Biderang}
\affiliation{Department of Physics, University of Toronto, Toronto, Ontario, M5S 1A7, Canada}
\author {Morteza NattaghNajafi}
\email{morteza.nattagh@gmail.com}
\affiliation{Department of Physics, University of Mohaghegh Ardabili, P.O. Box 179, Ardabil, Iran}


\begin{abstract}
We present a comprehensive quantum many-body theory for $kq$-deformed particles, offering a novel framework that relates particle statistics directly to effective interaction strength. Deformed by the parameters $k$ and $q$, these particles exhibit statistical behaviors that interpolate between conventional bosonic and fermionic systems, enabling us to model complex interactions via statistical modifications. We develop a generalized Wick’s theorem and extended Feynman diagrammatics tailored to $kq$-particles, allowing us to calculate two types of Green functions. Explicit expressions for these Green functions are derived in both direct and momentum spaces, providing key insights into the collective properties of $kq$-deformed systems. Using a random phase approximation (RPA), we estimate the dielectric function for $q$-fermion gas ($k=-1$), and analyze the Friedel oscillations, the plasmon excitations, and the energy loss function. Our results demonstrate that the effective interaction is tuned by the value of $q$, so that a non interacting limit is obtained as $q\to 0$, where the Friedel as well as the plasma oscillations disappear. There is an optimal value of $q$, namely $q^*$ the plasma freuency, as well as the energy loss function show an absolute maximum, and the effective interaction changes behavior. 
\end{abstract}
\maketitle
\tableofcontents

\section{INTRODUCTION}\label{1}
Almost four decades have elapsed since the introduction of the concepts of anyons~\cite{wilczek1982quantum,greiter2024fractional} and the generalized Pauli exclusion principle, which together paved the way for the broader framework of deformed (fractional) quantum statistics~\cite{haldane1991fractional}.
In recent years, the ability to model systems with unconventional statistics and nontrivial interaction effects triggered interest in deformed quantum particles has grown. 
This has inspired extensive studies on these systems, including efforts to deform the quantum occupation number by manipulating the commutation relations of fermionic or bosonic operators~\cite{chaichian1993statistics}.
Haldane further generalized the concept of fractional statistics to higher dimensions by redefining the Pauli principle for quasiparticles. This framework enables particles like spinons to exhibit a mixture of bosonic and fermionic characteristics, offering new insights into the behavior of multi-dimensional quantum systems~\cite{almeida2001remarks,haldane1991fractional}.
\\

The deformation of particles' statistics in condensed matter systems can also be viewed as a method to tune the effective interaction strength between particles~\cite{algin2024two,kuramoto2020quantum,pachos2018quantifying,greiter2024fractional,zhang2022interaction}. An important example is the interplay between strong correlations and statistics in anyonic (Laughlin) states observed in the fractional quantum Hall effect~\cite{wu1984multiparticle,laughlin1983anomalous} with vast applications, where strong interactions are effectively substituted by statistical properties. The interaction in this system can also lead to anyonic superconductivity~\cite{wilczek1992fractional,herrmann2010common,jeon2003fractional,laughlin1988relationship,lee1989anyon,lee1991anyon,chaichian1993statistics,lavagno2009deformed,hou2018q}. Topological field theories, particularly Chern-Simons models, are the tools to track the topological aspects of these system.  This analysis has been extended to study the fracton topological phases, and classify them using lattice gauge theory~\cite{shirley2019fractional,vijay2016fracton}. The connection between particle interactions and fractional exclusion statistics in Bose gases has been explored, revealing links to quantum criticality and strongly correlated systems~\cite{greiter2024fractional,barkeshli2010topological,zhang2022interaction}. This suggests that in some circumstances we are allowed to replace direct interactions with statistical interactions, which, in a large class of deformed particles, are encoded in the commutation relations of the corresponding quantum field theory which can be defined in all dimensions (not only two, as for the anyons)~\cite{jackson1909generalization}. 
The connection with quantum groups was originally discovered in Ref.~\cite{biedenharn1989quantum,macfarlane1989q}.
A relevant question is: what type of algebra would be best suited to describe a statistical theory that interpolates between Bose and Fermi statistics?~\cite{hou2018q}.

A number of foundational studies have specifically examined $kq$-deformed particles ($kq$-particles), which extend the concepts of bosonic and fermionic statistics by incorporating deformation parameters, 
$k$ and $q$~\cite{leinaas1977,goldin1980particle,medvedev1997properties,wilczek1982quantum,chaichian1993statistics,lee1992q,hou2020thermostatistics,lavagno2002generalized,tuszynski1993statistical,mirza2011thermodynamic} aiming to interpolate between different statistics by adjusting the underlying algebraic structure through modified commutation relations. 
This approach allows for a more flexible description of thermodynamic properties, such as entropy, and specific heat~\cite{lavagno2009deformed}. Studies have shown that such deformations also provide insight into behaviors like Bose-Einstein condensation, which is absent in traditional two-dimensional systems but can be explored through the $kq$-deformed models~\cite{lavagno2009deformed,algin2012high}. The so-called $kq$-deformed BCS superconductivity and the relativistic Fermi gas have been analyzed through the Green's function formalism and the generalized Matsubara frequencies~\cite{hou2018q,hou2020thermostatistics}. There are also some studies on the topological aspects of deformed particles and their implications for understanding exotic quantum phenomena~\cite{zhou2023probing,wachtler2024topological,forte1992relativistic}.\\

Despite significant focus and progress in studying $kq$-particles through the modulation of deformation parameters, comprehensive theories addressing the collective behavior of many $kq$-particles remain scarce. This paper aims to bridge this gap by developing a quantum many-body theory for a system of $kq$-particles. Specifically, we derive an $S$-matrix expansion, establish a generalized Wick's theorem, and formulate Feynman diagrams. Building on these foundations, we present explicit expressions for the dielectric function of a $kq$-fermion gas. \\

The paper is organized as follows: In the next section we descibe the main aim and achievments of the paper. SEC.~\ref{kqPARTICLE} is devoted to the algebra governing $kq$-particle and obtaining the proportional distribution function for $kq$-particle statistics.  The states used for the $kq$-particles are discussed in SEC.~\ref{kqALGEBRA} where, using a mode expansion, we build the framework for the system. The $S$-matrix theory and the Wick's theorem for generalized $kq$-particles will be discussed in this section. In SEC.~\ref{Green Functions} we derive and analyze the single particle Green's function for $kq$-particles. We develop a set of rules for constructing Feynman diagrams that represent contractions involving $kq$-particles in SEC.~\ref{DIAGRAMMATICkqParticles}. The results obtained for the response function and subsequent Friedel Oscillations and Plasmons and energy loss for $q$-Fermion gas are discussed in SEC.~\ref{Friedel-Oscillations}. We conclude the paper with a discussion in the SEC.~\ref{CONCLUD}.

\section{AIMS AND ACHIEVEMENTS OF THIS PAPER}\label{Aims}
In this paper, we develop a quantum many-body theory specifically for $kq$-deformed particles, establishing a correspondence between particle statistics and effective interaction strength. By treating the deformation parameters 
$k$ and 
$q$ as tunable quantities, we demonstrate that it is possible to explore a wide range of effective interactions through statistical modifications. The paper has two main bodies: Firstly the basic theory part for $kq$-particle systems in which we develop a general framework, and secondly the part where we use the concepts  to determine essential functions for being used in more practical situations. In the first part: 
\begin{itemize}
    \item Our framework introduces a generalized $S$-matrix expansion (SEC.~\ref{kqALGEBRA}).
    \item We generalize the Wick's theorem (SEC.~\ref{kqALGEBRA}).
    \item An extended Feynman diagrammatic approach is developed  (SEC.~\ref{DIAGRAMMATICkqParticles}), which forms the basis for computing Green functions specific to $kq$-particles.
\end{itemize}

The second part of the paper is started by defining two types of single particle Green's functions: the standard Green's function and the effective Green's function based on the $kq$ operator algebra and the poles of the occupation number function respectively. In this part:
\begin{itemize}
    \item We derive explicit expressions for these Green functions in both direct and momentum spaces (SEC.~\ref{Green Functions}). 
    \item This enables us to analyze key physical properties such as the dielectric response function within the random phase approximation (RPA) for $q$-fermion gas ($k=-1$), establishing connections to fundamental concepts in collective phenomena (SEC.~\ref{RESPONSE-FUNCTIONS}). 
    \item In SEC.~\ref{DIELECTRIC-FUNCTION} we discuss Friedel oscillations, plasmonic excitations, and energy loss functions (SEC.~\ref{Friedel-Oscillations}) within the RPA. 
\end{itemize}

Equation~\eqref{ٍEq:distribution-function} is the occupation number that is well-accepted in the literature, and Eq.~\eqref{Eq:DesiredState} is the quantum state in the occupation numbers representation. We generalize the time-ordering operator in Eq.~\eqref{TimeOrdering-kq}, and the generalized $S$-matrix expansion is shown in Eq.~\eqref{Eq:genSMatrix}. We present the generalized Wick's theorem in Eq.~\eqref{WickTheorem}. A Gorkov equation is presented for the standard single-particle Green's function in Eq.~\eqref{43} with the generalized Matsubara frequencies shown in Eq.~\eqref{Eq:Matsubara}. Using the poles of the occupation number we obtain the effective Green's function which satisfies Eqs.~\eqref{Eq:EffcGreen1} and~\eqref{Eq:EffecGreenGorkov}, and explicit form of which is given in Eq.~\eqref{Eq:Gwn}. The Feynman rules for the perturbative expansion is discussed in Sec.~\eqref{DIAGRAMMATICkqParticles}. Using these techniques, we obtain the formal form of the correlation energy in Eqs.~\eqref{Eq:EnergyCorr} and~\eqref{Gs19}. The random phase approximation analysis is presented in SEC.~\eqref{SEC:RPA}, the result of which is analytic expressions for various response functions. For example the dielectric function is given in Eq.~\eqref{dielectric-constant0}, as a result of which on is able to find the Friedel oscillations for the induced charge and the plasmon frequency represented in Eqs.~\eqref{Eq:Friedel} and~\eqref{Eq:PlasmonicOs} respectively. An effective interaction strength is proposed in terms of the $kq$ statistics in Eq.~\eqref{Eq:effectiveInteraction} which is one of the essential relations in this paper. Our results illustrate how quantum many-body effects in $kq$-particle systems can reveal new insights into the interplay between statistics and effective interactions, potentially opening up avenues for studying strongly correlated materials and other systems with nontrivial statistical properties.

\section{$kq$-PARTICLE STATISTICS, THE GENERAL CONSTRUCTION}\label{kqPARTICLE}
Let's start with the statistics of bosons and fermions in many-particle systems.
Mathematically, for indistinguishable particles, there can be a complete  set of commuting operators $\{N_1,N_2,...\}$, whose eigenvalues, i.e. the set of occupation numbers $\{n_1,n_2,...\}$, express how many particles occupy the quantum states $\alpha_1$, $\alpha_2$, and so on.
The state of a many-particle system is represented by  
\begin{align}
    |n_1,n_2,n_3,...\rangle,
\end{align}
which spans the Fock space corresponding to the many body system under consideration. These states are identified as the eigenvector of the number operator $N_{\nu}$;
\begin{align}
    N_{\nu}|n_1,n_2,n_3,...\rangle=n_{\nu}|n_1,n_2,n_3,...\rangle,\ {\nu}=1,2,...
\end{align}
One is able to construct these states out of the vacuum state
\begin{align}
|0\rangle=|n_1=0,n_2=0,...\rangle.
\end{align}
For this, one uses the fermionic and bosonic creation (annihilation) operator represented by $a^{\dagger(\text{X})}$ ($a^{(\text{X})})$, with $\text{X}=\text{B},\text{F}$ for bosons and fermions, respectively, such that:
\begin{equation}
\begin{aligned}
    &a^{\dagger(\text{X})}_{\nu}|n_1,n_2,...n_i,...\rangle\varpropto|n_1,n_2,...n_{\nu}+1,...\rangle,\\
    &a^{(\text{X})}_{\nu}|n_1,n_2,...n_{\nu},...\rangle\varpropto|n_1,n_2,...n_{\nu}-1,...\rangle,
\end{aligned}
\end{equation}
and
\begin{equation}
    a^{(\text{X})}_{\nu}|0\rangle=0.
\end{equation}
The bosonic and fermionic creation and annihilation operators satisfy commuting and aniticommuting algebras, respectively,
\begin{equation}
\begin{split}
&\left[a^{(X)}_{\nu},a^{(X)}_{\nu'}\right]_k=\left[a^{\dagger(X)}_{\nu},a^{\dagger(X)}_{\nu'}\right]_k=0 \ , \  \left[a^{(X)}_{\nu},a^{\dagger(X)}_{\nu'}\right]_k=\delta^{}_{\nu\nu'}
\label{Eq:classicAlgebra}
\end{split}
\end{equation}
with $k=+1$ for bosons and $k=-1$ for fernions, In addition, the commutation/anticommutation of two operators $O_1$ and $O_2$ are defined as
\begin{equation}
\left[O_1,O_2\right]_k=O_1O_2+kO_2O_1.
\end{equation}
Motivated by recent successful developments in the formulation of the theory concerning $q$-deformed bosons and fermions~\cite{chaichian1993statistics,lavagno2002generalized,lavagno2000thermostatistics}, we generalize the depicted algebra given
in Eq.~(\ref{Eq:classicAlgebra}) as follows
\begin{equation}
\begin{split}
&\left[a^{}_{\nu},a^\dag_{\nu}\right]_{kq}=a^{}_{\nu} a^\dag_{\nu}-kq a^\dag_{\nu} a^{}_{\nu}=q^{-N^{}_{\nu}}\\
&[a^{}_{\nu},a^{}_{\nu}]_{k}=[a^\dag_{\nu},a^\dag_{\nu}]_{k} =0
\end{split}
\label{E1}
\end{equation}
where $0\leq q\leq1$ is the generalization parameter, and $N_\nu$ is the number operator for $kq$-particles, satisfying
\begin{equation}\label{Anomber}
[N^{}_{\nu}]=a^\dag_{\nu} a^{}_{\nu},
\end{equation}
Moreover, the number (or operator) in the $q$ basis is defined as 
\begin{equation}
[x]=\frac{q^x - q^{-x}}{q-q^{-1}}.
\label{E3}
\end{equation}
in which $N_{\nu}$ satisfies the following relations
\begin{equation}
\begin{aligned}
[N_{\nu},a^\dag_{\nu}]_k=a^\dag_{\nu}~~,~~[N_{\nu},a^{}_{\nu}]_k=-a^{}_{\nu},\ 
a^{}_{\nu} a^\dag_{\nu}=[1+kN^{}_{\nu}].
\label{E2}
\end{aligned}
\end{equation}
More complicated commutators can be obtained using these fundamanetal relations. The examples are 
\eqref{Eq:Appcp}\eqref{Eq:Appcdp}:
\begin{equation}
\begin{aligned}
        &\left[a^\dagger_{\nu} a^{}_{\nu},a^{}_{\nu}\right]_{kq}=\left(k(1-q)[N^{}_{\nu}]-q^{-N^{}_{\nu}}\right)a^{}_{\nu},
        \\&\left[a^\dagger_{\nu} a^{}_{\nu},a^\dagger_{\nu}\right]_{kq}= a^\dagger_{\nu}\left(k(q-1)[N^{}_{\nu}]+q^{-N^{}_{\nu}}\right),\\
        & \left[a^{\dagger}_la^{}_l,a^{\dagger}_pa^{}_p\right]_{kq}=0
       \label{cp}
\end{aligned}
\end{equation}
Since the numbers in the $q$ basis revert to their standard definitions in the limit $q\to 1$, the algebra reduces to the conventional bosonic form for $k=+1$ and the fermionic form for $k=-1$.
Let us consider the following eigenvalue relation for $N_{\nu}$
\begin{equation}
    N^{}_{\nu}\left| \Phi \right\rangle = n^{}_{\nu}\left| \Phi \right\rangle.
\end{equation}
wherein $\left| \Phi \right\rangle$ represents the simultaneous eigenvector of all number operators $N_{\nu}$ with eigenvalue $n_{\nu}$. To calculate the average occupation number, denoted by $\bar{n}_{\nu}$, we calculate the expected value of the number operator as follows (see Eq.~\eqref{fk})
\begin{equation}
    [\bar{n}_{\nu}]=\frac{\text{Tr}\Big(e^{-\beta H}a^\dagger_{\nu} a^{}_{\nu}\Big)}{Z},
\end{equation}
where $Z= \text{Tr}\left[e^{-\beta H}\right]$ is the partition function. \\

To make the first steps towards recognizing this algebra and the corresponding thermodynamic behaviors, let us consider a free Hamiltonian of non-interacting $kq$-particles with the energy spectrum $\left\lbrace\epsilon_i \right\rbrace$, and chemical potential $\mu$ as follows
\begin{equation}\label{hamiltonian}
    H=\sum_{\nu} (\epsilon_{\nu}-\mu)N_{\nu}.
\end{equation}
For such a free system, one can readily show that (see Eqs.~\eqref{A1}, \eqref{knp1} and \eqref{knp2} of Appendix~\ref{Appendix})
\begin{equation}
   \begin{aligned}
        [\bar{n}_{\nu}]&=e^{-\beta(\epsilon_{\nu}-\mu)}[k\bar{n}_{\nu}+1],\\
        \left[k\bar{n}_{\nu}+1\right]&=kq\left[\bar{n}_{\nu}\right]+q^{-k\bar{n}_{\nu}},\\
        \left[k\bar{n}_{\nu}+1\right]&=(kq)^{-1}\left[\bar{n}_{\nu}\right]+q^{k\bar{n}_{\nu}}.
        \label{identities}
   \end{aligned}
\end{equation}
By combining these formulas, one arrives at the relation
\begin{equation}
  \begin{aligned}
\left[k\bar{n}_{\nu}+1\right]=\left[\bar{n}_{\nu}\right]\cosh\left(\ln{kq}\right)+\cosh\left(k\bar{n}_{\nu}\ln{q}\right),
     \label{np1}
  \end{aligned}
\end{equation}
or equivalently
\begin{equation}
\left[\bar{n}_{\nu}\right]=\frac{\cosh\left(k\bar{n}_{\nu}\ln{q}\right)}{e^{\beta(\epsilon_{\nu}-\mu)}-\cosh(\ln{kq})}.
\end{equation}
Solving the equation in terms of $\bar{n}_i$ reveals~\cite{lavagno2002generalized}
\begin{equation}\label{ٍEq:distribution-function}
    \bar{n}_{\nu}\equiv\bar{n}_{kq}(\zeta_{\nu})=\frac{1}{q-q^{-1}}\ln\left(\frac{e^{\beta\zeta_{\nu}}-kq^{-k}}{e^{\beta\zeta_{\nu}}-kq^{k}}\right),
\end{equation}
in which $\zeta_{\nu}\equiv \epsilon_{\nu}-\mu$. 
The above equation serves as the fundamental relation for the following sections, where more sophisticated relationships will be explored. In particular, we derive an effective Green function based on this relation.
\section{$kq$-ALGEBRA; A QUANTUM MANY BODY DESCRIPTION FOR THE DEFORMED PARTICLES}\label{kqALGEBRA}
A standard scheme for studying quantum systems is mode expansion, through which, for a translational invariant system, the quantum fields are expanded in terms of plane waves, defining creation and annihilation operators in momentum space. 

The examples of the mode expansion of fermions and bosons are given in Appendix~\ref{Mode-expansion}. Building upon the algebraic framework developed previously for $kq$-particles, we employ this algebraic structure to derive the commutation relations. We represent a $kq$-quantum field by $\hat{\psi}(x)$ and $\hat{\psi}^{\dagger}(x)$, where $x\equiv (t,\textbf{x})$ is space-time four-vector. Generally, a $kq$-field can be expanded in terms of creation and annihilation operators as follows
\begin{equation}
    \hat{\psi}(x)=\sum_{\nu} u_{\nu}(x)a_{\nu} \ , \ \hat{\psi}^\dagger(x)=\sum_{\nu} u_{\nu}^*(x)a^\dagger_{\nu},
    \label{Eq:expantion-kqParticle}
\end{equation}
where $u_{\nu}(x)$ is one-particle basis of the Hilbert space, and $u_{\nu}^*(x)$ is its complex conjugate. Note that in most applications there is an extra independent solution to be added in the above expansion. In the rest of this subsection we consider a translational invariant system where the momentum $p$ is a good quantum number, i.e. the quantum number $\nu$ is substituted by $p$, and $u_p(x)=\frac{1}{\sqrt{V}}e^{ip.x}$, where $p.x\equiv \epsilon_pt-\textbf{p}.\textbf{x}$, and $\epsilon_p$ is the energy of single particles. For this system the mode expansion follows Eq.~\eqref{Eq:expantion-kqParticle}. Explicitly we have
\begin{equation}
    \hat{\psi}(x)=\frac{1}{\sqrt{V}}\sum_p e^{ip.x} a_p \ , \ \hat{\psi}^\dagger(x)=\frac{1}{\sqrt{V}}\sum_p e^{-ip.x} a^\dagger_p.
    \label{mode-expantion-kqParticle}
\end{equation}
where the algebra of the mode components ($a_p$ and $a^{\dagger}_p$) are equivalent to Eqs.~\eqref{E1} and~\eqref{E2} as follows
\begin{equation}\label{Eq:algebra}
    \left[a^{}_p,a^\dagger_p\right]_{kq}=q^{-N_p} \ , \ \left[a^{\dagger}_p,a^{}_p\right]_{kq}=-kq^{N_p+1},
\end{equation}
and
\begin{equation}\label{Eq:algebra2}
\begin{split}
&\left[a^{}_p,a^\dagger_{p'}\right]_{k}=0 \ , \ \left[a^{}_p,a_{p'}\right]_{k}=0 \ \ \text{if} \ p\neq p',\\
&\left[a^{\dagger}_p,a^{}_{p'}\right]_{k}=0 \ , \ \left[a^{\dagger}_p,a^{\dagger}_{p'}\right]_{k}=0 \ \ \text{if} \ p\neq p'.   
\end{split}
\end{equation}
In these equations $N_p$ is the number operator given in Eq.~\eqref{Anomber}. We notice here that the mode expansion for pure fermions and complex bosons has an extra term relative to the Eq.~\eqref{mode-expantion-kqParticle} corresponding to the antiparticles. In the non-relativistic quantum many body fermionic systems this term corresponds to holes~\cite{fetter2012quantum,bruus2004many}, and the mode expansion is split to two parts: the annihilation (creation) of electrons (holes) above (below) the Fermi energy. In relativistic quantum systems the extra term is required to guarantee the completeness of the expansion~\cite{peskin2019concepts} and corresponds to creation or annihilation of antiparticles. Without loose of generality we drop this term to keep the formalism simple to follow, while the addition of anti-particles counterpart is straightforward. \\
Using the fact that
\begin{equation}\label{Eq:Algebrakq2}
    \left[ a_{p}, a^\dagger_{p'}\right]_{kq}=\left[ a_{p}, a^\dagger_{p'}\right]_{k}+k\left[ a_{p}, a^\dagger_{p'}\right]_{q}=0 \ \text{if} \ p\neq p',
\end{equation}
one finds
\begin{equation}\label{algebra-mode-expantion-kqParticle}
    \begin{aligned}
         \left[\hat{\psi}(x_1),\hat{\psi}^\dagger(x_2)\right]_{kq}
         &=\frac{1}{V}\sum_{p}e^{ip.(x_1-x_2)}q^{-N_{p}}\\
         \Big[\hat{\psi}(x_1),\hat{\psi}(x_2)\Big]_{k}&=\Big[\hat{\psi}^\dagger(x_1),\hat{\psi}^\dagger(x_2)\Big]_{k}=0,    
    \end{aligned}
\end{equation}
This relation is served as a generalized commutation relation for the fields in the real space. \\

To construct a Fock space for $kq$-particles as the first step towards defining a many-body theory, we need to define a $kq$-vacuum state as follows
\begin{equation}
a_p\left|0 \right\rangle_{kq} =0.
\end{equation}
Then a desired state is built upon applying the creation operator as mentioned in appendix~\ref{Appendix}
\begin{equation}\label{Eq:DesiredState}
\left|n_1,n_2,...,n_m \right\rangle_{kq}=\prod_{i=1}^{m}\frac{\left(a_{p_i}^{\dagger}\right)^{n_i}}{\sqrt{[n]!}}\left|0 \right\rangle_{kq}.
\end{equation}
Note that the factor in the denominator is in the $q$ basis to guarantee the correct $kq$-algebra. The restrictions we have for the occupation configurations depend on $k$ and $q$. For the fermionic limit $k=-1$, and $q=1$, we know that $n_i=0,1$, while in the bosonic limit, there are no restrictions. To construct a quantum many-body theory for translational invariant $kq$ particles, we begin with a non-interacting system described by the following Hamiltonian in the momentum space
\begin{equation}
\hat{K}^{\text{Free}}\equiv \hat{H}_{kq}^{\text{Free}}-\mu \hat{N}=\sum_p (\epsilon_p-\mu)\hat{N}_p.
\label{Eq:freeHamiltonian}
\end{equation}
Note that, $N_p$ in this relation is in the normal basis (not in the $q$-basis), which is expected for a free non-interacting system. The occupation number $n_p$ is its eigenvalue, the average of which is given by Eq.~\eqref{ٍEq:distribution-function}, i.e.
\begin{equation}\label{E29}
    \bar{n}_{kq}(\zeta_p)=\bar{n}_p=\frac{1}{q-q^{-1}}\ln\left(\frac{e^{\beta\zeta_{p}}-kq^{-k}}{e^{\beta\zeta_{p}}-kq^{k}}\right).
\end{equation}
In the next section we develope an $S$-matrix expansion for $kq$-particles and also derive the generalized Wick theorem which helps understanding the structure of interacting $kq$-particles.

\subsection{$S$-matrix expansion for $kq$-particles}\label{sMatrix}
For the future use, we present the details of the field expansion, $S$-matrix theory and the Wick's theorem in this section for the generalized $kq$ particles. We first concentrate on the $S$-matrix theory. The $S$-matrix expansion for $kq$-particles is a mathematical tool used to describe the interactions between particles in quantum field theory. It provides a systematic way to calculate the probability amplitudes for various processes involving these particles. The details of the calculations concerning the $S$-matrix expansion can be found in standard textbooks for quantum many body systems, and also in Appendix~\ref{s-matrix-theory}. Generally, the $S$ matrix is defined as an operator $S(t,t')$ which connects $\Psi(t)$ to $\Psi(t')$ ($t>t'$) as follows
\begin{equation}
\left|\Psi(t)\right\rangle=S(t,t')\left|\Psi(t')\right\rangle.
\end{equation} 
To develop a $S$-matrix theory for $kq$-particles, we consider a system of $kq$ free particles at $t\to-\infty$ which adiabatically evolves to a fully interacting system  at $t=0$, described by the Hamiltonian (see standard books):
\begin{equation}
H(t)=H_0+H_{\text{int}}e^{-\epsilon t},
\end{equation}
where $H_0$ in non-interaction part of the Hamitonian (with eigenfunction $\Phi_0$), and $H_{\text{int}}$ is the interaction term multiplied by an exponential term containing  $\epsilon=0^+$, guaranteeing that the system is free for $t\to-\infty$, so that $\Phi_0$ evolves adiabatically to the ground state of interaction Hamiltonian ($\Psi_0$) at $t=0$. 
Then one easily finds
\begin{equation}\label{s-matrix-psi}
\left| \Psi_0\right\rangle =S(0,-\infty)\left| \Phi_0\right\rangle,
\end{equation}
The aim of this section is to find an expansion for $S$. Before considering the details of expansion, we need to define the time ordering operators as follows:
\begin{equation}\label{kq-Time-ordering}
\begin{split}
T\left[\hat{O}_1(t)\hat{O}_2(t')\right]&\equiv\Theta(t-t')\hat{O}_1(t)\hat{O}_2(t')\\
&+k^{m}\Theta(t'-t)\hat{O}_2(t')\hat{O}_1(t),\\
 \mathcal{T}\left[\hat{O}_1(t)\hat{O}_2(t')\right]&\equiv \Theta(t-t')\hat{O}_1(t)\hat{O}_2(t')\\
 &+(kq)^{m}\Theta(t'-t)\hat{O}_2(t')\hat{O}_1(t),
\end{split}
\end{equation}
where $\hat{O}_i(t)$ is a string of $a$ and $a^{\dagger}$ operators, and $m$ is the number of single commutations required to commute $\hat{O}_1(t)$ and $\hat{O}_2(t)$. $T$ operator is the conventional time-ordering operator, while $\mathcal{T}$ is a $kq$ time ordering operator for the sake of later convenience. Note that these two identities are related through 
\begin{equation}\label{TimeOrdering-kq}
    \begin{aligned}
        &\mathcal{T}\left[\hat{O}_1(t)\hat{O}_2(t')\right]\\
        &=\Big(\Theta(t-t')+(kq)^{m}\Theta(t'-t)\Big)T\left[\hat{O}_1(t)\hat{O}_2(t')\right],
    \end{aligned}
\end{equation}
An important example is the Green function.
We can apply the generalized relation mentioned above to various examples, that a notable one is the zero-temperature Green's function for $kq$-particles, which is expressed as follows
\begin{equation}
iG(p;t,t')=\frac{\langle\Psi_0|\mathcal{T}\left[a_p(t)a^{\dagger}_p(t')\right]|\Psi_0\rangle}{\langle\Psi_0|\Psi_0\rangle}
\label{Eq:Green0}
\end{equation}
Using the above-mentioned identity, this
Green’s function is related to the conventional Green function:
\begin{equation}
\begin{split}
iG(p;t,t')&
=\left[\Theta(t-t')+kq\Theta(t'-t)\right]iG_T(p;t,t')
\end{split}
\end{equation}
where $iG_T(p;t,t')\equiv\frac{\langle\Psi_0|T\left[c_p(t)c^{\dagger}_p(t')\right]|\Psi_0\rangle}{\langle\Psi_0|\Psi_0\rangle}$ is based on the conventional time ordering, helping the calculations regarding the $S$-matrix expansions.

Starting from the definition of $\left| \Psi_0\right\rangle$~\eqref{s-matrix-psi}, we derive the expansion of $S$-matrix for the $kq$-particles using the conventional time ordering operator as follows (see Eq.~\eqref{smatrix})
\begin{equation}
    \begin{aligned}
        S(t,t')=&\sum^\infty_{n=0}\frac{(-i)^n}{n!}\int^t_{t'}dt_1\int^t_{t'}dt_2...\int^t_{t'}dt_n
        \\&\times T\left[H^{}_{\text{int}}(t_1)H^{}_{\text{int}}(t_2)...H^{}_{\text{int}}(t_n)\right].
    \end{aligned}
\end{equation}
where $\hat{O}_{\text{int}}$ represents the operator $\hat{O}$ in the interaction picture, and also we define the same operator in the Heisenberg representation as follows
\begin{equation}\label{representation}
\hat{O}_H\equiv e^{iHt}\hat{O}(0)e^{-iHt} ,\ \  \hat{O}_{\text{int}}\equiv e^{iH_0t}\hat{O}(0)e^{-iH_0t}
\end{equation}
Now, to proceed we consider the correlation of the general operators $\hat{O}_H^{(1)}(t)$ and $\hat{O}_H^{(2)}(t')$, The Eq.~\eqref{TimeOrdering-kq}, alongside with the Eqs.~\eqref{Wick2} and \eqref{representation} gives rise the following important identity
\begin{equation}
    \begin{aligned}
        &\frac{\langle\Psi_0|\mathcal{T}\left[\hat{O}_H^{(1)}(t)\hat{O}_H^{(2)}(t')\right]|\Psi_0\rangle}{\langle\Psi_0|\Psi_0\rangle}
        \\&
        ~=\frac{\Theta(t-t')}{\langle\Phi_0|\hat{S}|\Phi_0\rangle}\langle\Phi_0|\sum^\infty_{\nu=0}\left(-i\right)^\nu\frac{1}{\nu!}\int^\infty_{-\infty}dt_1...\int^\infty_{-\infty}dt_\nu \\&~~~~\times T\left[\hat{H}_{\text{int}}(t_1)...\hat{H}_{\text{int}}(t_\nu)\hat{O}_{\text{int}}^{(1)}(t)\hat{O}_{\text{int}}^{(2)}(t')\right]|\Phi_0\rangle\\&
        ~+\frac{(kq)^m\Theta(t'-t)}{\langle\Phi_0|\hat{S}|\Phi_0\rangle}\langle\Phi_0|\sum^\infty_{\nu=0}\left(-i\right)^\nu\frac{1}{\nu!}\int^\infty_{-\infty}dt_1...\int^\infty_{-\infty}dt_\nu
        \\&~~~~\times T\left[\hat{H}_{\text{int}}(t_1)...\hat{H}_{\text{int}}(t_\nu)\hat{O}_{\text{int}}^{(1)}(t')\hat{O}_{\text{int}}^{(2)}(t)\right]|\Phi_0\rangle.
    \end{aligned}
    \label{Eq:genSMatrix}
\end{equation}
This expression tends to the correct limit for conventional $S$-matrix expansion when $q\to 1$. We will use this identity for the Wick's theorem and to develop Feynman diagram techniques in the following sections.
\subsection{Generalized Wick's theorem} 
In this section, we introduce a generalization for the Wick's theorem, which is used for the diagrammatic expansion in the following sections. This is particularly important for calculating the many-body response functions for $kq$-particles.
As a standard approach, a general field $\hat{\Psi}(x)$ describing $kq$-particles is decomposed into two (annihilation and creation) parts as follows
\begin{equation}
\hat{\Psi}(x)\equiv A\hat{\psi}^+(x)+B\hat{\psi}^-(x),
\end{equation}
where $A$ and $B$ are constant complex coefficients.
Furthermore, $\hat{\psi}^-(x)$ and $\hat{\psi}^+(x)$ imply on the creation and annihilation parts of the field operator, respectively.
These operators are normally defined by their operation on the non-interacting ground state in quantum many body systems, or with respect to the non-interacting vacuum state in quantum field theories, such that:
\begin{equation}
\hat{\psi}^+(x)\left| \Phi_0\right\rangle=0,\ \{\hat{\psi}^-\}^{\dagger}(x)\left| \Phi_0\right\rangle=0.
\end{equation}
For the mode expansion represented in Eq.~\eqref{mode-expantion-kqParticle}, one finds
\begin{equation}
\begin{split}
&\hat{\psi}^-(x)=\frac{1}{\sqrt{V}}\sum_p e^{-ip.x} a_p^{\dagger}\\
&\hat{\psi}^+(x)=\frac{1}{\sqrt{V}}\sum_p e^{ip.x} a_p.
\end{split}
\end{equation}
Note also that for the general case, there is an extra term corresponding to creation or annihilation of anti-particles. The normal ordering of $n$ $kq$-particle fields is then defined based on these components, so that all the annihilation operators $\hat{\psi}^+(x)$ are brought to the right and all creation operators $\hat{\psi}^-(x)$ are brought to the left. Consider $n$ fields, for $m$ of them $\hat{\Psi}=\hat{\psi}^-$ ($A=0$ and $B=1$), while for the remaining $\hat{\Psi}=\hat{\psi}^+$ ($A=1$ and $B=0$). Therefore the normal ordering is defined as:
\begin{equation}
    \begin{aligned}
        &\mathcal{N}\{\hat{\Psi}_1(x_1)\hat{\Psi}_2(x_2)...\hat{\Psi}_n(x_n)\}\\&=(kq)^\mathcal{P}\{\hat{\psi}^-(x_{i_1})...\hat{\psi}^-(x_{i_m})\hat{\psi}^+(x_{j_1})...\hat{\psi}^+(x_{j_{n-m}})\}
    \end{aligned}
\end{equation}
where the subscript sets $(i_1,...,i_m)$, and $(j_1,...,j_{n-m})$ represent the positions of the creation and annihilation operators, respectively.
Besides, $\mathcal{P}$ denotes the number of permutations of $kq$-particles. A normal-ordered product of field operators is especially convenient because its expectation value in the unperturbed ground state vanishes 
identically.
\\
The other important operation needed for the Wick's theorem is contraction of two operators $\hat{\psi}(x_1)$ and $\hat{\psi}(x_2)$, defined as the difference between the $\mathcal{T}$ product and the $\mathcal{N}$ product:
\begin{equation}\label{UV}
    \overparent{\hat{\Psi}_1(x_1)\hat{\Psi}_2(x_2)}\equiv \mathcal{T}\left(\hat{\Psi}_1(x_1)\hat{\Psi}_2(x_2)\right)-\mathcal{N}\left(\hat{\Psi}_1(x_1)\hat{\Psi}_2(x_2)\right).
\end{equation}
This quantity is proven to be
\begin{equation}
    \begin{aligned}
        \overparent{\hat{\Psi}_1(x_1)\hat{\Psi}_2(x_2)}&=\langle\Phi_0|\mathcal{T}\left[\hat{\Psi}_1(x_1)\hat{\Psi}_2(x_2)\right]|\Phi_0\rangle.
    \end{aligned}
\end{equation}
Now we introduce the generalized Wick's theorem for $kq$-particles, which is obtained to be 
\begin{widetext}
    \begin{equation}\label{WickTheorem}
    \begin{aligned}
        \mathcal{T}\left\lbrace\hat{\Psi}_1(x_1)\hat{\Psi}_2(x_2)...\hat{\Psi}_n(x_n)\right\rbrace
        =&\mathcal{N}\Big\{\hat{\Psi}_1(x_1)\hat{\Psi}_2(x_2)...\hat{\Psi}_n(x_n)\Big\}+\overparent{\hat{\Psi}_1(x_1)\hat{\Psi}_2(x_2)}\mathcal{N}\Big\{\hat{\Psi}_3(x_3)...\hat{\Psi}_n(x_{n})\Big\}\\
        +kq\overparent{\hat{\Psi}_1(x_1)\hat{\Psi}_3(x_3)}
        \mathcal{N}\Big\{\hat{\Psi}_2(x_2)...\hat{\Psi}_n(x_{n})\Big\}&+...+(kq)^{n-2}\overparent{\hat{\Psi}_1(x_1)\hat{\Psi}_n(x_n)}\mathcal{N}\Big\{\hat{\Psi}_3(x_3)...\hat{\Psi}_n(x_{n-1})\Big\}+\text{all pair contractions}\\
        +...+(kq)^{\mathcal{P}_{i,j;i',j'}}\overparent{\hat{\Psi}_i(x_i)\hat{\Psi}_j(x_j)}&\overparent{\hat{\Psi}_{i'}(x_{i'})\hat{\Psi}_{j'}(x_{j'})}\mathcal{N}\Big\{\hat{\Psi}_1(x_1)...\hat{\Psi}_n(x_{n})\Big\}_{i,j,i',j'}+\text{all possible higher order contractions},
    \end{aligned}
\end{equation}
\end{widetext}
where $\mathcal{P}_{i,j;i',j'}$ is the number of permutations to pair $i$ and $j$, and also $i'$ and $j'$, and $\{\}_{i,j,i',j'}$ represents the set of all fields other than $i$, $j$, $i'$ and $j'$. We will use this theorem in the following sections, especially in designing the Feynman diagrammatics. 

\section{SINGLE-PARTICLE GREEN'S FUNCTION}\label{Green Functions}
In this section, we aim to perform the detailed calculation of single-particle Green's function of $kq$-particles.
The result is used in the construction of the Feynman diagrammatic rules, which are depicted in the following sections. The Green function can be defined either based on the $kq$-operators, or in terms of the $kq$-occupation number. These two methods converge to a same expression as $q\to 1$. The Hamiltonian of the free system can be considered as either Eq.~\eqref{Eq:freeHamiltonian} or the following one
\begin{equation}
\hat{K}_{kq}\equiv \hat{K}^{(0)}_{kq}+\hat{H}_{\text{int}}
\end{equation}
where $K^{(0)}_{kq}$ exhibits the Hamiltonian in non-interacting limit, while $\hat{H}_{\text{int}}$ represents the interaction part (including four $a$ and $a^{\dagger}$ fields).
In our notation,  $\left\langle ... \right\rangle $ implies either finite- or zero-temperature averaging.
More precisely, for finite temperature system the average of an operator $\hat{O}$ is expressed as
\begin{equation}
\left\langle \hat{O}\right\rangle\equiv\text{Tr}\left[\hat{O}e^{-\beta (\hat{K}_{kq}-\Omega_{kq})}\right],
\label{Eq:average1}
\end{equation}
while for a zero temperature one, we have
\begin{equation}
\left\langle \hat{O}\right\rangle=\frac{\langle\Psi_0|\hat{O}|\Psi_0\rangle}{\langle\Psi_0|\Psi_0\rangle}.
\label{Eq:average2}
\end{equation}
$\hat{K}^{(0)}_{kq}$ is either given by Eq.~\eqref{Eq:freeHamiltonian} (represented by $K_{kq}^{\text{free}}$) or the following expression
\begin{equation}
\begin{split}
\hat{K}^{(0)}_{kq}\equiv \hat{H}^{(0)}_{kq}-\mu [\hat{N}]&=\sum_{\nu}(\epsilon_{\nu}-\mu)[\hat{N}_{\nu}]\\
&=\sum_{\nu}\zeta_{\nu}a_{\nu}^{\dagger}a_{\nu},
\label{Eq:free2Hamiltonian}
\end{split}
\end{equation}
where $\epsilon_{\nu}$ represents the single particle energy. 
The key difference between these two choices lies in the basis of the occupation number operator: in Eq.~\eqref{Eq:freeHamiltonian}, it is expressed in the normal basis, whereas in the equation above, it is represented in the $q$ basis.
Some references use the above definition~\eqref{Eq:free2Hamiltonian} for the free particles systems, like~\cite{chaichian1993statistics}, while the others use Eq.~\eqref{Eq:freeHamiltonian}, see for instance~\cite{tuszynski1993statistical}. Although two Hamitonians converge to each other in the limit $q\to 1$, the $\hat{H}_{kq}^{\text{free}}$ describes \textit{real} free particles (dealing directly with $N_{\nu}$), while the Eq.~\eqref{Eq:free2Hamiltonian} is more complicated in the sense that the Hamiltonian is related to $N_{\nu}$ in a complicated way. Note that Eq.~\eqref{ٍEq:distribution-function} has been obtained based on the Hamiltonian Eq.~\eqref{Eq:freeHamiltonian} to consider both definitions for calculating the Green function. In the next subsection we use $\hat{K}_{kq}$ to define the Green function, while in Sec.~\eqref{SEC:effectiveG} we employ the definition given in Eq.~\eqref{Eq:freeHamiltonian}. In the former case we directly use Eq.~\eqref{Eq:free2Hamiltonian}  whose result is called the \textit{standard Green function}, while the latter case is based on the occupation number given in  Eq.~\eqref{ٍEq:distribution-function}, which is known as the \textit{effective Green function}.
\subsection{The standard single-particle Green's function}
The objective of this section is to analyze the Green's function for $kq$ particles sketched in Eq.~\eqref{Eq:Green0}, based on the Hamiltonian mentioned in Eq.~\eqref{Eq:free2Hamiltonian} using the annihilation and creation operators obeying the commutation relations(Eqs.~(\ref{Eq:algebra}) and~(\ref{Eq:algebra2})). From now on, we consider finite temperature Green's function with $\beta\equiv1/k_BT$.
It should be reminded that  Eq.~(\ref{Eq:Green0}) shows a zero-temperature Green's function.
The standard single $kq$-particle finite temperature Green's function is given by
\begin{equation}
	iG_{kq}^{(X)}(\nu;t,t')=\frac{1}{Z_{kq}}\text{Tr}\left[e^{-\beta \hat{K}_{kq}}a_{\nu}(t)a_{\nu}^{\dagger}(t')\right],
\end{equation}
with $a_{\nu}(t)\equiv e^{itH}a_{\nu}e^{-itH}$, and 
\begin{equation}
Z_{kq}\equiv e^{-\beta\Omega_{kq}}=\text{Tr}\left[e^{-\beta \hat{K}_{kq}}\right],
\end{equation}
is the partition function of the system, and $\Omega_{kq}$ is the Helmholtz free energy. The Matsubara Green function is defined by the following expression 
\begin{equation}
\begin{split}
&iG_{kq}^{(\mathcal{M},X)}(\nu;\tau,\tau') =-\left\langle \mathcal{T}_{\tau}a_{\nu}(\tau)a_{\nu}^{\dagger}(\tau')\right\rangle\\
&=-\text{Tr}\left[e^{-\beta (\hat{K}_{kq}-\Omega_{kq})}\mathcal{T}_{\tau}e^{\tau \hat{K}_{kq}}a_{\nu} e^{-(\tau-\tau')\hat{K}_{kq}}a_{\nu}^{\dagger}e^{-\tau' \hat{K}_{kq}}\right],
\end{split}
\label{Eq:greentau}
\end{equation}
in which $\tau\equiv it$ is the Wick's rotated (imaginary) time, and $\mathcal{T}_{\tau}$ is the time ordering operator which is in accordance with the definition Eq.~(\ref{kq-Time-ordering}).
In the notation presented above, the superscript $\mathcal{M}$ is used to emphasize our focus on the \textit{Matsubara} formalism.
Note that the retarded zero-temperature Green's function is obtained by a simple analytic continuation of the Matsubara green function.

In our notation, we introduce the bare single-particle Green's function of a non-interacting system by $G^{(\mathcal{M},X,0)}_{kq}$ using the Hamiltonian defined by Eq.~(\ref{Eq:free2Hamiltonian}).
Using the generalized Wick's theorem,  we can formulate the standard Dyson equation connecting the full single-particle Green's function to the bare one ($G^{(\mathcal{M},X,0)}_{kq}$).
The full single-particle Green's function is given by
(see Eq.~\eqref{kq-Time-ordering}):
\begin{equation}
\begin{aligned}
    iG^{(\mathcal{M},X,0)}_{kq}(\nu,\tau)
    =&-\Theta(\tau)\left\langle a_{\nu}(\tau)a^\dagger_{\nu}(0)\right\rangle\\
    &-kq\Theta(-\tau)\left\langle a^\dagger_{\nu}(0)a_{\nu}(\tau)\right\rangle.
    \label{GF}
\end{aligned}
\end{equation}
To this end, we employ the strategy of the Gorkov equation, which is derived from the imaginary-time derivative of the Green's function (see Eq.~\eqref{Eq:appGreen}) as follows:
\begin{equation}
    \begin{aligned}
        &i\frac{\partial}{\partial\tau} G^{(\mathcal{M},X,0)}_{kq}(\nu,\tau)
        \\&
        =-\delta(\tau)\left\langle \left[a_{\nu}(\tau),a^\dagger_{\nu}(0)\right]_{kq}\right\rangle-\left\langle \mathcal{T}^{}_{\tau} \frac{\partial a_{\nu}(\tau)}{\partial\tau}a^\dagger_{\nu}(0)\right\rangle,
        \label{38}
    \end{aligned}
\end{equation}
The first term can take a non-zero value only when $\tau=0$, which according to Eq.~(\ref{mode-expantion-kqParticle}), results in the value $q^{-N_{\nu}}$.
For the second term, the equation of motion approach, $d A(\tau)/d\tau = \left[H, A(\tau)\right]$ is applied.
Employing the time dependence of $a_{\nu}(\tau)$ using the Becker-Hausdorff theorem 
followed by the explicit expression for the  commutation of $\hat{H}_{kq}^{(0)}$ and $N_{\nu}$ (See Appendix \ref{Appendix:B}), one obtains
\begin{equation}
    \begin{aligned}
        i\frac{\partial}{\partial\tau} G^{(\mathcal{M},X,0)}_{kq}&(\nu,\tau)
        =-\delta(\tau)\left\langle q^{-N_{\nu}}\right\rangle\\&-\left\langle \mathcal{T}^{}_{\tau} \zeta_{\nu}\Big(k(1-q)[N_{\nu}]-q^{-N_{\nu}}\Big)a_{\nu}(\tau)a^\dagger_{\nu}(0)\right\rangle.
        \label{43}
    \end{aligned}
\end{equation}
Performing the calculations in the generalized Matsubara frequency space ($\omega_n^{(X,j)}$) would be facilitating the investigation.
In the next subsection we derive explicitly these frequencies using the deformed occupation numbers. This expansion is given by the following expresstion
\begin{equation}
iG^{(\mathcal{M},X)}_{kq}(\nu,\tau)=\frac{1}{\beta}\sum_{j,\omega_n^{(X,j)}}iG^{(\mathcal{M},X,0)}_{kq}(\nu,i\omega_n^{(X,j)})e^{-i\omega_n^{(X,j)}\tau},
\label{Eq:GreenTau}
\end{equation}
so that $\frac{\partial}{\partial \tau}\to i\omega_n^{(X,i)}$. Here, $j=1,2$, and the generalized Matsubara frequencies $i\omega_n^{(X,j)}$ are defined as 
  \begin{equation}\label{Eq:Matsubara}
  i\omega_n^{(X,j)}=i\omega_n^{(X)}+\frac{k_X(-1)^j}{\beta}\ln q,
  \end{equation}
  wherein $\omega_n^{B}\equiv \frac{2n\pi }{\beta}$, $\omega_n^{F}\equiv \frac{(2n+1)\pi}{\beta}$ are ordinary Matsubara frequencies for bosons and fermions respectively. 
  Consequently, the final form of the bare single-particle Matsubara Green's function is obtained by
\begin{equation}
   \begin{aligned}
       & iG^{(\mathcal{M},X,0)}_{kq}(\nu,i\omega_n^{(X,j)})
       \\&=\frac{\left([kn^{}_{\nu}+1]-kq[n^{}_{\nu}]\right)^{k}_{}}{i\omega_n^{(X,j)}+\zeta_{\nu}\Big(k(1-q)[n^{}_{\nu}]-\left([kn^{}_{\nu}+1]-kq[n^{}_{\nu}]\right)^{k}_{}\Big)}.
   \end{aligned}
    \label{eq29}
\end{equation}
\begin{figure}
    \centering
    \includegraphics[width=1\linewidth]{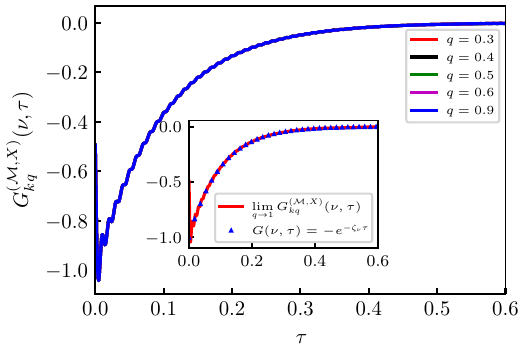}
    \caption{\justifying The standard single-particle Green’s function given in Eq.~\eqref{Eq:GreenTau} in terms of $\tau$ for various rates of $q$. The difference among the Green functions are small, but non-zero. Inset shows the result for $q=1$, and a comparison with the standard result for the normal fermions $G(\epsilon_{\nu},\tau)=-\exp(-\zeta_{\nu}\tau)$.}
    \label{fig:SecondGFqFermion}
\end{figure}
By simplifying the numerator of the Eq.~\eqref{eq29}, we find that $\left([kn^{}_{\nu}+1]-kq[n^{}_{\nu}]\right)^{k}_{}$ equals $q^{-n_\nu}$ for both values of $k=\pm 1$. This leads to the following simplified Green's function:
\begin{equation}
   \begin{aligned}
       iG_{kq}^{(\mathcal{M},X,0)}(\nu,i\omega_n^{(X,j)})
       =\frac{q^{-n^{}_\nu}_{}}{i\omega_n^{(X,j)}+\zeta_\nu\Big(k(1-q)[n^{}_\nu]-q^{-n^{}_\nu}_{}\Big)}.
   \end{aligned}
\end{equation}
We plotted the Green's function $G_{kq}^{(\mathcal{M},X)}(\nu,\tau)$ given by Eq.~\eqref{Eq:GreenTau} as a function of $\tau$ for various amounts of $q$ in Fig.~\ref{fig:SecondGFqFermion}. The difference among different values of $q$ is pretty small. In the inset we compare it with the Green's function of ordinary fermions with the analytic function $G^{(\mathcal{M},F)}_{-1,1}(\tau)=-e^{\zeta_{\nu}\tau}$ with a perfect match. Note that for a translational invariant system, it is more convenient to do calculations in the momentum space labelled by $\textbf{p}$.
\subsection{The effective single-particle Green's function}\label{2}
\subsubsection{Poles of the occupation number}\label{SEC:effectiveG}
There are various different ways of constructing the Green's function. For normal fermions and bosons these different methods result in the same expression, while for the $kq$-particles they are different. In the previous section we used the most direct one based on the annihilation and creation operators, appropriate for the Hamiltonian of the type mentioned in Eq.~\eqref{Eq:free2Hamiltonian}. In this section we use another technique, based on the occupation number defined in Eq.~\eqref{ٍEq:distribution-function}, corresponding to the non-interaction hamiltonin Eq.~\eqref{Eq:freeHamiltonian}.\\

In this method, we first consider the relevant distribution function and then we try to find the poles of this function using the pole expansion of meromorphic functions.
 Let us introduce
   \begin{equation}\label{fz}  		f(z)=\frac{1}{q-q^{-1}}\ln\left(\sigma_{kq}(z)\right),
  	\end{equation}
where $\sigma_{kq}(z)\equiv\frac{e^{\beta z}-k q^{-k}}{e^{\beta z}-k q^{k}}$ so that, according to Eq.~\eqref{ٍEq:distribution-function} $\bar{n}_{kq}(\zeta_{\nu})\equiv f(\epsilon^{}_\nu-\mu)$. To expand the function in terms of the simple roots, we use the Mittag-Leffler theorem~\cite{turner2013mittag},  which states that for a function $g(z)$ with simple poles at $z_n$, $n=1,...,N$ ($N$ may be infinity) we have the following expansion
  	\begin{equation}
  		g(z)=g(0)+\sum_{n=1}^{\infty}b_n\left(\frac{1}{z-z_n}+\frac{1}{z_n}\right),
        \label{pole}
   	\end{equation}
   	where $b_n=\lim\limits_{z\rightarrow z_n}(z-z_n)g(z)$. We use the following integral representation for the logarithm to expand $f(z)$:
    \begin{equation}\label{ln}
    \begin{split}
        f(z)=\int_{a}^z g_{kq}(z') dz'+\frac{1}{q-q^{-1}}\ln \left(\sigma_{kq}(a)\right),
    \end{split}
    \end{equation}
 where $a$ is an unimportant initial parameter, $g_{kq}(z')\equiv \left(\frac{1}{q-q^{-1}}\right)\frac{\sigma'_{kq}(z')}{\sigma_{kq}(z')}$, and $\sigma_{kq}'\equiv \frac{\text{d}}{\text{d}z'} \sigma_{kq}(z')$. One readily finds
    \begin{equation}
  	\begin{aligned}
  g_{kq}(z')= \left(\frac{1}{q-q^{-1}}\right)\frac{\beta e^{\beta z'}(-kq^k+kq^{-k})}{(e^{\beta z'}-kq^{-k})(e^{\beta z'}-kq^k)}.
  	\end{aligned}
    \label{Eq:integralExression}
  \end{equation}
    The poles of the $g_{kq}(z')$, defining the generalized Matsubara frequencies are obtained as follows
  \begin{equation}
  	(e^{\beta z'}-kq^{-k})(e^{\beta z'}-kq^k)=0,
  \end{equation}
which states that there are two independent groups or types of frequencies for each value of $k$ that we call $\omega_n^{(1)}$ and $\omega_n^{(2)}$, defined as the solutions of the following equations:
  \begin{equation}
  	    e^{\beta z_n^{(1)}}=e^{\beta i\omega_n^{(1)}}=kq^{-k}, e^{\beta z_n^{(2)}}=e^{\beta i\omega_n^{(2)}}=kq^{k}.
  \end{equation}
  The explicit forms of these generalized Matsubara frequencies then read as \begin{equation}\label{Eq:MatsubaraF1}
\begin{cases}
    i\omega_n^{(B,1)}=i\frac{(2n)\pi}{\beta}-\frac{\ln q}{\beta} &\text{if } k=1 \\
    i\omega_n^{(F,1)}=i\frac{(2n+1)\pi}{\beta}+\frac{\ln q}{\beta} &\text{if } k=-1
\end{cases} 
\end{equation}
and
\begin{equation}\label{Eq:MatsubaraF2}
\begin{cases}	i\omega_n^{(B,2)}=i\frac{(2n)\pi}{\beta}+\frac{\ln q}{\beta} &\text{if } k=1 \\
i\omega_n^{(F,2)}=i\frac{(2n+1)\pi}{\beta}-\frac{\ln q}{\beta} &\text{if } k=-1,
  	\end{cases} 
\end{equation}
which can be abbreviated in a single form as is given in Eq.~\eqref{Eq:Matsubara}.
We see that the generalized Matsubara frequencies are the classic ones endowed with an imaginary part. Notably, all the summations over the Matsubara frequencies should be performed over these two independent frequencies. Importantly, using Eq.~\eqref{pole}, $g_{kq}(z')$ can be expanded as	
	\begin{equation}
		g_{kq}(z')=\sum_{j,\omega_n^{(X,j)}}\frac{b_n^{(X,j)}}{z'-i\omega_n^{(X,j)}}+g^{(0)}_{kq},
		\label{Eq:It}
		\end{equation}
and
\begin{equation}
\begin{split}
g^{(0)}_{kq}&\equiv \tilde{g}^{(0)}_{kq}+\sum_{j,\omega_n^{(X,j)}}\frac{b_n^{(X,j)}}{i\omega_n^{(X,j)}},\\
\tilde{g}^{(0)}_{kq} &\equiv \frac{-\beta k (q^k-q^{-k})}{(q-q^{-1})(1-kq^{-k})(1-kq^k)}.
\end{split}
\end{equation}
   In these equations the residues read
   \begin{equation}       b_n^{(X,j)}=\lim\limits_{z'\rightarrow i\omega_n^{(X,j)}}\left(z'-i\omega_n^{(X,j)}\right)g_{kq}(z')=\frac{(-1)^{j+1}}{q-q^{-1}},
   \end{equation}
  which are independent of $k$. Substituting Eq.~\eqref{Eq:It} into Eq.~\eqref{ln}, and setting $z=\zeta_{\nu}$ one finds
  \begin{equation}\label{gw}
			\begin{aligned}
                \bar{n}_{kq}(\zeta_\nu)=&\frac{1}{q-q^{-1}}	\sum_{j,\omega_n^{(X,j)}}(-1)^{j+1}\ln\left[\frac{\zeta_{\nu}-i\omega_n^{(X,j)}}{a-i\omega_n^{(X,j)}}\right]\\
                &+g^{(0)}_{kq}(\zeta_{\nu}-a)+\frac{1}{q-q^{-1}}\ln\frac{e^{\beta a}-kq^{-k}}{e^{\beta a}-kq^{k}}. 
			\end{aligned}
  \end{equation}
  We notice that, as can be seen from Eq.~\eqref{ln}, the right hand side of the above equation is independent of $a$ which serves as the \textit{lowest energy scale} of the system. In the following sections we ignore the last unimportant constant. We use this relation to derive the effective Green function in the next section. 
  \subsubsection{Effective single particle Green's function}
  This subsection is devoted to the calculation of the effective green function $\mathcal{G}_{kq}^{(0)}(\epsilon_{\nu},\tau)$ defined by an analogy with the classical (fermionic or bosonic) definition. The superscript $(0)$ shows that we are considering a non-interacting system. Now we use the identity~\cite{fetter2012quantum}
  \begin{equation}
  \begin{aligned}
      \bar{n}_{kq}(\zeta_\nu)&=-\frac{1}{kq}\lim_{\tau\to 0^-}i\mathcal{G}^{(0)}_{kq}(\zeta_\nu,\tau)\\&
      =-\frac{1}{kq}\lim_{\tau\to 0^-}\sum_{j,\omega_n^{(X,j)}}i\tilde{\mathcal{G}}^{(0)}_{kq}(\zeta_\nu,i\omega_n^{(X,j)})e^{i\omega_n^{(X,j)}\tau},
  \end{aligned}
\label{Eq:green-directSpace}
  \end{equation}
where $\tilde{\mathcal{G}}^{(0)}_{kq}(\zeta_\nu,i\omega_n^{(X,j)})$ is a component of the single-particle effective Green's function in the frequency space. This relation helps much to read the bare Green's functions in the frequency space and gives rise to
 \begin{equation}\label{Eq:Gwn}
  \begin{split}
      i\tilde{\mathcal{G}}^{(0)}_{kq}(\zeta_\nu,i\omega_n^{(X,j)})&=\frac{(-1)^{j+1}}{q-q^{-1}}\log\left[\frac{\zeta_{\nu}-i\omega_n^{(X,j)}}{a-i\omega_n^{(X,j)}}\right] \\
      &+g^{(0)}_{kq}(\zeta_{\nu}-a).
  \end{split}	 
  \end{equation}
It is obvious that the bosonic and fermionic Green's functions can be retrieved in the limits $q\to1 , k\to1$ and $k\to-1$ respectively.

To find the effective single-particle Green's function in the direct space ($\mathcal{G}_{kq}(\epsilon_\nu,\tau)$), we should find the summation over the Matsubara frequencies according to the Eq.~\eqref{Eq:green-directSpace}. An alternative way for doing so is to use Eqs.~\eqref{Eq:integralExression} and~\eqref{Eq:It}, and multiply the integrand by $\exp[t\tau]$, so that after the integration and using the proper limits, $\exp[i\omega_n^{(X,j)}\tau]$ appears in the summand. The explicit form of the effective Green function is then obtained as follows 
\begin{equation}
\begin{aligned}
        i\mathcal{G}^{(0)}_{kq}(\epsilon^{}_\nu,\tau)=&\int^{\zeta_{\nu}}_{a} g_{kq}(t) e^{t \tau}dt\\
        =&\int^{\zeta_{\nu}}_{a} \frac{d}{dt}\left(\bar{n}_{kq}(t)e^{t\tau}\right) dt-\tau\int^{\zeta_{\nu}}_{a}\bar{n}_{kq}(t) e^{t\tau} dt\\
        =&\bar{n}_{kq}(t)e^{t\tau}\Biggr|_{a}^{\zeta_{\nu}} -\tau \mathscr{L}^{-1}_{a,\zeta_{\nu}}\left(\bar{n}_{kq}(t)\right),
\end{aligned}
\label{Eq:EffcGreen1}
\end{equation}
where $\mathscr{L}^{-1}_{a,b}\left(y\right)$ is the inverse Laplace transform with the bounds $a$ and $b$. Also, when we obtain the effective Green function by expanding it around the poles (Eq.~\eqref{Eq:Gwn}), we can write it as  
\begin{equation}\label{green0}          
  \begin{aligned}
      i\mathcal{G}^{(0)}_{kq}(\zeta_\nu,\tau)&=\sum_{j,\omega_n^{(X,j)}}i\tilde{\mathcal{G}}^{(0)}_{kq}(\zeta_\nu,i\omega_n^{(X,j)}) e^{-i\omega_n^{(X,j)}\tau}.
  \end{aligned}
\end{equation}
Note also that the effective Green's function show the following property as one under the transformation $\tau\to\tau+\beta$ for $\tau<0$ (see appendix~\ref{G-tau-beta})
\begin{equation}\label{taubeta}
    \mathcal{G}_{kq}(\nu,\tau+\beta)=k q^{k(-1)^j}_{}\mathcal{G}_{kq}(\nu,\tau),
\end{equation}
where $\mathcal{G}_{kq}(\nu,\tau)$ is the full effective Green's function for an arbitrary (interactive) $kq$-system. Although it is periodic (anti-periodic) for bosons (fermions), for the general case it depends on $k$ and $q$. This property is also valid for the stadrad Green's function $G^{(\mathcal{M},X)}_{kq}$.\\
\begin{figure}
    \centering
    \includegraphics[width=1\linewidth]{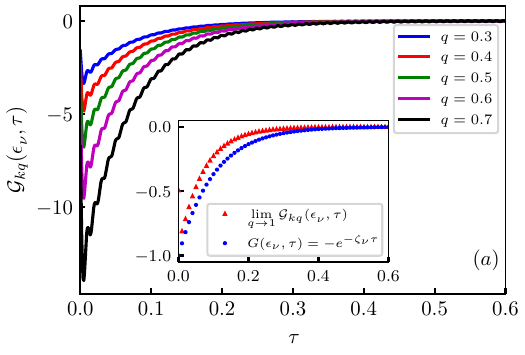}
    \includegraphics[width=1\linewidth]{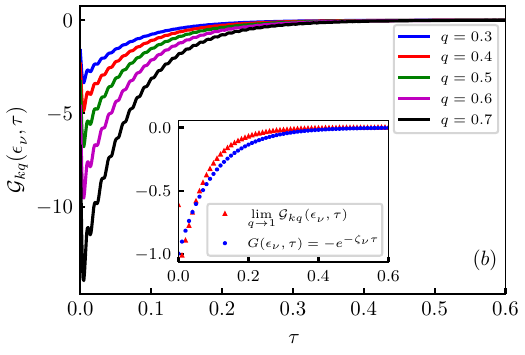}
    \caption{\justifying The effective single-particle Green’s function given in Eq.~\eqref{green0} in terms of $\tau$ for $\beta=1$ and (a) $k=1$ and (b) $k=-1$. Inset shows the result for $q=1$. There is a difference with the normal Green's function $G(\epsilon_{\nu},\tau)=-\exp(-\zeta_{\nu}\tau)$, which is due to neglecting the last constant term in Eq.~\eqref{gw}.}
    \label{fig:Green's function diagram}
\end{figure}
The Gorkov equation for the effective Green's function is obtained using a $\tau$ differentiation, which reads (see appendix~\ref{app:effective})
\begin{equation}
    \begin{aligned}
        \partial_\tau i\mathcal{G}^{(0)}_{kq}(\zeta^{}_\nu,\tau)+\zeta_\nu i\mathcal{G}^{(0)}_{kq}(\zeta^{}_\nu,\tau)&-\int_a^{\zeta^{}_\nu} i\mathcal{G}^{(0)}_{kq}(x,\tau)dx\\
        &=(\zeta^{}_\nu-a)\frac{q^{\tfrac{-\tau}{\beta}}-q^{\tfrac{\tau}{\beta}}}{q-q^{-1}}\delta(\tau), 
    \end{aligned}
    \label{Eq:EffecGreenGorkov}
\end{equation}
which is an integro-differential equation.
\section{DIAGRAMMATIC EXPANSION AND PERTURBATION THEORY FOR $kq$-particles}\label{DIAGRAMMATICkqParticles}
For developing diagrammatic techniques, we need to generalize Feynman rules using the the generalized Wick's teorem and the $S$-Matrix expansion we already introduced. In the following, we will present the relevant Feynman rules based on the $kq$-algebra.

The fundamental building block of the generalized Feynman diagrams is a solid line corresponding to a contraction of fields at two different space-time points $x_1$ and $x_2$, represented as follows:
\begin{center}
\begin{tikzpicture}[scale=1]
  \draw[black, thick] (0,0) -- (2,0);
  \draw[black, fill=black] (0,0) circle (0.08cm);
  \draw[black, fill=black] (2,0) circle (0.08cm);
  \node[above] at (0,0) {$x_1$};
  \node[above] at (2,0) {$x_2$};
\end{tikzpicture}    
\end{center}
which corresponds to the following contraction
\begin{equation}
    \begin{aligned}
        \langle \mathcal{T}\left\lbrace\hat{\psi}(x_1)\hat{\psi}(x_2)\right\rbrace\rangle=\overparent{\hat{\psi}(x_1)\hat{\psi}(x_2)}.
        \label{DF2}
    \end{aligned}
\end{equation}
The higher order contractions correspond to higher order diagrams composed of bold lines, representing fermionic or anti-fermionic propagators, and wavy lines representing the interactions. We consider the example of a four field term with all possible contractions as follows:
\begin{equation}
    \begin{aligned}
        &\langle \mathcal{T}\left\lbrace\hat{\psi}(x_1)\hat{\psi}(x_2)\hat{\psi}(x_3)\hat{\psi}(x_4)\right\rbrace\rangle
        \\&=\overparent{\hat{\psi}(x_1)\hat{\psi}(x_2)}\overparent{\hat{\psi}(x_3)\hat{\psi}(x_4)}+kq \overparent{\hat{\psi}(x_1)\hat{\psi}(x_3)}\overparent{\hat{\psi}(x_2)\hat{\psi}(x_4)}
        \\&+(kq)^2\overparent{\hat{\psi}(x_1)\hat{\psi}(x_4)}\overparent{\hat{\psi}(x_2)\hat{\psi}(x_3)},
    \end{aligned}
\end{equation}
We observe that the contractions resemble those of ordinary fermions and bosons, with an additional factor of $kq$ appearing when the fields are exchanged. The diagrammatic expansion of this term reads respectively:
\begin{equation}
\begin{aligned}
  \begin{tikzpicture}[scale=1]
  \draw[black, thick] (0,0) -- (2,0);
  \draw[black, fill=black] (0,0) circle (0.08cm);
  \draw[black, fill=black] (2,0) circle (0.08cm);
  \node[above] at (0,0) {$x_1$};
  \node[above] at (2,0) {$x_2$};
  \draw[black, thick] (0,-1.3) -- (2,-1.3);
  \draw[black, fill=black] (0,-1.3) circle (0.08cm);
  \draw[black, fill=black] (2,-1.3) circle (0.08cm);
  \node[below] at (0,-1.3) {$x_3$};
  \node[below] at (2,-1.3) {$x_4$};
   \end{tikzpicture}~~~~
\begin{tikzpicture}[scale=1]
  \draw[black, thick] (1,1) -- (1.3,1);
  \draw[black, thick] (1.15,1.15) -- (1.15,0.85);
  \draw[white, thick] (1.15,.1) -- (1.15,-.1);
\end{tikzpicture}
   \begin{tikzpicture}
  \draw[black, thick] (0,0.3) -- (0,-1.2);
  \draw[black, fill=black] (0,0.3) circle (0.08cm);
  \draw[black, fill=black] (0,-1.2) circle (0.08cm);
  \node[left] at (0,0.3) {$x_1$};
  \node[left] at (0,-1.2) {$x_3$};
  \draw[black, thick] (2,0.3) -- (2,-1.2);
  \draw[black, fill=black] (2,0.3) circle (0.08cm);
  \draw[black, fill=black] (2,-1.2) circle (0.08cm);
  \node[right] at (2,0) {$x_2$};
  \node[right] at (2,-1.2) {$x_4$};
\end{tikzpicture}~~~~\\
\begin{tikzpicture}[scale=1]
  \draw[black, thick] (1,1) -- (1.3,1);
  \draw[black, thick] (1.15,1.15) -- (1.15,0.85);
  \draw[white, thick] (1.15,.1) -- (1.15,-.1);
\end{tikzpicture}
  \begin{tikzpicture}
  \draw[black, thick] (0,0) -- (2,-1.3);
  \draw[black, fill=black] (0,0) circle (0.08cm);
  \draw[black, fill=black] (0,-1.3) circle (0.08cm);
  \node[left] at (0,0) {$x_1$};
  \node[left] at (0,-1.3) {$x_3$};
  \draw[black, thick] (2,0) -- (0,-1.3);

  \draw[black, fill=black] (2,0) circle (0.08cm);
  \draw[black, fill=black] (2,-1.3) circle (0.08cm);
  \node[right] at (2,0) {$x_2$};
  \node[right] at (2,-1.3) {$x_4$};
\end{tikzpicture}~~~~
\end{aligned}
\end{equation}
Now we include an interaction term containing wavy line. The general form of the interaction for spinless $kq$-particles is
\begin{equation}
    \begin{aligned}
\hat{H}_{\text{int}}=\frac{1}{2}\int \text{d}\textbf{x}\text{d}\textbf{x}'v_0(x,x')\hat{\psi}^\dagger(x)\hat{\psi}^\dagger(x')\hat{\psi}(x')\hat{\psi}(x),
\label{Eq:Interaction}
    \end{aligned}
\end{equation}
where for our later use we defined $x=(\tau,\textbf{x})$, where $\tau\equiv 0$ above, but is a free parameter in the perturbative expansion, and $v_0(x,x')$ is the bare, simultaneous interaction potential between the spatial points $\textbf{x}$ and $\textbf{x}'$ (proportional to $\delta(\tau-\tau')$). The spin degree of freedom should be included for spinful particles. The perturbative expansion of the Green's function reads (Appendix~\ref{Appendix:DF})
\begin{equation}
    \begin{aligned}
        iG^{(\mathcal{M},X)}_{kq}(x,y)=\sum_{m=0}^{\infty}\frac{(-i)^m}{m!}\int_{0}^{\beta}d\tau_1...\int_{0}^{\beta}d\tau_m 
        &\\ \times\left\langle\mathcal{T}[\hat{H}_{\text{int}}(\tau_1)...\hat{H}_{\text{int}}(\tau_m)\hat{\psi}_\alpha(x)\hat{\psi}^\dagger_\beta(y)]\right\rangle.&
    \end{aligned}\label{Greenfunction_S}
\end{equation} 
A prototypical example is a transnational invariant system (jellenium model), in which the expansion of the interaction potential in the momentum space reads
\begin{equation}
    \hat{H}_{\text{int}}=\sum_{PP'Q}v_0(Q)a^{\dagger}_{P+Q}a^{\dagger}_{P'-Q}a^{}_{P}a^{}_{P'}.
\end{equation}
where $P,P'$ and $Q$ are some momentums, and $v(Q)$ is the interaction in the Fourier space and $Q$ momentum. As the interaction term connects two fermionic nodes, the abve equation guarantees the momentum conservation. One has to insert the following component in the diagrammatic expansion:
\begin{equation}
	\begin{tikzpicture}[baseline={(current bounding box.center)}]
		\begin{feynman}
			\vertex (a1) at (-0.8,1.2) {$P$};
			\vertex (a2) at (-0.2,0) {};
			\vertex (a3) at (-0.8,-1.2) {$P+Q$};
			\vertex (a4) at (1.4,0) {};
			\vertex (a5) at (2,1.2) {$P'$};
			\vertex (a6) at (2,-1.2) {$P'-Q$};
			\vertex (b1) at (0,-0.13) {};
			\vertex (b3) at (0,0.13) {};
			\vertex (b5) at (1.2,-0.13) {};
			\vertex (b6) at (1.2,0.13) {};
			\diagram{(b1) -- [fermion,edge , line width=0.8pt] (a1)};
			\diagram{(a3) -- [fermion,edge , line width=0.8pt] (b3)};
			\diagram{(a2) -- [photon,edge , line width=0.8pt,edge label'=$Q$] (a4)};
			\diagram{(b5) -- [fermion,edge , line width=0.8pt] (a5)};
			\diagram{(a6) -- [fermion,edge , line width=0.8pt] (b6)};
		\end{feynman}
	\end{tikzpicture}
\end{equation}
where there are two in-states and two out-states, and the wavy line shows the interaction potential $v(Q)$.\\

In calculations related to the Green's function, where contraction terms appear along with the Green's function operators, we will have the terms like
\begin{equation}
    \langle\mathcal{T}\{\underbrace{\hat{\psi}^\dagger(x_1)\hat{\psi}^\dagger(x'_1)\hat{\psi}(x'_1)\hat{\psi}(x_1)}_{\text{Int}}\underbrace{\hat{\psi}(x)\hat{\psi}^\dagger(y)}_{\text{GF}}\}\rangle,
\end{equation}
where ``Int'' shows an interaction term in the perturbation theory, and ``GF'' represents the operators appearing in the definition of the Green's function. To relate these contributions to a Feynmann diagram, we use different contractions, each of which has an extra prefactor $(kq)^m$ with respect to the ordinary fermionic or bosonic expansions, where $m$ is the exchange of fields required for the contractions. Examples can be found in the appendix~\ref{Appendix:DF}. In the calculations regarding the Green's function $G^{(\mathcal{M},X)}_{kq}(x,y)$, one can check by inspection (see Eq.~\eqref{FDaigram} for example) that for each loop we have a prefactor $kq$, and the contraction between vertices $\textbf{x}$ and $\textbf{y}$ with internal vertices ($x_1\to x_1'$ for example) and also the contraction between the internal vertices themselves ($x_1\to x_1'$) results in an extra $kq$ multiplication. We denote the total number of these operations as $m'$ and the total number of loops as $m$.\\

The following rules describe the $n$th-order contribution to the 
single-particle Green's function 
\begin{itemize}
\item We draw all topologically distinct connected diagrams with $n$ interaction lines (like $v(Q)$ in the momentum space) and $2n+1$ directed Green's functions $G^{(\mathcal{M},X,0)}_{kq}$. 
 To each particle contraction we attribute a bold (particle) line, while each interaction is represented by a wavy line.
\item Label each vertex with a four-dimensional space-time point $x^{}_{i}$ in the real space (for the momentum space other quantum labels are attributed to vertices, and to each line a momentum is attributed).
\item Each solid line is a path from $x$ to $y$, showing the Green's function $G^{(\mathcal{M},X,0)}_{kq}(x,y)$. Every wavy line shows an instantaneous interaction $v_0(x,y)=v_0(\textbf{x},\textbf{y})\delta(t_x-t_y)$, where $v_0(\textbf{x},\textbf{y})$ is the bare interaction potential between two spatial points $\textbf{x}$ and $\textbf{y}$.

\item Integrate all internal variables over space and time.
\item To compute $G(x,y)$ affix a sign factor $(kq)^{m+m'}$ to each term, and assign a factor $(i/\hbar)^n$ to each $n$th-order term.
\end{itemize}

\section{AVERAGE ENERGY AND RESPONSE FUNCTIONS}\label{RESPONSE-FUNCTIONS}
This section is devoted to the perturbative expansion of the average energy, and also the dielectric constant in the random phase approximation (RPA). We consider translational invariant system, where the momentum is a good quantum number.
\subsection{The average energy}

To calculate the average energy of a finite temperature system, and also the ground state energy of the system, we need to calculate the (finite temperature and ground state) average of the interaction term Eq.~\eqref{Eq:Interaction}. We emphasis again that we use the notation $\left\langle \right\rangle $ for both the finite temperature average, and the ground state expectation value, having in mind that for a finite temperature system the average of the operator $\hat{O}$ is given in Eq.~\eqref{Eq:average1},
while for a zero temperature system we use Eq.~\eqref{Eq:average2}.
The average potential energy, as defined in Eq.~\eqref{Eq:Interaction}, is given by: (see Eq.~\eqref{Eq:P-Energy})
\begin{equation}\label{EnergyPotantio}
    \begin{aligned}
\langle\hat{H}_{\text{int}}\rangle &=\frac{1}{2}\int \text{d}\textbf{x}\text{d}\textbf{x}'v_0(\textbf{x},\textbf{x}')\langle\hat{\psi}^\dagger(x)\hat{\psi}^\dagger(x')\hat{\psi}(x')\hat{\psi}(x)\rangle
\\&=\frac{1}{2q}\int d\textbf{x} d\textbf{x}'v_0(\textbf{x},\textbf{x}')\\
&\times\left[\langle\hat{n}(x)\hat{n}(x')\rangle-\frac{1}{V}\sum_{\textbf{p}}e^{i\textbf{p}.(\textbf{x}-\textbf{x}')}q^{-n_{\textbf{p}}}\langle\hat{\psi}^\dagger(x)\hat{\psi}(x')\rangle\right]
\end{aligned}
\end{equation}
where $x$ and $x'$ are considered in equal times, and in the second line we used Eq.~\eqref{algebra-mode-expantion-kqParticle} in the real and momentum spaces.
In this equation
$\hat{n}(x)=\hat{\psi}^{\dagger}(x)\hat{\psi}(x)$, and in the limit $q\to 1$ , the expression $(1/V)\sum_{\textbf{p}}\exp{(i\textbf{p}.(\textbf{x}-\textbf{x}'))}$ reduces to the Dirac delta function, so that the operator $\hat{\psi}^\dagger(x)\hat{\psi}(x')$ simplifies to the number operator, $\hat{n}(x)$ giving rise to the standard limits. Note that $\langle\hat{n}(x)\rangle\equiv n=N/V$. The average energy of a $kq$-particle system is found to be (See appendix~\ref{LFanction})
\begin{equation}
    \begin{aligned}
        \langle\hat{H}_{\text{int}}\rangle&=\langle\hat{H}_{\text{int}}\rangle_0+\frac{1}{2q}\int d\textbf{x} d\textbf{x}'v_0(\textbf{x}-\textbf{x}')\Big(i\delta D_{kq}(\tau\textbf{x},\tau\textbf{x}')\Big),
    \end{aligned}
    \label{Eq:averageEnergy}
\end{equation}
where 
\begin{equation}
\begin{split}
\langle\hat{H}_{\text{int}}\rangle_0=\frac{1}{2q}\int d\textbf{x} d\textbf{x}'v_0(\textbf{x}-\textbf{x}')\Big[ & iD^0_{kq}(\tau\textbf{x},\tau\textbf{x}')+n^2_{}\\
&-\delta^{(kq)}(\textbf{x}-\textbf{x}')\Big],
\end{split}
\label{Eq:averageEnergy0}
\end{equation}
and
\begin{equation}
\delta^{(kq)}(\textbf{x})\equiv \frac{1}{V}\sum_{\textbf{p}}e^{i\textbf{p}.(\textbf{x})}q^{-n_{\textbf{p}}}\langle\hat{\psi}^\dagger(\tau\textbf{x})\hat{\psi}(\tau \textbf{0})\rangle.
\end{equation}
In these relations $i\delta D_{kq}(x,x')\equiv iD_{kq}(x,x')-iD_{kq}^0(x,x')$, where the time-ordered correlation function is defined as
\begin{equation}
iD_{kq}(x,x')=\langle\mathcal{T}[\tilde{n}^{}_H(x)\tilde{n}^{}_H(x')]\rangle,
\end{equation}
and for a zero temperature system is 
\begin{equation}
    iD_{kq}^{(T=0)}(x,x')=\frac{\langle\Psi_0|\mathcal{T}[\tilde{n}^{}_H(x)\tilde{n}^{}_H(x')]|\Psi_0\rangle}{\langle\Psi_0|\Psi_0\rangle}.
\end{equation}
$D_{kq}^0(x',x)$ is defined as the correlation function for a non-interacting system, and the deviation operator is as follows
\begin{equation}\label{average(n)}
    \tilde{n}(x)\equiv \hat{n}(x)-\langle\hat{n}(x)\rangle.
\end{equation}
We see from Eq.~\eqref{Eq:averageEnergy0} that for the determination of the average energy of a non-interacting system, we need to calculate $D^0(x,x')$, and for the average energy of a full interacting system we need the second part of Eq.~\eqref{Eq:averageEnergy}, called the correlation energy ($E_{\text{corr}}$):
\begin{equation}
    \begin{aligned}
        E_{\text{corr}}&=\frac{1}{2q}\int^1_0 \frac{d\lambda}{\lambda}\int d\textbf{x} d\textbf{x}'\lambda v_0(\textbf{x}-\textbf{x}')\Big(i\delta D_{kq}^\lambda(\tau\textbf{x},\tau\textbf{x}')\Big)\\
        &=\frac{V}{2q}(2\pi)^{-4}\int_0^1 \frac{d\lambda}{\lambda}\int d^4 Q \lambda v_0(Q)\Big(i\delta D_{kq}^\lambda(\textbf{Q},\omega)\Big).
    \end{aligned}
    \label{Eq:EnergyCorr}
\end{equation}
In the second line, we have written the equations in the momentum space, $Q\equiv (\textbf{Q},\omega)$ represents the Fourier coordinates, $v(Q)$ denotes the Fourier transform of $v(x)$, and $\delta D^{\lambda}(\textbf{Q},\omega)$ is the Fourier component of $\delta D_{kq}^{\lambda}(x,x')$ (the superscript $\lambda$ shows the interaction strength).
For the calculation of $D^0(x,x')$ we use the Wick's theorem, which gives
\begin{equation}\label{Eq:D0&noninteractingsystem}
    \begin{aligned}
    iD^0_{kq}(x,x')&=\left\langle \mathcal{T}\big[\hat{n}(x')\hat{n}(x)\big]\right\rangle_0
    -\left\langle\hat{n}(x')\right\rangle_0\left\langle\hat{n}(x)\right\rangle_0
     \\&=(kq)^2_{}iG_{kq}^{\mathcal{M}}(x,x)iG_{kq}^{\mathcal{M}}(x',x')\\
     &+(kq)iG_{kq}^{\mathcal{M}}(x,x')iG_{kq}^{\mathcal{M}}(x',x)-n^2\\
     &=(kq)iG_{kq}^{\mathcal{M}}(x,x')iG_{kq}^{\mathcal{M}}(x',x)-\Big(1-(kq)^2\Big)n^2.
    \end{aligned}
\end{equation}
This product is represented as the following bubble diagram:
\begin{equation}\label{Eq:bubble}
	\begin{tikzpicture}
		\begin{feynman}
			\vertex (a) at (0,0);
			\vertex (b) at (2.5,0);
			\vertex (a1) at (-0.3,0){$x$};
			\vertex (b1) at (2.8,0){$x'$};
			\diagram{
				(a) -- [fermion, quarter right] (b);
				(b) -- [fermion, quarter right] (a);
			};
		\end{feynman}
		\node [below, yshift=-0.8cm] at (1.2,-0.3) {(a)};
	\begin{feynman}
	\vertex (c) at (4,0);
	\vertex (d) at (6.5,0);
	\vertex (c1) at (3.8,0){$Q$};
	\vertex (d1) at (6.8,0){$Q$};
	\diagram{
		(c) -- [fermion, quarter right,edge label'=$p+Q$] (d);
		(d) -- [fermion, quarter right,edge label'=$p$] (c);
	};
   \end{feynman}
   \node [below, yshift=-0.8cm] at (5.2,-0.3) {(b)};
	\end{tikzpicture}
\end{equation}
in the coordinate and momentum space respectively, and $p=(\epsilon,\textbf{p})$ is internal four-momentum (to be integrated), and $Q=(E,\textbf{Q})$ is external four-momentum. \\

The full interaction line $v_{kq}(x,x')$ can be found in a same way as above. We firstly define the irreducible polarization function, represented by
the proper polarization function $\Pi_{kq}^*(x,x')$ as
\begin{equation}
    \begin{tikzpicture}
    \begin{feynman}
        \vertex[blob ,shape=ellipse,minimum height=0.8cm,minimum width=1.8cm] (m) at (-2.4, 0) {};
        \vertex[plain ,shape=ellipse,minimum height=0.8cm,minimum width=1.8cm] (n) at (0, 0) {};
        \vertex[plain ,shape=ellipse,minimum height=0.8cm,minimum width=1.8cm] (p) at (2.5, 0) {};
        \vertex[blob ,shape=rectangle,minimum height=0.8cm,minimum width=0.43cm] (pp) at (2.5, 0) {};
        \vertex (a) at (-2.6,0) {};
        \vertex (c) at (2.6, 0) {};
        \vertex (aa) at (-2.5, 0) {};
        \vertex (cc) at (1, 0) {};
        \vertex (aaa) at (-0.9, 0) {};
        \vertex (ccc) at (-1.5, 0) {};
        \vertex (up) at (2.3, 0.5) {};
        \vertex (down) at (2.3, -0.5) {};
        \vertex (up1) at (2.7, 0.5) {};
        \vertex (down1) at (2.7, -0.5) {};
        \vertex (ud) at (3.5, 0) {};
        \vertex (f1) at (0, 0.4) {};
        \vertex (f2) at (0, -0.4) {};
        \vertex (f3) at (2, 0.33) {};
        \vertex (f4) at (2, -0.33) {};
        \vertex (f5) at (3, 0.33) {};
        \vertex (f6) at (3, -0.33) {};
        \diagram* { (m),};
        \diagram* { (p),};
        \diagram* { (pp),};
   \node [right] at (ccc) {$=$};
   \node [right] at (cc) {$+$};
   \node [right] at (ud) {$+~...$};
   \node  at (f1) {$\blacktriangleright$};
   \node  at (f2) {$\blacktriangleleft$};
   \node [rotate=20] at (f3) {\scriptsize $\blacktriangleright$};
   \node [rotate=-20] at (f4) {\scriptsize $\blacktriangleleft$};
   \node [rotate=-20] at (f5) {\scriptsize $\blacktriangleright$};
   \node [rotate=20] at (f6) {\scriptsize $\blacktriangleleft$};
    \end{feynman}
\end{tikzpicture}
\end{equation}
Then the Dyson expansion of the full interaction line $v(x,x')$ reads:
\begin{equation}
\begin{aligned}
    \begin{tikzpicture}
        \begin{feynman}
        \vertex (a) at (0,0) {};
        \vertex (b) at (1.8,0) {};
        \vertex (aa) at (0.1,0) {};
        \vertex (bb) at (1.7,0) {};
        \diagram* {
    (a) -- [photon,edge , line width=1.3pt] (b),};
        \node [draw, circle, fill=black, inner sep=1pt] at (aa) {};
         \node [below] at (aa.south) {$x$};
         \node [draw, circle, fill=black, inner sep=1pt] at (bb) {};
         \node [below] at (bb.south) {$x'$};
        \node [right] at (bb) {$~~~=$};
        \end{feynman}
    \end{tikzpicture}\hspace{5.4cm}
    \\
    \begin{tikzpicture}
        \begin{feynman}
        \vertex (a) at (0,0) {};
        \vertex (b) at (1.8,0) {};
        \vertex (aa) at (0.1,0) {};
        \vertex (bb) at (1.7,0) {};
        \diagram* {
    (a) -- [photon,edge , line width=0.6pt] (b),};
        \node [draw, circle, fill=black, inner sep=1pt] at (aa) {};
         \node [below] at (aa.south) {$x$};
         \node [draw, circle, fill=black, inner sep=1pt] at (bb) {};
         \node [below] at (bb.south) {$x'$};
        \node [right] at (bb) {$~~~+$};
        \end{feynman}
    \end{tikzpicture}
    \begin{tikzpicture}
    \begin{feynman}
        \vertex[blob,shape=ellipse,minimum height=0.8cm,minimum width=1.8cm] (m) at (0, 0) {};
        \vertex (a) at (-2.6,0) {};
        \vertex (c) at (2.6, 0) {};
        \vertex (aa) at (-2.5, 0) {};
        \vertex (cc) at (2.5, 0) {};
        \vertex (aaa) at (-0.9, 0) {};
        \vertex (ccc) at (0.9, 0) {};
        \diagram* {
    (a) -- [photon,edge , line width=0.6pt] (m) -- [photon,edge ,line width=1.3pt] (c),
  };
        \node [draw, circle, fill=black, inner sep=1pt] at (aa) {};
         \node [below] at (aa.south) {$x$};
         \node [draw, circle, fill=black, inner sep=1pt] at (aaa) {};
         \node [below] at (aaa.south) {$x_1$};
        \node [draw, circle, fill=black, inner sep=1pt] at (cc) {};
         \node [below] at (cc.south) {$x'$};
         \node [draw, circle, fill=black, inner sep=1pt] at (ccc) {};
         \node [below] at (ccc.south) {$x'_1$};
    \end{feynman}
\end{tikzpicture}
\end{aligned}
\end{equation}
the mathematical expression of which is 
\begin{equation}
    \begin{aligned}
        &v_{kq}(x,x')
        =v_0^{}(x,x')
        \\&+\int d^4x_1 d^4x'_1 v_0^{}(x,x_1)\Pi^*_{}(x_1,x'_1)v_{kq}^{}(x'_1,x'),
    \end{aligned}
\end{equation}
Defining the polarization function $\Pi_{kq}(x,x')$ as the bubble expansion of $\Pi_{kq}^*(x,x')$, we have 
\begin{equation}
\Pi_{kq}(x,x')=D_{kq}(x,x')-i\left(1-(kq)^2\right)n^2.
\end{equation}
The correlation energy of the system can be written in terms of the irreducible polarization function 
\begin{equation}
    \begin{aligned}        
    E^{}_{\text{corr}}&=i\frac{V}{2q}(2\pi)^{-4}\int^1_0\frac{d\lambda}{\lambda}\int d^4Q\\
    &\times\Big[v_{kq}^\lambda(Q)\Pi_{kq}^{*\lambda}(Q)-\lambda v^{}_{0}(Q)\Pi^{0}_{kq}(Q)
    \\&~~~+i\left(v^\lambda_{kq}(Q)-\lambda v_0(Q)\right)\left(1-(kq)^2\right)n^2\Big],
    \label{Gs19}
    \end{aligned}
\end{equation}
where $v^\lambda(Q)=\lambda v(Q)$. The dielectric constant $\varepsilon(Q)=\varepsilon(\textbf{Q},\omega)$ can be calculated as 
\begin{equation}
v_{kq}(Q)=\frac{v_0(Q)}{\varepsilon_{kq}(Q)},
\end{equation}
which gives $\varepsilon_{kq}(Q)=1-v_0(Q)\Pi^*_{kq}(Q)$.
This gives further
\begin{equation}\label{Gs19corr}
   \begin{aligned}
        v_{kq}(Q)\Pi^*_{kq}(Q)=\frac{1}{\varepsilon_{kq}(Q)}-1,
   \end{aligned}
\end{equation}
giving rise to the following expression equation for the correlation energy
\begin{equation}
    \begin{aligned}
        E^{}_{\text{corr}}&=i\frac{1}{2q}V(2\pi)^{-4}
        \int^1_0\frac{d\lambda}{\lambda}\int d^4Q
    \\&\times\Big[\frac{1}{\varepsilon^\lambda_{kq}(Q)}-1-\lambda v_{0}(Q)\Pi^{0}_{kq}(Q)
    \\&~~~+i\left(v^\lambda_{kq}(Q)-\lambda v_0(Q)\right)\left(1-(kq)^2\right)n^2\Big].
    \end{aligned}
\end{equation}
\subsection{Evaluation of $\Pi^0$: Random Phase Approximation (RPA)}\label{SEC:RPA}
To conduct a more in-depth analysis, we need to examine the lowest-order polarization insertion, denoted by $\Pi^0_{}$ defined as a single bubble represented in Eq.~\eqref{Eq:bubble}, and mathematically is
\begin{equation}
   \Pi^0_{kq}(x,x')=-i(kq)iG_{kq}^{(\mathcal{M},X,0)}(x,x')iG_{kq}^{(\mathcal{M},X,0)}(x',x),
\end{equation}
which in the momentum space is expressed as
\begin{equation}\label{LindhardF}
    \begin{aligned}
    &\Pi^0_{k,q}(Q,\omega^{(j)}_n)=-\frac{kq}{\beta}\sum_{j',m}\int \frac{d^3p}{(2\pi)^3}\\&\times G^{(\mathcal{M},X,0)}(p,i\epsilon_m^{(j')})G^{(\mathcal{M},X,0)}(p+Q,i\omega_n^{(j)}+i\epsilon^{(j')}),
    \end{aligned}
\end{equation}
which is represented in the following graph:
\begin{equation}
        \begin{tikzpicture}
        \begin{feynman}
            \vertex (a) ;
            \vertex [right=of a] (b) ;
            \vertex [right=of b] (c) ;
            \vertex [right=of c] (d) ;
            
            \diagram{
            (b) -- [fermion, quarter right,edge label'=$p~\text{,}~i\epsilon_m^{(j')}$] (c);
            (c) -- [fermion, quarter right,edge label'=$p+Q~\text{,}~i\epsilon_m^{(j')}+i\omega_n^{(j)}$] (b);
            (a) -- [boson] (b);
            (c) -- [boson] (d);
            };
        \node [left] at (a.south) {$Q~\text{,}~i\omega_m^{(j)}$};
         \node [draw, circle, fill=black, inner sep=1pt] at (b) {};
        \node [draw, circle, fill=black, inner sep=1pt] at (c) {};
         \node [right] at (d.south) {$Q~\text{,}~i\omega_m^{(j)}$};
        \end{feynman}
    \end{tikzpicture}
\end{equation}
The response function  given by Eq.~\eqref{lindhard-function} can be expressed as follows:
\begin{equation}\label{Eq.Li1}
    \begin{aligned}
        &\Pi_{k,q}^0(\omega^{(j)}_n,Q)
        \\&=-kq\int \frac{d^3p}{(2\pi)^3}\frac{q^{-(\bar{n}_{p+Q}^{}+\bar{n}_p^{})}\left[S^{}_{kq}(\zeta^{}_{p+Q})-S^{}_{kq}(\zeta^{}_{p})\right]}{i\omega_n^{(j)}-f_{kq}(\zeta^{}_{p+Q})+f_{kq}(\zeta^{}_{p})}.
    \end{aligned}
\end{equation}
in which
\begin{equation}
    S_{kq}(\zeta_{p})=\frac{1}{\beta}\sum_{j',m}\frac{f_{kq}(\zeta_{p})}{\epsilon_m^{(j')2}+f_{kq}(\zeta_{p})^2_{}},
\end{equation}
In this equation, we have $f_{kq}(\zeta^{}_p)=C_{kq}(p)\zeta_p^{}
$, $\zeta_{p}=p^2/2m$, and $C_{kq}(p)\equiv\Big(q^{-\bar{n}^{}_p}_{}-k(1-q)[\bar{n}^{}_p]\Big)$. An straightforward calculation shows that 
\begin{equation}
S_{kq}(\zeta^{}_p)=\frac{1}{2}\left[\coth \frac{x-\log q}{2}\right]^k+\frac{1}{2}\left[\coth \frac{x+\log q}{2}\right]^k,
\end{equation}
where $x=\beta f(\zeta_p)$. One can also express $S_{kq}(\zeta_p)$ in terms of the distribution function(\ref{S_{kq}-np1},\ref{S_{kq}-np2}). Note that
\begin{equation}
    \lim_{q\to 1}S_{kq}(\zeta_p)=1+2k\bar{n}_p^{},
\end{equation}
giving rise to the classical limits. One may reform the Eq.~\eqref{Eq.Li1} using the transformation $p+Q\to -p'$, and then renaming $-p'$ back to $-p$ results in
\begin{equation}\label{Eq.Li2}
    \begin{aligned}
        &\Pi_{k,q}^0(\omega^{(j)}_n,Q)
        =-kq\int \frac{d^3p}{(2\pi)^3}q^{-(\bar{n}_{p+Q}^{}+\bar{n}_p^{})}S^{}_{kq}(\zeta^{}_{p})\\&\times\Big(\frac{1}{f(\zeta^{}_{p+Q})-f(\zeta^{}_{p})-i\omega_n^{(j)}}+\frac{1}{f(\zeta^{}_{p+Q})-f(\zeta^{}_{p})+i\omega_n^{(j)}}\Big),
    \end{aligned}
\end{equation}
which is a general expression. We will use this expression in the following section to capture the physics of the system in the zero temperature limit.
\section{ZERO-TEMPERATURE DIELECTRIC FUNCTION FOR $q$-FERMION GAS ($k=-1$)}\label{DIELECTRIC-FUNCTION}
\begin{figure}
    \centering
    \includegraphics{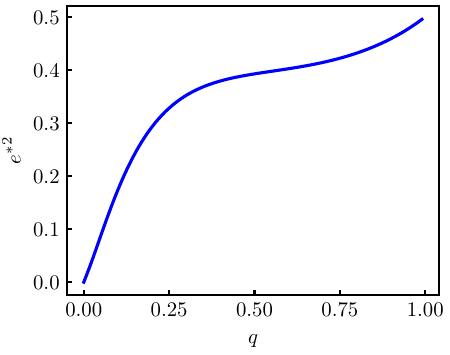}
    \caption{\justifying The effecttive columb interaction strength $\big(\frac{e^*}{e}\big)^2$ in terms of $q$, see Eq.~\eqref{Eq:effectiveInteraction}.}
    \label{fig:Estar}
\end{figure}

We now evaluate Eq.~\eqref{Eq.Li1} at $T=0$ and $k=-1$ limit, where the occupation function is expressed as:
\begin{equation}\label{density-particle}
    \lim_{T\to 0}\bar{n}_p^{}=\bar{n}(\zeta_p,T=0) = \begin{cases}
   \frac{2}{q-q^{-1}}\ln{q} &\text{for  } \zeta_p\leq 0 \\
   0 &\text{for  } \zeta_p>0
\end{cases}.
\end{equation}
For  the sake of convenience we define $S^F_q(\zeta_p)\equiv S_{-1,q}(\zeta_p)$, and $C_q^F(p)\equiv C_{-1,q}(p)$. Given that we are often interested in the small $Q$ values, the difference $S_{q}^F(\zeta_{p+Q})-S_{q}^F(\zeta_{p})$ is non-zero only in a thin layer around $\zeta=0$, i.e. in the vicinity of the generalized Fermi surface $\epsilon_{p_F^{(q)}}\equiv\epsilon_F^{(q)}\equiv \mu(T=0)$ for a non-interacting $q$-Fermion gas, where
$\epsilon_F^{(q)}\equiv\frac{{p_F^{(q)}}^2}{2m}$. Within this scenario, the total energy is given by
\begin{equation}
E_{\text{total}}^{\text{NI}}=\frac{3}{5}N\epsilon_F^{(q)}.
\end{equation}
In these relations the Fermi momentum is calculated to be
\begin{equation}
p_F^{(q)}=\left(\frac{3\pi^2n(q-q^{-1})}{\ln q}\right)^{\frac{1}{3}},
\end{equation}
For detailed analysis of the generalized Fermi energy, and the total energy of a non-interacting $q$-Fermion gas see Appendix~\ref{LFanction}. In this case the main contribution of the integrand comes from $S_{q}^F(0)$. Therefore, we can approximate $S_{q}^F(\zeta_p)$ with $S_{q}^F(0)$. This allows us to take $S_{q}^F(0)$ and $\bar{n}_p^{}+\bar{n}_{p+Q}^{}\approx 2\bar{n}_p^{}\to\bar{n}(\zeta=0)$ out of the integral. The first term of Eq.~\eqref{Eq.Li2} is then approximated to be:
\begin{equation}\label{Eq.Li}
    \begin{aligned}
        R^{{(q)}}_{1}=\frac{q^{-(2\bar{n}(0)-1)} S^{F}_{q}(0)}{(2\pi)^3}\int_0^{p_F^{(q)}} \frac{\text{d}^3p}{f(\zeta^{}_{p+Q})-f(\zeta^{}_{p})-i\omega_n^{(j)}}.
    \end{aligned}
\end{equation}
One can evaluate this integral noting that for a parabolic dispersion  $f(\zeta_{p+Q})-f(\zeta_{p})=\frac{\hbar^2}{2m}C_{q}^F(p)(\textbf{Q}^2+2\textbf{p}.\textbf{Q})$. After an analytic continuation $i\omega^{(j)}_n=\omega+i\delta$. These relations lead to
\begin{equation}
    \begin{aligned}     
    R^{{(q)}}_{1}=&-\frac{mf_q{p_F^{(q)}}^2}{2\pi^2\hbar^2Q}\left[Z^{}_1+\frac{1-Z_1^2}{2}\ln{\frac{Z^{}_1+1}{Z^{}_1-1}}\right],
    \end{aligned}
\end{equation}
where 
\begin{equation}
Z_1\equiv \frac{Q}{2p_F^{(q)}}-\frac{m(i\omega_n^{(j)})}{\hbar^2 C_{q}^F(0)p_F^{(q)} Q},
\end{equation}
and
\begin{equation}
f_q\equiv \frac{-q^{-(2\bar{n}(0)-1)}S^{F}_{q}(0)}{2C_{q}^F(0)}.
\end{equation}
It is notable that for spinfull particles, we have an extra factor of $2$ which corresponds to spin degree of freedom. Here our particles are considered to be spinless, so that
\begin{equation}
    \lim_{q\to 1}f_q=\frac{1}{2}.
    \label{Eq:limitF_q}
\end{equation}
The second term of Eq.~\eqref{Eq.Li2} ($R^{kq}_{2}$) can be expressed in a similar way with $Z_2$
\begin{equation}
    \begin{aligned}
        Z_2\equiv \frac{Q}{2p_F^{(q)}}+\frac{m(i\omega_n^{(j)})}{\hbar^2 C_{q}^F(0)p_F^{(q)}Q}.
    \end{aligned}
\end{equation}
The final expression for $\Pi_{q}^0(\omega^{(j)}_n,Q)=R^{(q)}_{1}+R^{(q)}_{2}$ now allows the computation of the dielectric constant in RPA: 
\begin{equation}
\label{dielectric-constant0}
    \begin{aligned}
        \varepsilon^{{\text{RPA}}}(\omega^{(j)}_n&,Q)=1-v_0(Q)\Pi_{q}^0(\omega^{(j)}_n,Q)\\
        =1+&\frac{2m e^2f_q{p_F^{(q)}}}{\pi\hbar^2 Q^2}\Bigg(1+\frac{p_F^{(q)}}{2Q}\sum_{i=1}^2(1-Z_i^2)\ln{\frac{Z^{}_i+1}{Z^{}_i-1}}\Bigg).
    \end{aligned}
\end{equation}
where we considered the bare Coulomb interaction $v_0(Q)=\frac{4\pi e^2}{Q^2}$. Showing the real part and imaginary part of the dielectric function as
\begin{equation}
\varepsilon^{{\text{RPA}}}=\varepsilon_1^{{\text{RPA}}}+i\varepsilon_2^{{\text{RPA}}},
\end{equation}
one finds the static dielectric function ($\omega^{(j)}_n\to 0$) to be 
    \begin{equation}
    \begin{aligned}
        \varepsilon_s^{{\text{RPA}}}(Q)\equiv \varepsilon_1^{{\text{RPA}}}(0,Q)=   1+\Big(\frac{Q_{\text{TF}}^{(q)}}{Q}\Big)^2F(x_q).
    \end{aligned}
    \end{equation}
    where
    \begin{equation}
        F(x)\equiv \frac{1}{2}+\frac{1-x^2}{4x}\ln\left|\frac{x+1}{x-1}\right|,
    \end{equation}
    and $x_q\equiv \frac{Q}{2p_F^{(q)}}$. In these relations:
    \begin{equation}
    Q_{\text{TF}}^{(q)}\equiv \sqrt{\frac{4m e^2f_qp_F^{(q)}}{\pi\hbar^2}},
    \end{equation}
    is the generalized Thomas-Fermi (TF) wave number. Here, we observe that the overall form of the functions are similar to that for the standard Fermionic limit~\cite{fetter2012quantum,sadovskii2006diagrammatics,mahan2013many}, and the parameters are renormalized (deformed) accordingly. In that sense, $q$-Fermions are included in the Fermi liquid scenario. Noting that in small $x_q$ limit $F(x_q)\to 1$, i.e. one retrieves a generalized TF model:
    \begin{equation}
    \begin{aligned}
        \varepsilon^{{\text{TF}}}(Q)\equiv\lim_{\text{small }Q}\varepsilon_s^{{\text{RPA}}}(Q)=1+\frac{{Q_{\text{TF}}^{(q)}}^2}{Q^2},
    \end{aligned}
    \end{equation}
    Moreover, we find that $Q_{\text{TF}}^{(q)}$ can be interpreted as the tuning parameter to find the best $q$ value compatible with the system under study. 
\begin{figure}
    \centering
    \includegraphics[width=\linewidth]{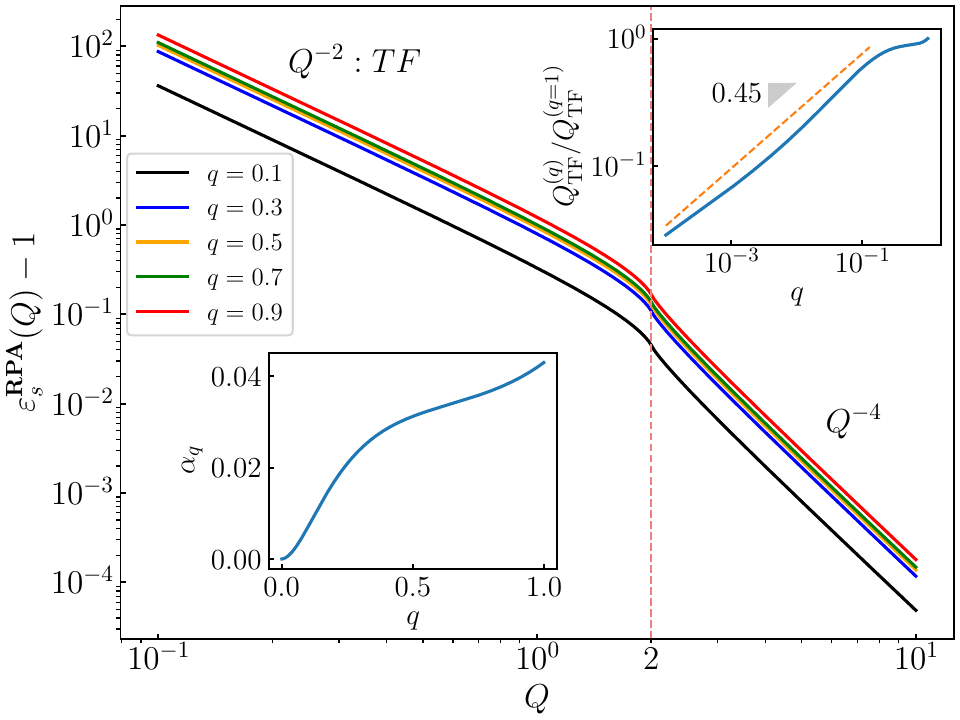}
    \caption{\justifying The main panel is a log-log plot of $\varepsilon_1^{\text{RPA}}(0,Q)$ in terms of $Q/p_F^{(q)}$ for $q$-Fermions. Inset is $|\alpha_q|$, i.e. the strength of Friedel oscillations in terms of $q$. For $Q\ll 2p_F^{(q)}$ the slope is $-2$ according to the TF theory, while in the opposite limit the slope is $-4$ according to Eq.~\eqref{Eq:largeQ}.}
    \label{fig:RPAfunction}
\end{figure}
\begin{figure}
    \centering
    \includegraphics[width=\linewidth]{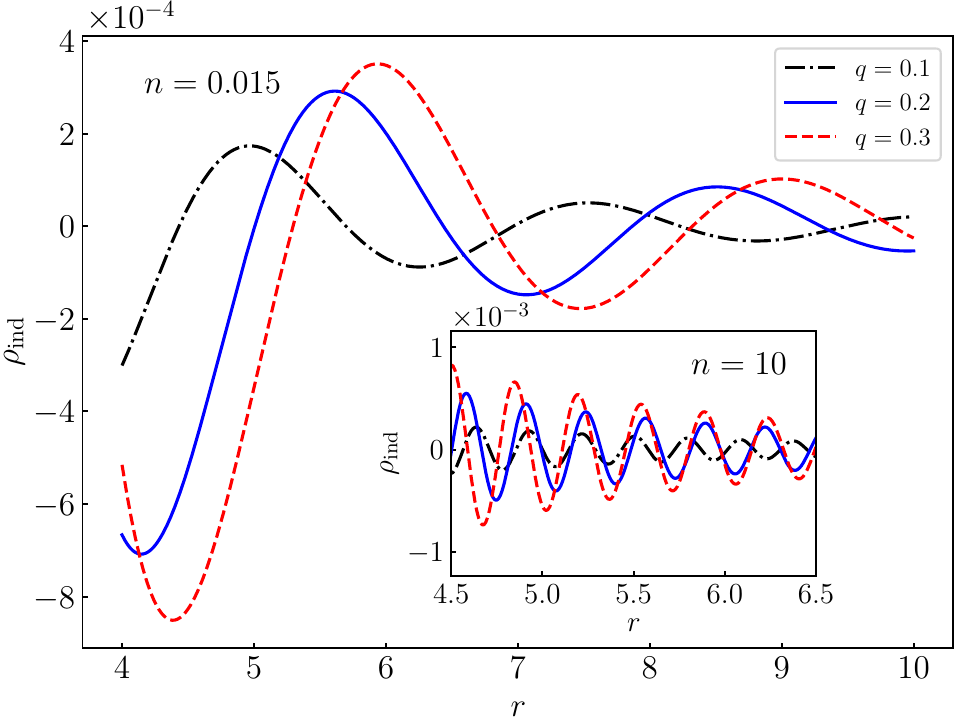}
    \caption{\justifying The Friedel oscillations for $n=0.015$ (main panel), $n=10$ (inset) for $q=0.1, 0.2$ and $0.3$. Vertical axis is the induced charge density, while the horizontal axis is the normalized distance $r/a_B$.}
    \label{fig:Friedel}
\end{figure}
In the large $Q$ limit we have $F(x)\to \frac{1}{3x_q^2}$, which leads to
    \begin{equation}
    \begin{aligned}
        \lim_{Q\gg p_F^{(q)}}\varepsilon_s^{{\text{RPA}}}(Q)=
        1+\frac{4}{3}\frac{{Q_{\text{TF}}^{(q)}}^2{p_{F}^{(q)}}^2}{Q^4}.
    \end{aligned}
    \label{Eq:largeQ}
    \end{equation}
 The static dielectric function is singular at $Q\to 2p_F^{(q)}$. This singularity is logarithmic:
    \begin{equation}
  \varepsilon_s^{{\text{RPA}}}(Q)\approx \varepsilon_s^{{\text{RPA}}}(2p_F^{(q)}) +\alpha_q(Q-2p_F^{(q)})\ln \left|\frac{Q-2p_F^{(q)}}{4p_F^{(q)}}\right|,
    \end{equation}
    where 
    \begin{equation}
    \varepsilon_s^{{\text{RPA}}}(2p_F^{(q)})=1+2\alpha_qp_F^{(q)},\ \alpha_q\equiv \frac{{Q_{\text{TF}}^{(q)}}^2}{16{p_F^{(q)}}^3}.
    \end{equation}
    This relation shows manifestly that $\frac{\text{d}\varepsilon_1^{{\text{RPA}}}}{\text{d}Q}\to -\infty$ as $Q\to 2p_F^{(q)}$. Fig.~\ref{fig:RPAfunction} shows $\varepsilon^{{\text{RPA}}}_s(Q)-1$ in terms of $Q$, wherein the insets are $\alpha_q$ and $\frac{Q_{\text{TF}}^{(q)}}{Q_{\text{TF}}^{(q=1)}}$ in terms of $q$. This figure shows two power-law regimes: TF regime in small $Q$s, where $\varepsilon_s(Q)$ varies as $Q^{-2}$, and the large $Q$ regime where it varies as $Q^{-4}$, with a singular point in between (at $Q=2p_F^{(q)}$).
    It is interestingly observed that the singularity strength $\alpha_q$ (that is related to the amplitude of the Friedel oscillations as becomes clear in the next subsection) and the normalized TF wave number $\frac{Q_{\text{TF}}^{(q)}}{Q_{\text{TF}}^{(q=1)}}$ grow with $q$. This shows that as $q$ decreases, i.e. $q$-fermions becomes farther away from the standard fermions, and the screening length $\lambda_{\text{TF}}^{(q)}=\frac{2\pi}{Q_{\text{TF}}^{(q)}}$ increases which can be interpreted as a result of changing an \textit{effective interaction} between the $q$-Fermions. The observed divergence of screening length as $q\to 0$ implies that the system becomes less effective at screening, and the $q$-fermions become weakly interacting.  Given that for the standard ($q=1$) fermions ${Q_{\text{TF}}^{(q=1)}}^2=\frac{4e^2mp_F^{(q=1)}}{\pi \hbar^2}$, one may implement the following substitution, in order to absorb the statistics in an effective interaction:
    \begin{equation}
    e^2\to{e^*}^2\equiv f_q\frac{p_F^{(q)}}{p_F^{(q=1)}}e^2.
    \label{Eq:effectiveInteraction}
    \end{equation}
    Resulting dielectric function $e^*$ is shown in Fig.~\ref{fig:Estar}. Note that due to the limit Eq.~\eqref{Eq:limitF_q}, $\lim_{q\to 1}e^*_q=\frac{1}{2}e$, while if we consider the spin degree of freedom, we should multiply $f_q$ by a factor $2$, resulting to the expected result. We observe that when we decrease $q$, $e^*$ is almost constant (slowly-varying function) down to $q^*\approx 0.2$, below which it falls off rapidly. Therefore, for $q\lesssim q^*$ one expects dramatic deviations from the standard fermionic system.
    
    Substituting this relation into Eq.~\eqref{dielectric-constant0} shows that although the first coefficient of the expansion corresponds to the standard fermionic dielectric function with the effective charge $e^*$, the second term does not. One therefore concludes that the $q$-fermion system does not match completely with a standard fermionic system with renormalized charge.
    \subsection{Friedel Oscillations and Plasmons and Energy Loss for $q$-Fermion gas}\label{Friedel-Oscillations}   
    As we saw in the previous section, the response of the $q$-Fermion gas to an external charge depends on the amount of $q$. The singularity at $Q=2p_F^{(q)}$ results in Friedel oscillations~\cite{grosso2013solid}. To be more precise, let us show the induced charge density around a point impurity with a charge $Ze$, which follows the relation~\cite{mahan2013many}:
    \begin{equation}
    \begin{split}
\rho_{\text{ind}} &\equiv\frac{Ze}{(2\pi)^3}\int \bigg(\frac{1}{\varepsilon_s(Q)}-1\bigg)e^{i\textbf{Q}.\textbf{r}}\text{d}\textbf{Q}\\
&=\frac{Ze}{r^3}\int_0^{\infty}g''(Q)\sin (Qr) \text{d}Q,
    \end{split}
    \end{equation}
where 
\begin{equation}
g(Q)\equiv\frac{Q}{2\pi^2}\frac{\varepsilon_s(Q)-1}{\varepsilon_s(Q)},
\end{equation}
and $g''\equiv \frac{\text{d}^2 g}{\text{d}Q^2}=\frac{\alpha_q}{Q-2p_F^{(q)}}$ (note that $\lim_{Q\to 0}g(Q)=\lim_{Q\to \infty}g(Q)=0$). Given the fact that the main contribution of the integral comes from the region in the close vicinity of the singular point, we consider a window $\delta$ around the singular point  $p_F^{(q)}$ inside which the integrand has a considerable contribution, as a result of which one ends up with:
\begin{equation}
\begin{split}
\rho_{\text{ind}}\approx & \frac{Ze\alpha_q}{r^3}\int_{2p_F^{(q)}-\delta}^{2p_F^{(q)}+\delta}\frac{\sin(Qr)}{Q-2p_F^{(q)}}\text{d}Q\\
&=\frac{Ze\alpha_q\cos(2p_F^{(q)}r)}{r^3}\int_{-r\delta}^{r\delta}\frac{\sin y}{y}\text{d}y\\
&\longrightarrow \frac{\pi Ze\alpha_q\cos(2p_F^{(q)}r)}{r^3},  
\end{split}
\label{Eq:Friedel}
\end{equation}
where in the last step we let $r\to \infty$. This realizes the Friedel oscillations, where the wave length and the amplitude depends on $q$. Fig.~\ref{fig:Friedel} demonstrates how $\rho_{\text{ind}}$ oscillates around the impurity located at the origin for two densities $n=0.015$ (low density regime) and $n=10$ (high density regime). We now observe $\alpha_q$ controls the strength of the oscillations. We conclude that the oscillations diminishing to zero indicates that the response of the fermions to an impurity becomes very weak, consistent with a reduction in interaction strength. In the non-interacting limit (very small $q$ values), the fermions would not exhibit significant Friedel oscillations because there’s minimal collective density response.\\
    
\begin{figure}
    \centering
    \includegraphics[width=\linewidth]{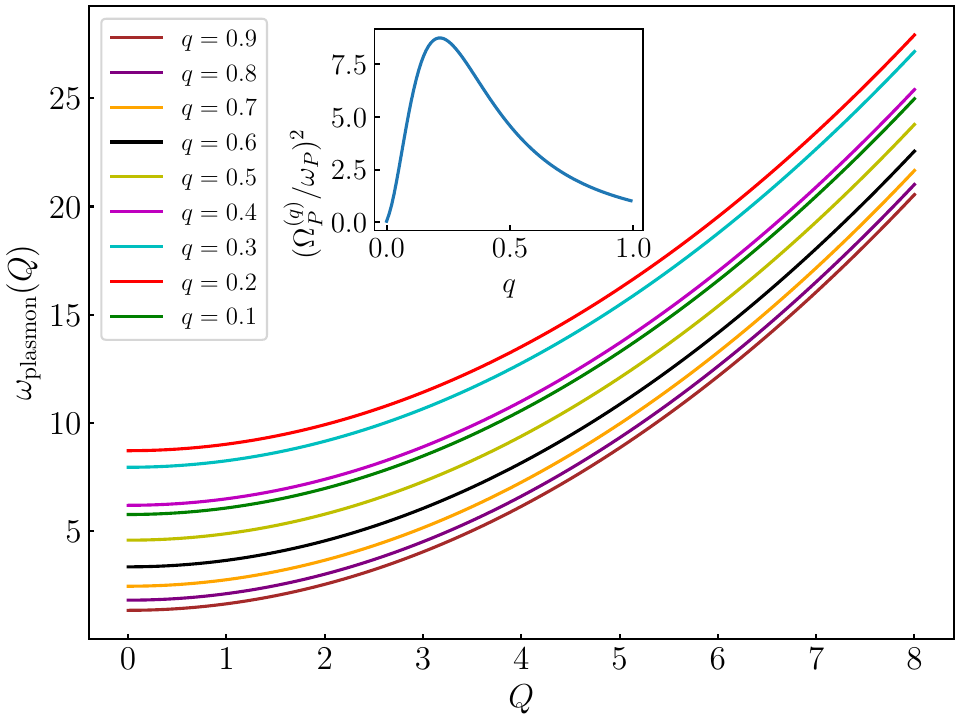}
    \caption{\justifying The dispersion relation Eq.~\eqref{wplasmon} in terms of $Q$ for different values of $q$. The inset shows the behavior of the function $\Omega_P^{(q)}/\omega_P$ in terms of $q$ with a maximum at $q^*\approx 0.2$.}
    \label{fig:epsilon2}
\end{figure}

At very high frequencies, one can expand Eq.~\eqref{dielectric-constant0} to reach to the following approximation by ignoring $Q$ with respect to the frequency
    \begin{equation}
\varepsilon(\omega^{(j)}_n,0)=1-\frac{{\Omega_P^{(q)}}^2}{(\omega^{(j)}_n+i\delta)^2_{}},
    \end{equation}
  where  $\Omega^{}_p$ is the plasma frequency for $q$-fermions:
  \begin{align}
      {\Omega_P^{(q)}}^2&=\frac{(q-1)q^{-\frac{6q\ln{q}}{q^2_{}-1}}_{}\left(1+2q-q^{1+\frac{4q\ln{q}}{q^2_{}-1}}_{}\right)}{2\ln{q}}\omega^2_P,
      \label{Eq:PlasmonicOs}
  \end{align}
  and $\omega_P=\frac{4\pi ne^2}{m}$ is the plasma frequency for standard fermions ${\Omega_P^{(q=1)}}^2$.
  Replacing $2\delta\to 1/\tau$ leads to
\begin{equation}
\begin{split}
&\varepsilon_1(\omega,0)=1-\frac{{\Omega_P^{(q)}}^2}{\omega^2_{}+1/\tau^2},\\
&\varepsilon_2(\omega,0)=\frac{{\Omega_P^{(q)}}^2}{\tau\omega\big(\omega^2+1/\tau^2\big)}
\end{split}
\end{equation}
\begin{figure}
    \centering
    \includegraphics[width=\linewidth]{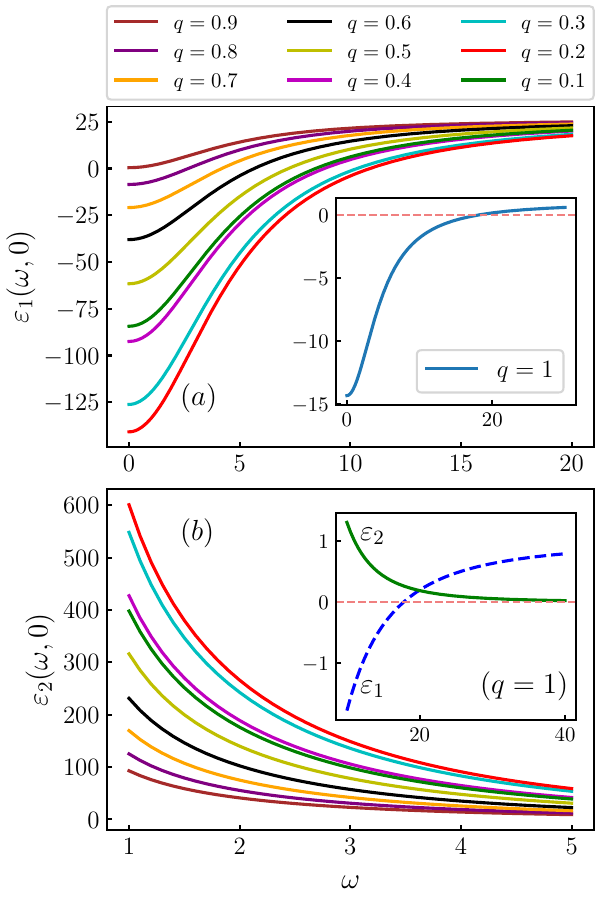}
    \caption{\justifying
    (a) The frequency dependence of (a) $\varepsilon_1$ and (b) $\varepsilon_1$ for $q$-fermions. The inset of (a) highlights the specific point where the  $\varepsilon_1$ becomes for $q=1$. The inset of (b) provides a zoomed-in view of $\varepsilon_1$ and $\varepsilon_2$ for $q = 1$. }
    \label{fig:epsilon12}
\end{figure}
$(\Omega_P^{(q)}/\omega_P)^2$ is shown in Fig.~\ref{fig:epsilon2} as a function of $q$. We see that the frequency goes to zero in the limit $q\to 0$, and goes to unity as $q\to 1$. This frequency is directly tied to the strength of the fermion-fermion interactions, as plasmon modes arise from correlated charge density fluctuations. A reduction in the plasmonic frequency to zero as $q\to 0$ suggests that, in the weak interaction limit, collective charge density oscillations lose coherence and essentially vanish. This is again indicative of $q$ controlling the interaction strength; a small $q$ corresponds to a weak collective behavior and thus a lower plasmon frequency. Contrary to the other cases, for an intermediate $q^*\approx 0.2$ the frequency shows a maximum, and the collective mode is strongest. \\

The poles of the interaction potential identify the plasmon modes, i.e. $\varepsilon(\omega_n^{(j)},Q) = 0$, which determines the dispersion relation for the plasmons.
By expanding Eq.~\eqref{dielectric-constant} to second order with respect to $\omega_n^{(j)}$ and $Q$, we can determine the plasmonic mode as follows:
\begin{equation}\label{Expand}
    1-\frac{{\Omega^{(q)}_P}^2}{\omega^2}-\frac{3}{5}\frac{{\Omega^{(q)}_P}^2}{\omega^4}{\mathrm{v}_F^{(q)}}^2 Q^2=0,
\end{equation}
where $\mathrm{v}_F^{(q)}=\hbar p_F^{(q)}/m$.
Solving Eq.~\eqref{Expand} yields the dispersion relation for plasmon modes at small wavevectors $Q$, which reads
\begin{equation}\label{wplasmon}
    \omega_{\text{plasmon}}(Q)=\Omega^{(q)}_P\left[1+\frac{3}{10}\frac{{\mathrm{v}_F^{(q)}}^2 Q^2}{{\Omega^{(q)}_P}^2}+...\right].
\end{equation}
This is the plasmonic dispersion relation. In the Fig.~\ref{fig:epsilon2} we show this function for various $q$ values, and in the inset we show $\Omega_P^{(q)}/\omega_P$ in terms of $q$. We see that the frequency increases as $q$ increases up to $q=q^*$, and the decreases. The imaginary part of $\varepsilon$ has an important physical interpretation: the energy dissipation. More precisely $-\text{Im}(1/\varepsilon)$, is proportional the energy loss of fast charged $q$-fermions. The energy-loss function provides important insights into plasmon excitations. The following relation holds
\begin{figure}
    \centering
    \includegraphics[width=1\linewidth]{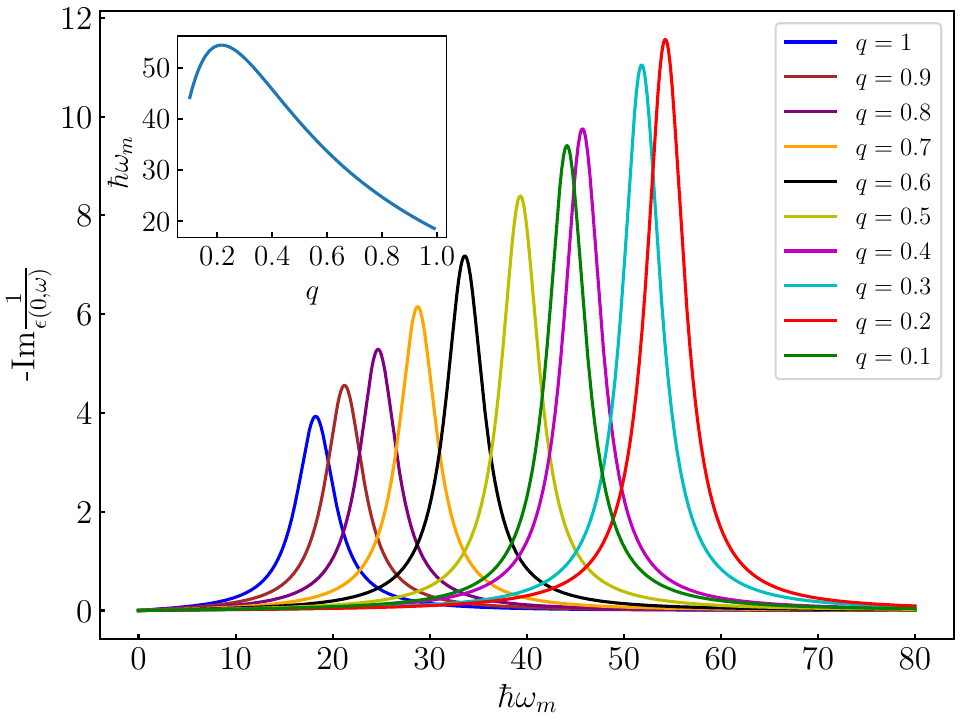}
    \caption{\justifying Energy-loss function for $q$-fermion gas. Inset shows the peak point in terms of $q$ with a maximum at $q\approx 0.2$.}
    \label{fig:energy-loss-function-q-fermions}
\end{figure}
   \begin{equation}
       -\text{Im}\frac{1}{\varepsilon(\omega,0)}=\frac{\tau^{-1}\omega{\Omega^{(q)}_P}^2}{\bigg(\omega^2_{}-{\Omega^{(q)}_P}^2\bigg)^2_{}+\omega^2_{}/\tau^2_{}}.
   \end{equation}
   This function is shown in Fig.~\ref{fig:energy-loss-function-q-fermions} for $q$-fermions. The inset shows the peak point $\Omega_P^{(q)}$ in terms of $q$, which is consistent with Fig.~\ref{fig:energy-loss-function-q-fermions}. The maximal energy loss occurs in $q=q^*$. This highlights the fact especial role of $q^*$, where the $q$-fermions enter a new mode of activity, i.e. shows dramatic differences with respect to standard fermions.

   \section{CONCLUDING REMARKS}\label{CONCLUD}
   In this work, we have developed a quantum many-body theory for the system of particles with deformed statistics, namely the $kq$-particles, offering a new framework to analyze systems where particle statistics and effective interaction strengths are deeply intertwined. By leveraging deformation parameters $k$ and $q$, we constructed a theoretical structure in which many-body interactions could be effectively captured through modified particle statistics. This study not only presents the standard quantities for quantum many body $kq$-particles and simplifies the treatment of complex interactions but also offers flexibility in modeling a broad spectrum of correlated systems by adjusting statistical parameters rather than relying on explicit interaction potentials.

Our findings include the derivation of single particle Green's functions for $kq$-particles in both direct and momentum spaces, allowing us to quantify physical quantities and collective behaviors unique to these systems. To this end we generalized Wick's theorem and also the $S$-matrix expansion of quantum many body systems. As a result, we ere able to develop a Feynman diagrammatic approach, to extend standard techniques to this deformed statistical regime, demonstrating how $kq$-particles differ fundamentally from standard bosonic or fermionic systems. The dielectric function was calculated within the random phase approximation for $q$-fermions, illuminating the statistical nature of screening effects. Up to the authors knowledge, this is the first time the dielectric function is calculated for a deformed quantum many body system.

Additionally, the framework enabled us to explore essential physical phenomena such as Friedel oscillations, plasma frequency, and the energy loss function for $q$-fermions. An effective interaction strength was proposed as a function of the deformation parameter $q$ (Eq.~(\ref{Eq:effectiveInteraction})), which serves as the key relation for the statistical interaction as a function of the deformation parameter. These results show how deformation-based statistical approaches can provide insights into various collective effects, potentially applicable in fields like condensed matter physics, where non-standard statistics and strong correlations play crucial roles.

This work opens several avenues for future research. Our formalism invites extensions to higher-order interactions and correlations, potentially enabling studies of more sophisticated many-body effects within $kq$-deformed systems. Moreover, the concept of statistically-driven interactions suggests applications in fields where direct control of particle interactions is challenging, offering a novel perspective for experimental and theoretical investigations alike. By establishing a foundation for $kq$-deformed many-body theory, we hope this research will contribute to a broader understanding of nontraditional quantum systems, encouraging further exploration into the rich interplay between quantum statistics and interaction strengths. We conclude that $kq$-deformation can serve as a versatile tool for understanding the interaction-driven behavior of many-body systems, with potential applications in fields where nonstandard quantum statistics and strong correlations are essential.

\bibliography{refs}
\newpage
\appendix
\section{THE $kq$-STATISTICS}\label{Appendix}
To derive the commutation relation in Eq.~\eqref{cp}, we proceed step by step as follows:\\
First, we examine the commutation between the operator product \(a^\dagger_l a_l\) and the annihilation operator \(a_\nu\)
\begin{equation}
\begin{aligned}
        \left[a^\dagger_l a^{}_l,a_\nu\right]&=a^\dagger_l \underleftrightarrow{a^{}_l a^{}_\nu}-a^{}_\nu a^\dagger_l a^{}_l,
\end{aligned}
\end{equation}
By rewriting the order of multiplication and using commutation terms, we arrive at:
\begin{equation}
    \left[a^\dagger_l a^{}_l,a_\nu\right]
        =k a^\dagger_l a^{}_\nu a^{}_l-a^{}_\nu a^\dagger_l a^{}_l
        =\left(k a^\dagger_l a^{}_\nu -a^{}_\nu a^\dagger_l \right)a^{}_l,
\end{equation}
If \(\nu = l\), applying the modified commutation relation \(\left[a_\nu, a^\dag_\nu\right]_{kq} = q^{-N_\nu}\), we get:
\begin{equation}
    \left(k a^\dagger_\nu a^{}_\nu -kq a^\dagger_\nu a^{}_\nu-q^{-N^{}_\nu}\right)a^{}_\nu,
\end{equation}
By substituting \(a^\dagger_\nu a_\nu = [N_\nu]\), we simplify to:
\begin{equation}
    \left(k(1-q)[N^{}_\nu]-q^{-N_\nu}\right)a_\nu,
\end{equation}
Thus, the commutation relation simplifies as:
\begin{equation}
    \left[a^\dagger_\nu a^{}_\nu,a^{}_\nu\right]=\left(k(1-q)[N^{}_\nu]-q^{-N_\nu}\right)a_\nu,
       \label{Eq:Appcp}
\end{equation}
Now we look at the commutation relation for \(a^\dagger_l a_l\) and \(a^\dagger_\nu\)
\begin{equation}
    \begin{aligned}
        \left[a^\dagger_l a^{}_l,a^\dagger_\nu\right]&=a^\dagger_l a^{}_l a^\dagger_\nu-\underleftrightarrow{a^\dagger_\nu a^\dagger_l} a^{}_l
        \\&=a^\dagger_l a^{}_l a^\dagger_\nu-ka^\dagger_la^\dagger_\nu a^{}_l
        =a^\dagger_l\left( a^{}_l a^\dagger_\nu -ka^\dagger_\nu a^{}_l \right),
    \end{aligned}
\end{equation}
Again, if \(\nu = l\), substituting \(\left[a_\nu, a^\dag_\nu\right]_{kq} = q^{-N_\nu}\) and \(a^\dagger_\nu a_\nu = [N_\nu]\), we find:
\begin{equation}
    a^\dagger_l\left( kq a^\dagger_\nu a^{}_\nu+q^{-N^{}_\nu} -ka^\dagger_\nu a^{}_\nu \right)=a^\dagger_\nu\left(k(q-1)[N^{}_\nu]+q^{-N^{}_\nu}\right),
\end{equation}
So,
\begin{equation}
    \left[a^\dagger_\nu a^{}_\nu,a^\dagger_\nu\right]= a^\dagger_\nu\left(k(q-1)[N^{}_\nu]+q^{-N^{}_\nu}\right).
       \label{Eq:Appcdp}
\end{equation}
We need to show that the commutation relation \(\left[a^{\dagger}_l a_l, a^{\dagger}_p a_p\right] = 0\) holds. Expanding it, we have:
\begin{equation}
\begin{aligned}
      \left[a^{\dagger}_la^{}_l,a^{\dagger}_pa^{}_p\right]=\left[a^{\dagger}_la^{}_l,a^{\dagger}_p\right]a^{}_p+a^{\dagger}_p\left[a^{\dagger}_la^{}_l,a^{}_p\right],
\end{aligned}
\end{equation}
For the first term, we calculate:
\begin{equation}
    \left[a^{\dagger}_la^{}_l,a^{\dagger}_p\right]=a^{\dagger}_la^{}_la^{\dagger}_p-\underleftrightarrow{a^{\dagger}_pa^{\dagger}_l}a^{}_l,
\end{equation}
Applying the commutation conditions \(\left[a^{\dagger}_l, a^{\dagger}_p\right]_k = 0\) and \(\left[a_l, a_p\right]_k = 0\), we have:
\begin{equation}
    \begin{aligned}
        a^{\dagger}_la^{}_la^{\dagger}_p-ka^{\dagger}_la^{\dagger}_Pa^{}_l&=
    a^{\dagger}_l(a^{}_la^{\dagger}_p-ka^{\dagger}_pa^{}_l)\\&=ka^{\dagger}_l(ka^{}_la^{\dagger}_p-a^{\dagger}_pa^{}_l),
    \end{aligned}
\end{equation}
Thus
\begin{equation}
    \left[a^{\dagger}_la^{}_l,a^{\dagger}_p\right]=-ka^{\dagger}_l\left[a^{\dagger}_p,a^{}_l\right]_k.
\end{equation}
Similarly, for the second term, we calculate:
\begin{equation}
    \begin{aligned}
        \left[a^{\dagger}_la^{}_l,a^{}_p\right]&=a^{\dagger}_l\underleftrightarrow{a^{}_la^{}_p}-a^{}_pa^{\dagger}_la^{}_l \\&=ka^{\dagger}_la^{}_pa^{}_l-a^{}_pa^{\dagger}_la^{}_l 
        =k\left[a^{\dagger}_l,a^{}_p\right]_ka^{}_l,
    \end{aligned}
\end{equation}
Combining the two terms, we get:
\begin{equation}
\begin{aligned}
      \left[a^{\dagger}_la^{}_l,a^{\dagger}_pa^{}_p\right]&=-ka^{\dagger}_l\left[a^{\dagger}_p,a^{}_l\right]_ka^{}_p+ka^{\dagger}_p\left[a^{\dagger}_l,a^{}_p\right]_ka^{}_l,
\end{aligned}
\end{equation}
with replace $p$ and $l$ in the second term
\begin{equation}
\begin{aligned}
      \left[a^{\dagger}_la^{}_l,a^{\dagger}_pa^{}_p\right]&=-ka^{\dagger}_l\left[a^{\dagger}_p,a^{}_l\right]_ka^{}_p+ka^{\dagger}_l\left[a^{\dagger}_p,a^{}_l\right]_ka^{}_p=0,
      \label{Eq:Appcccc}
\end{aligned}
\end{equation}
Thus, we have shown that \(\left[a^{\dagger}_l a_l, a^{\dagger}_p a_p\right] = 0\) as required.\\
Here we explain and prove a number of identities in the text of the article in detail.
The number of particles can be expressed as follows:
\begin{equation}
    \begin{aligned}
        [\bar{n}_\nu]&=\frac{q^{\bar{n}_\nu}-q^{-\bar{n}_\nu}}{q^{}-q^{-1}}
        =\frac{1}{Z}\frac{tr(e^{-\beta H}q^{+N_\nu})-tr(e^{-\beta H}q^{-N_\nu})}{q^{}-q^{-1}}\\&
        =\frac{1}{Z}tr\bigg(e^{-\beta H}\frac{q^{+N_\nu}-q^{-N_\nu}}{q^{}-q^{-1}}\bigg)=\frac{1}{Z}tr\Big(e^{-\beta H}[N_\nu]\Big).
        \label{fk}
    \end{aligned}
\end{equation}
To prove the identity in Eq.~\eqref{identities}, we can directly apply the expression derived above. This leads to the following result:
\begin{equation}
   \begin{aligned}
       [\bar{n}_\nu]&=\frac{1}{Z}tr\big(e^{-\beta H}a^\dagger_\nu a^{}_\nu\big)=\frac{1}{Z}tr\big(a^{}_\nu e^{-\beta H}a^\dagger_\nu \big)\\&
       =\frac{1}{Z}tr\big(e^{-\beta H}e^{\beta H}a^{}_\nu e^{-\beta H}a^\dagger_\nu \big)=e^{-\eta}\frac{1}{Z}tr\big(e^{-\beta H}a^{}_\nu a^\dagger_\nu \big)
       \\&=e^{-\eta}[k\bar{n}_\nu+1],
       \label{A1}
   \end{aligned}
\end{equation}
where $e^{\beta H}a^{}_\nu e^{-\beta H}=e_{}^{-\eta}a_\nu$ and $\eta=\beta(E_\nu-\mu)$.
Continuing the proof of Eq.~\eqref{identities}:
\begin{equation}
    \begin{aligned}
        \left[k\bar{n}_\nu+1\right]&=\frac{q^{k\bar{n}_\nu+1}-q^{-k\bar{n}_\nu-1}}{q-q^{-1}}
        \\&=\frac{q^{k\bar{n}_\nu+1}-q^{-k\bar{n}_\nu+1}}{q-q^{-1}}+\frac{q^{-k\bar{n}_\nu+1}-q^{-k\bar{n}_\nu-1}}{q-q^{-1}}\\&
        =q\Big(\frac{q^{k\bar{n}_\nu}-q^{-k\bar{n}_\nu}}{q-q^{-1}}\Big)+q^{-k\bar{n}_\nu}\Big(\frac{q-q^{-1}}{q-q^{-1}}\Big)
        \\&=kq\left[\bar{n}_\nu\right]+q^{-k\bar{n}_\nu}.
        \label{knp1}
    \end{aligned}
\end{equation}
Alternatively, it can be expressed as follows:
\begin{equation}
    \begin{aligned}
        \left[k\bar{n}_\nu+1\right]&=\frac{q^{k\bar{n}_\nu+1}-q^{-k\bar{n}_\nu-1}}{q-q^{-1}}
        \\&=\frac{q^{k\bar{n}_\nu-1}-q^{-k\bar{n}_\nu-1}}{q-q^{-1}}+\frac{q^{k\bar{n}_\nu+1}-q^{k\bar{n}_\nu-1}}{q-q^{-1}}\\&
        =q^{-1}\Big(\frac{q^{k\bar{n}_\nu}-q^{-k\bar{n}_\nu}}{q-q^{-1}}\Big)+q^{k\bar{n}_\nu}\Big(\frac{q-q^{-1}}{q-q^{-1}}\Big)
        \\&=kq^{-1}\left[\bar{n}_\nu\right]+q^{k\bar{n}_\nu}.
        \label{knp2}
    \end{aligned}
\end{equation}
Summing the two terms in Eq.~\eqref{identities}, we obtain
\begin{equation}
  \begin{aligned}
     2\left[k\bar{n}_\nu+1\right]&=\left(kq+kq^{-1}\right)\left[\bar{n}_\nu\right]+\left(q^{-k\bar{n}_\nu}+q^{k\bar{n}_\nu}\right)\\
     \left[k\bar{n}_\nu+1\right]&=\frac{\left(kq+(kq)^{-1}\right)}{2}\left[\bar{n}_\nu\right]+\frac{\left(q^{-k\bar{n}_\nu}+q^{k\bar{n}_\nu}\right)}{2}\\
    \left[k\bar{n}_\nu+1\right]&=\left[\bar{n}_\nu\right]\cosh\left(\ln{kq}\right)+\cosh\left(k\bar{n}_\nu\ln{q}\right).
     \label{Eq:appnp1}
  \end{aligned}
\end{equation}
To derive the commutation relation in Eq.~\eqref{Eq:DesiredState}, we proceed as follows (for $k=1$).
We start by applying the number operator \( N_\nu \) and the annihilation operator \( a_\nu \) to the state \( |n_\nu\rangle \):
\begin{equation}
    \begin{aligned}
        N_\nu a_\nu\left| n_\nu \right\rangle=\left(a_\nu N_\nu-a_\nu\right)\left| n_\nu \right\rangle=(n_\nu-1)a_\nu\left| n_\nu \right\rangle,
    \end{aligned}
\end{equation}
where the commutation relation \( [N_\nu, a_\nu] = -a_\nu \) has been used. This expression demonstrates that acting with \( N_\nu \) on the annihilation operator state \( a_\nu |n_\nu\rangle \) reduces the eigenvalue \( n_\nu \) by one. Next, we express \( a_\nu |n_\nu\rangle \) as:
\begin{equation}
     a_\nu\left| n_\nu \right\rangle=c_{-}^n\left| n_\nu-1 \right\rangle,
\end{equation}
where \( c_{-}^n \) is a coefficient that may depend on \( n \) and serves as a normalization factor.\\
Similarly, we apply the number operator \( N_\nu \) and the creation operator \( a^\dagger_\nu \) to \( |n_\nu\rangle \):
\begin{equation}
    \begin{aligned}
        N_\nu a^\dagger_\nu\left| n_\nu \right\rangle&=\left(a^\dagger_\nu N_\nu+a^\dagger_\nu\right)\left| n_\nu \right\rangle=(n_\nu+1)a^\dagger_\nu\left| n_\nu \right\rangle,
    \end{aligned}
\end{equation}
using \( [N_\nu, a^\dagger_\nu] = a^\dagger_\nu \). This operation increases the eigenvalue \( n_\nu \) by one.
We can thus express \( a^\dagger_\nu |n_\nu\rangle \) as:
\begin{equation}
    a^\dagger_\nu\left| n_\nu \right\rangle=c_{+}^n\left| n_\nu+1 \right\rangle,
\end{equation}
where \( c_{+}^n \) is another normalization coefficient.
We evaluate the inner products for the operators \( a_\nu^\dagger a_\nu \) and \( a_\nu a_\nu^\dagger \) acting on \( |n_\nu\rangle \):
\begin{equation}
    \begin{aligned}
        \left\langle n_\nu \right|a^\dagger_\nu a_\nu\left|n_\nu\right\rangle=c_{-}^{n}c_{+}^{n-1}~~~,~~~
        \left\langle n_\nu \right|a_\nu a^\dagger_\nu\left|n_\nu\right\rangle=c_{+}^{n}c_{-}^{n+1},
    \end{aligned}
\end{equation}
Given the relations \( a_\nu^\dagger a_\nu = [N_\nu] \) and \( a_\nu a_\nu^\dagger = [N_\nu + 1] \), we rewrite the inner products as follows:
\begin{equation}
    \begin{aligned}
        \left\langle n_\nu \right|a^\dagger_\nu a_\nu\left|n_\nu\right\rangle&=\left\langle n_\nu \right|[N_\nu]\left|n_\nu\right\rangle=[n_\nu]=c_{-}^{n}c_{+}^{n-1},
        \\
        \left\langle n_\nu \right|a_\nu a^\dagger_\nu\left|n_\nu\right\rangle&=\left\langle n_\nu \right|[N_\nu+1]\left|n_\nu\right\rangle=[n_\nu+1]=c_{+}^{n}c_{-}^{n+1},
    \end{aligned}
\end{equation}
By setting \( c_{-}^n = \sqrt{[n_\nu]} \) and \( c_{+}^n = \sqrt{[n_\nu + 1]} \), we obtain:
\begin{equation}
    \begin{aligned}
        a_\nu\left| n_\nu \right\rangle&=\sqrt{[n_\nu]}\left| n_\nu-1 \right\rangle,
        \\a^\dagger_\nu\left| n_\nu \right\rangle&=\sqrt{[n_\nu+1]}\left| n_\nu+1 \right\rangle,
    \end{aligned}
\end{equation}
As a result of the commutation relations and normalization choices, we can express \( |n_\nu\rangle \) in terms of the creation operator \( a_\nu^\dagger \) as:
\begin{equation}\label{Eq:desiredstateApp}
    \left| n_\nu \right\rangle=\frac{\left(a_\nu^{\dagger}\right)^{n}}{\sqrt{[n_\nu]!}}\left| 0 \right\rangle.
\end{equation}
This formulation provides a compact representation of the eigenstate \( |n_\nu\rangle \), constructed by applying the creation operator \( a_\nu^\dagger \) to the vacuum state \( |0\rangle \) \( n \) times, properly normalized.\\

\subsection{Mode expansion}\label{Mode-expansion}
Here we begin by examining the standard mode expansions for fermionic and bosonic systems.
Let $\hat{\psi}_X(x)$ represents an ordinary fermionic (X=F) or bosonic (X=B) field in the real space (represented by $x\equiv (\textbf{r},t)$ and $x'\equiv (\textbf{r}',t')$), subjected to the commutation or anti-commutation relations. More precisely 
\begin{equation}
    \begin{aligned}
       \left[\hat{\psi}^{}_F(x),\hat{\psi}_F^\dagger(x')\right]_+=\delta^d(\textbf{r}-\textbf{r}')\delta(t-t'),
    \end{aligned}
\end{equation}
where $\delta^d(\textbf{r})$ is $d$-dimensional delta function. Then the mode expansion for fermions is given by
\begin{equation}
    \begin{aligned}
    \label{WT}
        \hat{\psi}^{}_F(x)&=\frac{1}{\sqrt{V}}\sum_{p}e^{ip.x} \left(a^{(F)}_{p}+ b^{\dagger(F)}_{-p}\right),
        \\
        \hat{\psi}_F^\dagger(x)&=\frac{1}{\sqrt{V}}\sum_{p}e^{-ip.x} \left(a^{\dagger(F)}_{p}+ b^{(F)}_{-p}\right) .
    \end{aligned}
\end{equation}
where $V$ is the system size, $a_p^{(F)}$ and $a^{\dagger(F)}_{p}$ ($b_p^{(F)}$ and $b^{\dagger(F)}_{p}$) annihilate and create ferminonis particles (antiparticles) at $p$th mode. In the above formula we defined $p.x\equiv \textbf{p}.\textbf{r}-\omega_pt$, where $\omega_p=\epsilon_p/\hbar$ is the fermionic dispersion relation. A similar formulation applies for bosonic systems, where commutation relations are employed
\begin{equation}
    \begin{aligned}
        &\left[\hat{\psi}^{}_B(x),\dot{\hat{\psi}}_B^\dagger(x')\right]_-=i\hslash c^2\delta^d(\textbf{r}-\textbf{r}')\delta(t-t'),
    \end{aligned}
\end{equation}
and the bosonic mode expansion is as follows
\begin{equation}
    \begin{aligned}
        \hat{\psi}^{}_B(x)&=\sum_k\left(\frac{\hslash c^2}{2V\omega_p}\right)^{1/2}\left(a^{(B)}_p e^{-ip.x}+b^{\dagger(B)}_p e^{ip.x}\right),\\
         \hat{\psi}_B^\dagger(x)&=\sum_k\left(\frac{\hslash c^2}{2V\omega_p}\right)^{1/2}\left(b^{(B)}_p e^{-ip.x}+a^{\dagger(B)}_pe^{ip.x}\right),
    \end{aligned}
\end{equation}
where $a_p^{(B)}$ and $a^{\dagger(B)}_{p}$ ($b_p^{(B)}$ and $b^{\dagger(B)}_{p}$) annihilate and create boson particles (antiparticles) at $p$th mode.\\

We use the mode expansion  presented in Eq.~\eqref{mode-expantion-kqParticle}, resulting to the following commutation relation:
\begin{equation}\label{modeexpantionkq}
    \begin{aligned}
        &\left[\hat{\psi}(x_1),\hat{\psi}^\dagger(x_2)\right]_{kq}
        \\&=\left[\frac{1}{\sqrt{V}}\sum_p e^{ip^{}_1.x^{}_1} a_p , \frac{1}{\sqrt{V}}\sum_p e^{-ip^{}_2.x^{}_2} a^\dagger_p\right]_{kq}
        \\&=\frac{1}{V}\sum_{p^{}_1 p^{}_2}e^{ip^{}_1x^{}_1-ip^{}_2x^{}_2}\left[ a_{p^{}_1}, a^\dagger_{p^{}_2}\right]_{kq},
    \end{aligned}
\end{equation}
To accurately examine the commutation relation~\eqref{modeexpantionkq}, we add a term to the right-hand side of the equation that does not alter its value.
\begin{equation}\label{Eq:modekq}
    \begin{aligned}
        &\left[\hat{\psi}(x^{}_1),\hat{\psi}^\dagger(x^{}_2)\right]_{kq}
        \\&~=\frac{1}{V}\sum_{p^{}_1 p^{}_2}e^{ip^{}_1x^{}_1-ip^{}_2x^{}_2}\left(\delta^{}_{p^{}_1p^{}_2}+1-\delta^{}_{p^{}_1p^{}_2}\right)\left[ a^{}_{p^{}_1}, a^\dagger_{p^{}_2}\right]_{kq},        
    \end{aligned}
\end{equation}
when $p^{}_1$ is equal to $p^{}_2$, according to definition~\eqref{E1}, we can write $q^{-N_{p^{}_1}}$. Otherwise, we use this relation, which gives us a result of zero
\begin{equation}
    \left[ a^{}_{p^{}_1}, a^\dagger_{p^{}_2}\right]_{kq}=\left[ a^{}_{p^{}_1}, a^\dagger_{p^{}_2}\right]_{k}+k\left[ a^{}_{p^{}_1}, a^\dagger_{p^{}_2}\right]_{q}=0 \ \text{if} \ p^{}_1\neq p^{}_2,
\end{equation}
Therefore, Eq.~\eqref{Eq:modekq} can be rewritten as follows using the above definitions:
\begin{equation}
    \begin{aligned}
         \left[\hat{\psi}(x^{}_1),\hat{\psi}^\dagger(x^{}_2)\right]_{kq}
         &=\frac{1}{V}\sum_{p^{}_1}e^{ip^{}_1(x^{}_1-x^{}_2)}q^{-N_{p^{}_1}}
         \\&+\frac{1}{V}\sum_{p^{}_1\neq p^{}_2}e^{i(p^{}_1x^{}_1-p^{}_2x^{}_2)}\left[ a^{}_{p^{}_1}, a^\dagger_{p^{}_2}\right]_{kq},    
    \end{aligned}
\end{equation}
As a result, since the second term becomes zero~\eqref{Eq:Algebrakq2}, the final result will be as follows:
\begin{equation}
    \begin{aligned}
         \left[\hat{\psi}(x^{}_1),\hat{\psi}^\dagger(x^{}_2)\right]_{kq}
         =\frac{1}{V}\sum_{p^{}_1}e^{ip^{}_1(x^{}_1-x^{}_2)}q^{-N_{p^{}_1}}.    
    \end{aligned}
\end{equation}
\section{THE $S$-MATRIX THEORY}\label{s-matrix-theory}
 In this section We concentrate on the $S$-matrix theory. These details can be found in standard textbooks for quantum many body systems. Suppose that 
\begin{equation}
H(t)=H_0+H_{\text{int}}e^{-\epsilon t},
\end{equation}
where $H_0$ in non-interaction part of the Hamitonian (with eigenfunction $\Phi_0$), and $H_{\text{int}}$ is the interaction term which is multiplied by an exponential term containing  $\epsilon=0^+$, guaranteeing that the system is free for $t\to-\infty$, so that $\Phi_0$ evolves adiabaticaly to the ground state of interaction Hamiltonian and $t=0$. We define $S$-matrix as follows:
\begin{equation}
\left| \Psi_0\right\rangle =S(0,-\infty)\left| \Phi_0\right\rangle ,
\end{equation}
$S$ Matrix as the operator $S(t,t')$ which changes the wave function $\psi(t')$ into $\psi(t)$:
\begin{equation}
    \psi(t)=S(t,t')\psi(t')
\end{equation}
So here we are dealing with a time evolution operator $U(t)$ which $S(t,t')$ is a combination of  these operators in the interaction representation
\begin{equation}
    \psi(t)=U(t)\psi(0)=S(t,t')U(t')\psi(0),
\end{equation}
where $S(t,t')=U(t)U^(t')$. 
Thus, the expression for $S(t,t')$ in terms of $H_{\text{int}}$ is given by:
\begin{equation}\label{Stt}
   \begin{aligned}
        &\frac{\partial}{\partial t}S(t,t')
        \\&=\frac{\partial}{\partial t}U(t)U(t')=\left(ie^{iH_0t}\left(H_0-H\right)e^{-iHt}\right)U(t')
        \\&=-ie^{iH_0t}H^{}_{\text{int}}\left(e^{iH_0t}e^{-iH_0t}\right)e^{-iHt}U(t')=-iH^{}_{\text{int}}S(t,t'),
   \end{aligned}
\end{equation}
where $U(t)=e^{iH_0t}e^{iHt}$.
To solve this equation, we perform integration on both sides with respect to the time variable
\begin{equation}
    \begin{aligned}
        S(t,t')=1-i\int^t_{t'}dt_1H^{}_{\text{int}}(t_1)S(t_1-t'),
    \end{aligned}
\end{equation}
By repeatedly applying this equation, we obtain the following result:
\begin{equation}\label{S}
    \begin{aligned}
        S(t,t')=\sum^\infty_{n=0}(-i)^n\int^t_{t'}dt_1\int^{t_1}_{t'}dt_2...\int^{t_{n-1}}_{t'}dt_n
        &\\\times H^{}_{\text{int}}(t_1)H^{}_{\text{int}}(t_2)...H^{}_{\text{int}}(t_n)&,
    \end{aligned}
\end{equation}
Continuing these calculations, according to the definition of the ordinary time order operator, we have
\begin{equation}
   \begin{aligned}
        &T\left[H^{}_{\text{int}}(t_1)H^{}_{\text{int}}(t_2)\right]
        \\&=\Theta(t_1-t_2)H^{}_{\text{int}}(t_1)H^{}_{\text{int}}(t_2)+\Theta(t_2-t_1)H^{}_{\text{int}}(t_2)H^{}_{\text{int}}(t_1).
   \end{aligned}
\end{equation}
Now we rearrange this integral using the identity above:
\begin{equation}
    \begin{aligned}
    \frac{1}{2!}&\int^t_{t'}dt_1\int^t_{t'}dt_2T\left[H^{}_{\text{int}}(t_1)H^{}_{\text{int}}(t_2)\right]
    \\&=\frac{1}{2!}\int^t_{t'}dt_1\int^{t_1}_{t'}dt_2H^{}_{\text{int}}(t_1)H^{}_{\text{int}}(t_2)\\&+\frac{1}{2!}\int^t_{t'}dt_2\int^{t_2}_{t'}dt_1H^{}_{\text{int}}(t_2)H^{}_{\text{int}}(t_1),
    \end{aligned}
\end{equation}
The second term on the right-hand side is equal to the first, which is easy to see by just  redefining the integration variables $t_1\to t_2$ , $t_2\to t_1$. This process yields: 
\begin{equation}\label{2factoril}
    \begin{aligned}
        &\frac{1}{2!}\int^t_{t'}dt_1\int^t_{t'}dt_2T\left[H^{}_{\text{int}}(t_1)H^{}_{\text{int}}(t_2)\right]\\&=\int^t_{t'}dt_1\int^{t_1}_{t'}dt_2H^{}_{\text{int}}(t_1)H^{}_{\text{int}}(t_2).
    \end{aligned}
\end{equation}
By utilizing Eq.~\eqref{2factoril}, we can derive the expansion form of $S(t,t')$(Eq.~\eqref{S}) as follows:
\begin{equation}\label{smatrix}
    \begin{aligned}
        S(t,t')&=1+\sum^\infty_{n=1}\frac{(-i)^n}{n!}\int^t_{t'}dt_1\int^t_{t'}dt_2...\int^t_{t'}dt_n\\&\times T\left[H^{}_{\text{int}}(t_1)H^{}_{\text{int}}(t_2)...H^{}_{\text{int}}(t_n)\right]\\
        &=T\exp\left[-i\int_{t'}^tdt_1H^{}_{\text{int}}(t_1)\right].
    \end{aligned}
\end{equation}
We want to prove a basic theorem that relates the matrix element of a Heisenberg operator $\hat{O}_H(t)$ to the matrix element of the corresponding interaction operator $\hat{O}_{\text{int}}(t)$: 
\begin{equation}\label{ssmatrix}
    \begin{aligned}
        &\frac{\langle\Psi_0|\hat{O}_H(t)|\Psi_0\rangle}{\langle\Psi_0|\Psi_0\rangle}=\frac{1}{\langle\Phi_0|\hat{S}|\Phi_0\rangle}\langle\Phi_0|\sum^\infty_{\nu=0}\left(-i\right)^\nu\frac{1}{\nu!}\\&\times \int^\infty_{-\infty}dt_1...\int^\infty_{-\infty}dt_\nu T\left[\hat{H}_{\text{int}}(t_1)...\hat{H}_{\text{int}}(t_\nu)\hat{O}_{\text{int}}(t)\right]|\Phi_0\rangle
    \end{aligned}
\end{equation}
where $\hat{S}=\hat{U}(\infty,-\infty)$, and this definition is already known
 \begin{equation}\label{defU}
    \begin{aligned}
        &\hat{U}(t,t_0)
        =\sum^\infty_{n=0}\frac{\left(-i\right)^n}{n!}\int^t_{t_0}dt_1...\int^t_{t_0}dt_n T\left[\hat{H}_{\text{int}}(t_1)...\hat{H}_{\text{int}}(t_n)\right]
    \end{aligned}
\end{equation}
The proof is as follows: The Gell-Mann and Low theorem expresses the ground state of the interacting system in the interaction picture 
\begin{equation}
    |\Psi_0\rangle=\hat{U}(0,-\infty)|\Phi_0\rangle
\end{equation}
We write the denominator on the left side of Eq.~\eqref{ssmatrix} as follows
\begin{equation}
    \begin{aligned}
        \langle\Psi_0|\Psi_0\rangle&=\langle\Phi_0|\hat{U}(0,\infty)^\dagger\hat{U}(0,-\infty)|\Phi_0\rangle\\&=\langle\Phi_0|\hat{U}(\infty,-\infty)|\Phi_0\rangle=\langle\Phi_0|\hat{S}|\Phi_0\rangle
    \end{aligned}
\end{equation}
where $\hat{U}(t_1,t_2)\hat{U}(t_2,t_3)=\hat{U}(t_1,t_3)$ and $\hat{U}(t_0,t)=\hat{U}^\dagger(t,t_0)$ have been used.
We must pay attention to the fact that the time dependence of the operators in the Heisenberg representation and interaction representation are different as follows 
\begin{equation}
  \hat{O}(t) = \begin{cases}
   \hat{O}_H(t)=e^{iHt}\hat{O}(0)e^{-iHt}  \\
   \hat{O}_{\text{int}}(t)=e^{iH_0t}\hat{O}(0)e^{-iH_0t} 
\end{cases}  
\end{equation}
Therefore, the relation between $\hat{O}_H(t)$ and $\hat{O}_{\text{int}}(t)$ can be written as follows
\begin{equation}\label{U}
    \begin{aligned}
        &\hat{O}_H(t)=e^{iHt}\left(e^{-iH_0t}e^{iH_0t}\right)\hat{O}(0)\left(e^{-iH_0t}e^{iH_0t}\right)e^{-iHt}\\&=e^{iHt}e^{-iH_0t}\hat{O}_{\text{int}}(t)e^{iH_0t}e^{-iHt}=\hat{U}(0,t)\hat{O}_{\text{int}}\hat{U}(t,0)
    \end{aligned}
\end{equation}
In a similar way, the numerator on the left side of Eq.~\eqref{ssmatrix} is obtained using Eq.~\eqref{U}
\begin{equation}
    \begin{aligned}
        &\langle\Psi_0|\hat{O}_H(t)|\Psi_0\rangle\\&=\langle\Phi_0|\hat{U}(\infty,0)\hat{U}(0,t)\hat{O}_{\text{int}}(t)\hat{U}(t,0)|\hat{U}(0,-\infty)|\Phi_0\rangle\\&=\langle\Phi_0|\hat{U}(\infty,t)\hat{O}_{\text{int}}(t)\hat{U}(t,-\infty)|\Phi_0\rangle
    \end{aligned}
\end{equation}
As a result, the left side of Eq.~\eqref{ssmatrix} is of this form
\begin{equation}
    \frac{\langle\Psi_0|\hat{O}_H(t)|\Psi_0\rangle}{\langle\Psi_0|\Psi_0\rangle}=\frac{\langle\Phi_0|\hat{U}(\infty,t)\hat{O}_{\text{int}}(t)\hat{U}(t,-\infty)|\Phi_0\rangle}{\langle\Phi_0|\hat{S}|\Phi_0\rangle}
    \label{Ag}
\end{equation}
In Eq.~\eqref{Ag}, we can rewrite the numerator of the right side 
\begin{equation}
    \begin{aligned}
        &\hat{U}(\infty,t)\hat{O}_{\text{int}}(t)\hat{U}(t,-\infty)\\&=\sum^\infty_{n=0}\frac{\left(-i\right)^n}{n!}\int^\infty_{t}dt_1...\int^\infty_{t}dt_n T\left[\hat{H}_{\text{int}}(t_1)...\hat{H}_{\text{int}}(t_n)\right]\hat{O}_{\text{int}}(t)\\&~~~\sum^\infty_{m=0}\frac{\left(-i\right)^m}{m!}\int^t_{-\infty}dt_1...\int^t_{-\infty}dt_m T\left[\hat{H}_{\text{int}}(t_1)...\hat{H}_{\text{int}}(t_m)\right].
        \label{Ag1}
    \end{aligned}
\end{equation}
In the $\nu$th term of the sum in Eq.~\eqref{ssmatrix}, divide the integration variables into $n$ factors with $t_i>t$ and $m$ factors with $t_i<t$ , where $m+n=\nu$. The right side of Eq.~\eqref{ssmatrix} becomes
\begin{equation}
    \begin{aligned}
        &\sum^\infty_{\nu=0}\frac{\left(-i\right)^\nu}{\nu!}\int^\infty_{-\infty}dt_1...\int^\infty_{-\infty}dt_\nu T\left[\hat{H}_{\text{int}}(t_1)...\hat{H}_{\text{int}}(t_\nu)\hat{O}_{\text{int}}(t)\right]\\&
        =\sum^\infty_{\nu=0}\frac{\left(-i\right)^\nu}{\nu!}\sum^\infty_{n=0}\sum^\infty_{m=0}\delta^{}_{\nu,m+n}\frac{\nu!}{m!n!}\\&\times\underbrace{\int^\infty_{t}dt_1...\int^\infty_{t}dt_n T\left[\hat{H}_{\text{int}}(t_1)...\hat{H}_{\text{int}}(t_n)\right]}_{t_i>t}\\&\times\hat{O}_{\text{int}}(t)\underbrace{\int^t_{-\infty}dt_1...\int^t_{-\infty}dt_mT\left[\hat{H}_{\text{int}}(t_1)...\hat{H}_{\text{int}}(t_m)\right]}_{t_i<t},
        \label{Result}
    \end{aligned}
\end{equation}
The Kronecker delta here ensures that $m+n=\nu$, but it also can be used to perform  the summation over $\nu$, which proves the theorem because Eq.~\eqref{Result} then reduces  to Eq.~\eqref{Ag1}.

In a similar manner, the expectation value of time-ordered Heisenberg  operators may be written as 
\begin{equation}
    \begin{aligned}
        &\frac{\langle\Psi_0|T\left[\hat{O}_H(t)\hat{O}_H(t')\right]|\Psi_0\rangle}{\langle\Psi_0|\Psi_0\rangle}=\frac{1}{\langle\Phi_0|\hat{S}|\Phi_0\rangle}
        \\&\langle\Phi_0|\sum^\infty_{\nu=0}\left(-i\right)^\nu\frac{1}{\nu!}\int^\infty_{-\infty}dt_1...\int^\infty_{-\infty}dt_\nu \\&\times T\left[\hat{H}_{\text{int}}(t_1)...\hat{H}_{\text{int}}(t_\nu)\hat{O}_{\text{int}}(t)\hat{O}_{\text{int}}(t')\right]|\Phi_0\rangle.
        \label{Wick2}
    \end{aligned}
\end{equation}
\section{THE WICK'S THEOREM FOR FERMIONS AND BOSONS}
Our purpose in stating this case is how to deal with time ordering like this expression
\begin{equation}
    \langle\phi_0|T\{\hat{\psi}^\dagger...\hat{\psi}\hat{\psi}_\alpha(x)\hat{\psi}_\beta^\dagger(y)\}|\phi_0\rangle
\end{equation}
Proof of Wick's theorem: In the following, we first derive Wick's theorem for two fields, and for a large number of fields we will prove this theorem according to mathematical induction. We want to obtain the following relation
\begin{equation}\label{2Feild}
    \begin{aligned}
        &T\left\lbrace\hat{\psi}(x_1)\hat{\psi}(x_2)\right\rbrace\equiv N\left\lbrace\hat{\psi}(x_1)\hat{\psi}(x_2)+\overparent{\hat{\psi}(x_1)\hat{\psi}(x_2)}\right\rbrace\\&=N\left\lbrace\hat{\psi}(x_1)\hat{\psi}(x_2)\right\rbrace+\left[\hat{\psi}^{(+)}(x_1),\hat{\psi}^{(-)}(x_2)\right],
    \end{aligned}
\end{equation}
where $\overparent{\psi(x_1)\psi(x_2)}$ is the contraction between two fields, which is actually the same propagator $D(x_1,x_2)$.
The expanded form of Eq.~\eqref{2Feild} is as follows:
\begin{equation}
    \begin{aligned}
        T&\left\lbrace\hat{\psi}(x_1)\hat{\psi}(x_2)\right\rbrace\xlongequal{t_1>t_2}\hat{\psi}(x_1)\hat{\psi}(x_2)
        \\&=\left(\hat{\psi}^{(+)}(x_1)+\hat{\psi}^{(-)}(x_1)\right)\left(\hat{\psi}^{(+)}(x_2)+\hat{\psi}^{(-)}(x_2)\right)\\&
        =\hat{\psi}^{(+)}(x_1)\hat{\psi}^{(+)}(x_2)+\hat{\psi}^{(-)}(x_2)\hat{\psi}^{(+)}(x_1)
        \\&~~~+\hat{\psi}^{(-)}(x_1)\hat{\psi}^{(+)}(x_2)+\hat{\psi}^{(-)}(x_1)\hat{\psi}^{(-)}(x_2)\\&
        ~~~+\left[\hat{\psi}^{(+)}(x_1),\hat{\psi}^{(-)}(x_2)\right]
        \label{W2}
    \end{aligned}
\end{equation}
where $\hat{\psi}^{(+)}(x)=\sum_{k}(1/\sqrt{V})e^{i(k.x-\omega_kt)} a^{(F)}_{k} \ , \ \hat{\psi}^{(-)}(x)=\sum_{k}(1/\sqrt{V})e^{i(k.x-\omega_kt)} b^{\dagger(F)}_{-k}$.
The first four terms of the above expression~\eqref{W2} are in normal ordering, and the last term is related to the contraction between the fields. This means that the expected value of $ T\left\lbrace\hat{\psi}(x_1)\hat{\psi}(x_2)\right\rbrace$ in the basic state is equal to the contraction between the fields
$\langle 0|T\left\lbrace\hat{\psi}(x_1)\hat{\psi}(x_2)\right\rbrace|0\rangle=D(x_1,x_2)$.
To prove Wick's theorem for a large number of fields, we act in such a way that if this theorem is true for two fields~\eqref{2Feild}, we can prove Wick's theorem for three fields as well 
\begin{equation}
    \begin{aligned}
        &T\left\lbrace\hat{\psi}(x_1)\hat{\psi}(x_2)\hat{\psi}(x_3)\right\rbrace
        \\&=N\left\lbrace\hat{\psi}(x_1)\hat{\psi}(x_2)\hat{\psi}(x_3)+ \text{all possible contractions}\right\rbrace,
    \end{aligned}
\end{equation}
So we will get to
\begin{equation}
    \begin{aligned}
        &T\left\lbrace\hat{\psi}(x_1)\hat{\psi}(x_2)\hat{\psi}(x_3)\right\rbrace\\&=N\{\hat{\psi}(x_1)\hat{\psi}(x_2)\hat{\psi}(x_3)+ \overparent{\hat{\psi}(x_1)\hat{\psi}(x_2)}\hat{\psi}(x_3)\\&+\overparent{\hat{\psi}(x_1)\hat{\psi}(x_3)}\hat{\psi}(x_2)+\hat{\psi}(x_1)\overparent{\hat{\psi}(x_2)\hat{\psi}(x_3)}\}.
        \label{3feild}
    \end{aligned}
\end{equation}
With this method, Wick's theorem can be proved for $n$ fields.
\section{THE GENERALIZED WICK'S THEOREM FOR $kq$-PARTICLES}
\subsection {A Proof for the Wick's theorem for $kq$-particles}

Here we use the usual time order, considering that only our fields obey $kq$-deformed Algebra. Since $t_1>t_2$ and the $T_\epsilon$ operation arranges things so that the earlier time is on the right, this expression is already time-ordered, so we have
\begin{equation}\label{Tu}
    \begin{aligned}
        &\mathcal{T}\left\lbrace\hat{\Psi}_1(x_1)\hat{\Psi}_2(x_2)\right\rbrace=\hat{\Psi}_1(x_1)\hat{\Psi}_2(x_2)
        \\&
        =\hat{\psi}^{(+)}(x_1)\hat{\psi}^{(+)}(x_2)+kq\hat{\psi}^{(-)}(x_2)\hat{\psi}^{(+)}(x_1)\\&~~~+\left[\hat{\psi}^{(+)}(x_1),\hat{\psi}^{(-)}(x_2)\right]_{kq}+\hat{\psi}^{(-)}(x_1)\hat{\psi}^{(+)}(x_2)\\&~~~+\hat{\psi}^{(-)}(x_1)\hat{\psi}^{(-)}(x_2),
    \end{aligned}
\end{equation}
where $\hat{\Psi}(x)=\hat{\psi}^{(+)}(x)+\hat{\psi}^{(-)}(x)$. We have
\begin{equation}
    \begin{aligned}
        \mathcal{T}\left\lbrace\hat{\Psi}_1(x_1)\hat{\Psi}_2(x_2)\right\rbrace&=\mathcal{N}\left\lbrace\hat{\Psi}_1(x_1)\hat{\Psi}_2(x_2)\right\rbrace\\&+\left[\hat{\psi}^{(+)}(x_1),\hat{\psi}^{(-)}(x_2)\right]_{kq},
    \end{aligned}
\end{equation}
The last term is an example of a contraction. Contractions are defined for $kq$-particles as
\begin{equation}\label{Contraction}
    \overparent{\hat{\Psi}_1(x_1)\hat{\Psi}_2(x_2)} \equiv \begin{cases}
   \left[\hat{\psi}^{(+)}(x_1),\hat{\psi}^{(-)}(x_2)\right]_{kq} &\text{if ~~} t_1>t_2, \\
   \\
   kq\left[\hat{\psi}^{(+)}(x_2),\hat{\psi}^{(-)}(x_1)\right]_{kq} &\text{if ~~} t_2>t_1,
\end{cases}
\end{equation}
Thus we have, for $t_1>t_2$
\begin{equation}
    \mathcal{T}\left\lbrace\hat{\Psi}_1(x_1)\hat{\Psi}_2(x_2)\right\rbrace=\mathcal{N}\left\lbrace\hat{\Psi}_1(x_1)\hat{\Psi}_2(x_2)\right\rbrace+\overparent{\hat{\Psi}_1(x_1)\hat{\Psi}_2(x_2)}.
\end{equation}
\begin{widetext}
For three fields, considering all the contractions between the fields, we will have
\begin{equation}
    \begin{aligned}
        &\mathcal{T}\left\lbrace\hat{\Psi}_1(x_1)\hat{\Psi}_2(x_2)\hat{\Psi}_3(x_3)\right\rbrace
        \\&=\mathcal{N}\left\lbrace\hat{\Psi}_1(x_1)\hat{\Psi}_2(x_2)\hat{\Psi}_3(x_3)+ \overparent{\hat{\Psi}_1(x_1)\hat{\Psi}_2(x_2)}\hat{\Psi}_3(x_3)+\hat{\Psi}_1(x_1)\overparent{\hat{\Psi}_2(x_2)\hat{\Psi}_3(x_3)}+kq\overparent{\hat{\Psi}_1(x_1)\hat{\Psi}_3(x_3)}\hat{\Psi}_2(x_2)\right\rbrace
    \end{aligned}
\end{equation}
As a result wick's theorem for $n$  fields  is as follows:
 \begin{equation}\label{Wick's theorem}
    \begin{aligned}
        \mathcal{T}\left\lbrace\hat{\Psi}_1(x_1)\hat{\Psi}_2(x_2)...\hat{\Psi}_n(x_n)\right\rbrace
        =&\mathcal{N}\Big\{\hat{\Psi}_1(x_1)\hat{\Psi}_2(x_2)...\hat{\Psi}_n(x_n)\Big\}+\overparent{\hat{\Psi}_1(x_1)\hat{\Psi}_2(x_2)}\mathcal{N}\Big\{\hat{\Psi}_3(x_3)...\hat{\Psi}_n(x_{n})\Big\}\\
        +kq\overparent{\hat{\Psi}_1(x_1)\hat{\Psi}_3(x_3)}
        \mathcal{N}\Big\{\hat{\Psi}_2(x_2)...\hat{\Psi}_n(x_{n})\Big\}&+...+(kq)^{n-2}\overparent{\hat{\Psi}_1(x_1)\hat{\Psi}_n(x_n)}\mathcal{N}\Big\{\hat{\Psi}_3(x_3)...\hat{\Psi}_n(x_{n-1})\Big\}+\text{all pair contractions}\\
        +...+(kq)^{\mathcal{P}_{i,j;i',j'}}\overparent{\hat{\Psi}_i(x_i)\hat{\Psi}_j(x_j)}&\overparent{\hat{\Psi}_{i'}(x_{i'})\hat{\Psi}_{j'}(x_{j'})}\mathcal{N}\Big\{\hat{\Psi}_1(x_1)...\hat{\Psi}_n(x_{n})\Big\}_{i,j,i',j'}+\text{all possible higher order contractions}.
    \end{aligned}
\end{equation}
\end{widetext}
\section{SOME IDENTITIES FOR THE STANDARD GREEN FUNCTION}\label{Appendix:B}
In this section, we write how to calculate the Eq.~\eqref{38}
\begin{equation}
    \begin{aligned}
        &i\frac{\partial}{\partial\tau} G_{kq}^{(\mathcal{M},X,0)}(\nu,\tau)=\frac{\partial}{\partial\tau}\Big\lbrace-\left\langle \mathcal{T} a_\nu(\tau)a^\dagger_\nu(0)\right\rangle\Big\rbrace\\&=\frac{\partial}{\partial\tau}\Big\lbrace-\Theta(\tau)\left\langle a_\nu(\tau)a^\dagger_\nu(0)\right\rangle-kq\Theta(-\tau)\left\langle a^\dagger_\nu(0)a_\nu(\tau)\right\rangle\Big\rbrace
        \\&
        =-\delta(\tau)\left\langle \left[a_\nu(\tau),a^\dagger_\nu(0)\right]_{kq}\right\rangle-\left\langle \mathcal{T} \frac{\partial a_\nu(\tau)}{\partial\tau}a^\dagger_\nu(0)\right\rangle,
        \label{Eq:appGreen}
    \end{aligned}
\end{equation}
The Becker-Hausdorff theorem is expressed in the following way, we will use this theorem
\begin{equation}
\begin{aligned}
     &e^A C e^{-A}
     =C+\left[A,C\right]+\frac{1}{2!}\left[A,\left[A,C\right]\right]+...
     \\
      &e^{\tau H} a_\nu e^{-\tau H}=a_\nu+\left[\tau H,a_\nu\right]
      +\frac{1}{2!}\left[\tau H,\left[\tau H,a_\nu\right]\right]+...
\end{aligned}
\label{BF}
\end{equation}
according to $H=\sum_{\nu} (\epsilon_{\nu}-\mu)\left[N_{\nu}\right]$ , Eq.~\eqref{Anomber} and Eq.~\eqref{cp} the first and second terms are obtained in this way 
\begin{equation}
\begin{aligned}
        \left[a^\dagger_\nu a^{}_\nu,a^{}_\nu\right]_{kq}& =R a_\nu,
\end{aligned}
\end{equation}
where $R=(k(1-q)[N^{}_\nu]-q^{-N_\nu})$. Given that $\left[H,N_\nu\right]_k=0$, then we have
\begin{equation}
\begin{aligned}
        \left[N_\nu,R a_\nu\right]
        =R \left[N_\nu,a_\nu\right]=R^2 a_\nu,
\end{aligned}
\end{equation}
And according to the Becker-Hausdorff theorem~\eqref{BF}, we have
\begin{equation}
 \begin{aligned}
     a_\nu(\tau)&=e^{\tau H} a_\nu e^{-\tau H}
     \\&
     =a_\nu+\tau\zeta_\nu R a_\nu+\frac{1}{2!}{\tau\zeta_\nu}^2 R^2 a_\nu+\frac{1}{3!}{\tau\zeta_\nu}^3 R^3 a_\nu+...
     \\&=(\sum_{n=0}^\infty\frac{(\tau\zeta_\nu R)^n}{n!})a_\nu=e^{\tau\zeta^{}_\nu R}a_\nu,
 \end{aligned}   
\end{equation}
By using the above definitions, the Eq.~\eqref{Eq:appGreen} comes in this form
\begin{equation}\label{E6}
    \begin{aligned}
        &i\frac{\partial}{\partial\tau} G_{kq}^{(\mathcal{M},X,0)}(\nu,\tau)
        \\&=-\delta(\tau)\left\langle \left[a_\nu(\tau),a^\dagger_\nu(0)\right]_{kq}\right\rangle-\left\langle \mathcal{T} \frac{\partial a_\nu(\tau)}{\partial\tau}a^\dagger_\nu(0)\right\rangle
        \\&=-\delta(\tau)\left\langle q^{-N^{}_\nu}\right\rangle-\left\langle \mathcal{T} \zeta_\nu R a_\nu(\tau)a^\dagger_\nu(0)\right\rangle,
    \end{aligned}
\end{equation}
By substituting $q^{-N^{}_\nu}=\left([kN^{}_\nu+1]-kq[N^{}_\nu]\right)^{k}_{}$, into the Eq.~\eqref{E6}, we can rewrite it in a form where the eigenstates of the operators are explicitly shown, and we have:
\begin{equation}
    \begin{aligned}
        &i\frac{\partial}{\partial\tau} G_{kq}^{(\mathcal{M},X,0)}(\nu,\tau)=-\delta(\tau)\left([kn^{}_\nu+1]-kq[n^{}_\nu]\right)^{k}_{}
        \\&+\zeta_\nu\Big(k(1-q)[n^{}_\nu]-\left([kn^{}_\nu+1]-kq[n^{}_\nu]\right)^{k}_{}\Big)G_{kq}^{(\mathcal{M},X,0)}(\nu,\tau),
    \end{aligned}
\end{equation}
In order to determine the Green's function in the frequency domain, we need to perform a Fourier transform. To achieve this, we will use the following definitions:
\begin{equation*}
\begin{cases}
    \begin{aligned}
        &G_{kq}^{(\mathcal{M},X,0)}(\nu,\tau)=\frac{1}{\beta}\sum_{i\omega_n^{(X,j)}}e^{-i\omega_n^{(X,j)}\tau}_{}G_{kq}^{(\mathcal{M},X,0)}(\nu,i\omega_n^{(X,j)}),\\&
        \frac{d}{d\tau}\to -i\omega_n^{(X,j)},\\&
        \delta(\tau)=\frac{1}{\beta}\sum_{i\omega_n^{(X,j)}}e^{-i\omega_n^{(X,j)}\tau}_{}.
    \end{aligned}
\end{cases}
\end{equation*}
Upon Fourier transforming the Green's function, we obtain the following expression:
\begin{equation}
    \begin{aligned}
    &iG_{kq}^{(\mathcal{M},X,0)}(\nu,i\omega_n^{(X,j)})
    \\&=\frac{\left([kn^{}_\nu+1]-kq[n^{}_\nu]\right)^{k}_{}}{i\omega_n^{(X,j)}+\zeta_\nu\Big(k(1-q)[n^{}_\nu]-\left([kn^{}_\nu+1]-kq[n^{}_\nu]\right)^{k}_{}\Big)}.
    \end{aligned}
\end{equation}
\section{THE GORKOV EQUATION FOR THE EFFECTIVE GREEN FUNCTION}\label{app:effective}
In this section, the details of calculating the Gorkov Eq.~\eqref{Eq:EffecGreenGorkov} for the effective Green's function are given. We start with Eq.~\eqref{green0}:
\begin{equation}
\begin{aligned}
    &i\mathcal{G}_{kq}^{(0)}(\zeta^{}_\nu,\tau)=\sum_{j,\omega_n^{(X,j)}}b_n^{(X,j)}\log\left[\frac{\zeta^{}_\nu-i\omega_n^{(X,j)}}{a-i\omega_n^{(X,j)}}\right] e^{-i\omega_n^{(X,j)}\tau},
\end{aligned}
\end{equation}
The derivative of the Green's function yields the following expression:
\begin{equation}\label{derivativeGF}
    \begin{aligned}
       &i
    \frac{\partial}{\partial\tau} \mathcal{G}_{kq}^{(0)}(\zeta^{}_\nu,\tau)\\&
    =\sum_{j,\omega_n^{(X,j)}} b_n^{(X,j)}(-i\omega_n^{(X,j)})\log\left[\frac{\zeta^{}_\nu-i\omega_n^{(X,j)}}{a-i\omega_n^{(X,j)}}\right] e^{-i\omega_n^{(X,j)}\tau}\\&
    =\sum_{j,\omega_n^{(X,j)}} b_n^{(X,j)}(i\omega_n^{(X,j)}) e^{-i\omega_n^{(X,j)}\tau}\times
    \\&~~~~\times\bigg((\zeta^{}_\nu-i\omega_n^{(X,j)})\log(\zeta^{}_\nu-i\omega_n^{(X,j)})
    \\&~~~~~~~~-(a-i\omega_n^{(X,j)})\log(a-i\omega_n^{(X,j)})
    \\&~~~~~~~~-\zeta^{}_\nu\log(\zeta^{}_\nu-i\omega_n^{(X,j)})+a\log(a-i\omega_n^{(X,j)})\bigg),
    \end{aligned}
\end{equation}
We also know that
\begin{equation}
    \begin{aligned}
            \int_1^x \log\eta d\eta=x\log(x)-x+1
    \end{aligned}
\end{equation}
And using the above definition, Eq.~\eqref{derivativeGF} can be written in a different form:
\begin{equation}
    \begin{aligned}
        i\frac{\partial}{\partial\tau}& \mathcal{G}_{kq}^{(0)}(\zeta^{}_\nu,\tau)=
        \sum_{j,\omega_n^{(X,j)}} b_n^{(X,j)} e^{-i\omega_n^{(X,j)}\tau}\\&\times\Bigg(\zeta^{}_\nu-i\omega_n^{(X,j)}-1+\int_1^{\zeta^{}_\nu-i\omega_n^{(X,j)}}\log\eta d\eta\\&
        ~~~~-a+i\omega_n^{(X,j)}+1-\int_1^{a-i\omega_n^{(X,j)}}\log\eta d\eta\\&
        ~~~~-\zeta_\nu\log(\zeta^{}_\nu-i\omega_n^{(X,j)})+\zeta^{}_\nu\log(a-i\omega_n^{(X,j)})
        \\&~~~~-\zeta^{}_\nu\log(a-i\omega_n^{(X,j)})+a\log(a-i\omega_n^{(X,j)})\Bigg),
    \end{aligned}
\end{equation}
This yields:
\begin{equation}\label{eqGreen5}
    \begin{aligned}
        i\frac{\partial}{\partial\tau} \mathcal{G}_{kq}^{(0)}(\zeta^{}_\nu,\tau)&=\sum_{j,\omega_n^{(X,j)}} b_n^{(X,j)} e^{-i\omega_n^{(X,j)}\tau}\Bigg(\zeta^{}_\nu-a\\&+\int_{a-i\omega_n^{(X,j)}}^{\zeta^{}_\nu-i\omega_n^{(X,j)}}\log\eta d\eta
        -\zeta_\nu^{}\log\frac{\zeta^{}_\nu-i\omega_n^{(X,j)}}{a-i\omega_n^{(X,j)}}\\&-(\zeta^{}_\nu-a)\log(a-i\omega_n^{(X,j)})\Bigg),
    \end{aligned}
\end{equation}
To simplify the calculations, we will introduce a change of variables and add some additional terms
\begin{equation}
    \begin{aligned}
        &\int_{a-i\omega_n^{(X,j)}}^{\zeta^{}_\nu-i\omega_n^{(X,j)}}\log\eta d\eta ~:~ \eta\equiv x-i\omega_n^{(X,j)} \\&
        ~\to~
        \int_{a}^{\zeta^{}_\nu}\log(x-i\omega_n^{(X,j)})dx=
        \int_{a}^{\zeta^{}_\nu}\bigg(\log(x-i\omega_n^{(X,j)})\\&~~~~~~~~~~~~~~~~-\log(a-i\omega_n^{(X,j)})+\log(a-i\omega_n^{(X,j)})\bigg) dx\\&=\int_{a}^{\zeta^{}_\nu}\log\left[\frac{x-i\omega_n^{(X,j)}}{a-i\omega_n^{(X,j)}}\right]dx+(\zeta^{}_\nu-a)\log(a-i\omega_n^{(X,j)}),
    \end{aligned}
\end{equation}
Substituting the above relation into Eq.~\eqref{eqGreen5}, we obtain the following:
\begin{equation}
    \begin{aligned}
        \partial_\tau i\mathcal{G}^{(0)}_{kq}(\zeta^{}_\nu,\tau)
        &=\frac{(\zeta^{}_\nu-a)}{q-q^{-1}}(q^{\tfrac{-\tau}{\beta}}-q^{\tfrac{\tau}{\beta}})\delta(\tau)\\&-\zeta^{}_\nu i\mathcal{G}^{(0)}_{kq}(\zeta^{}_\nu,\tau)+\int_a^{\zeta^{}_\nu} i\mathcal{G}^{(0)}_{kq}(x,\tau)dx.
    \end{aligned} 
\end{equation}
To demonstrate the validity of Eq.~\eqref{taubeta}, we will follow the steps outlined below:
\begin{equation}\label{G-tau-beta}
    \begin{aligned}
        \mathcal{G}_{kq}^{}(\nu,\tau+\beta)
        =\frac{1}{\beta}\sum_{j,\omega_n^{(X,j)}}e^{-i\omega_n^{(X,j)}(\tau+\beta)} \mathcal{G}_{kq}^{}(\nu,i \omega_n^{(X,j)})
        &\\=\frac{1}{\beta}\sum_{j,\omega_n^{(X,j)}}e^{-i\omega_n^{(X,j)}\tau}e^{-i\omega_n^{(X,j)}\beta} \mathcal{G}_{kq}^{}(\nu,i \omega_n^{(X,j)})&,
    \end{aligned}
\end{equation}
Using the definition of generalized Matsubara frequencies in Eq.~\eqref{Eq:Matsubara}, we have:
\begin{equation}
    \begin{aligned}
        e^{-i\omega_n^{(X,j)}\beta}=kq^{k(-1)^j}_{},
    \end{aligned}
\end{equation}
Therefore
\begin{equation}
    \mathcal{G}_{kq}(\nu,\tau+\beta)=k q^{k(-1)^j}_{}\mathcal{G}_{kq}(\nu,\tau).
\end{equation}
\begin{widetext}
\section{FEYNMAN DIAGRAMS AND THE PERTURBATION THEORY}\label{Appendix:DF}
An illustrative example is the exact Green's function, which can be formulated as
\begin{equation}
    \begin{aligned}
        iG^{(\mathcal{M},X)}(x,y)=\sum_{m=0}^{\infty}\frac{(-i)^m)}{m!}\int_{0}^{\beta}d\tau_1...\int_{0}^{}d\tau_m 
        \left\langle\mathcal{T}[\hat{H}_{\text{int}}(\tau_1)...\hat{H}_{\text{int}}(
        \tau_m)\hat{\psi}(x)\hat{\psi}^\dagger(y)]\right\rangle,
    \end{aligned}\label{Greenfunction_S}
\end{equation}
It's helpful to express the interparticle potential in $\hat{H}_{\text{int}}$ as $    u(x_1,x_2)\equiv v(x_1,x_2)\delta(t_1-t_2), 
$, which allows us to write the integrations symmetrically.
For example, the  Eq.~\eqref{Greenfunction_S}, which we will denote by $i\tilde{G}$, becomes 
\begin{equation}
      i\tilde{G}^{(\mathcal{M},X)}(x,y)\equiv iG^{(\mathcal{M},X)}(x,y)+i\tilde{G}^{(\mathcal{M},X)}_{(1)}(x,y)
      \label{Eq:I-G-function},
\end{equation}
where
\begin{equation}
    \begin{aligned}
     iG^{(\mathcal{M},X)}(x,y)&=\langle \mathcal{T}[\hat{\psi}(x)\hat{\psi}^\dagger(y)]\rangle,
     \\
        i\tilde{G}^{(\mathcal{M},X)}_{(1)}(x,y)&=\left(-i\right)\frac{1}{2}\int d^4x_1 d^4x'_1u(x_1,x'_1)          
      \left\langle \mathcal{T}[\hat{\psi}^\dagger(x_1)\hat{\psi}^\dagger(x'_1)\hat{\psi}(x'_1)\hat{\psi}(x_1)\hat{\psi}(x)\hat{\psi}^\dagger(y)]\right\rangle.
    \end{aligned}
\end{equation}
All possible contractions the first-order term of Eq.~\eqref{Eq:I-G-function}  can be written
\begin{equation}
    \begin{aligned}
        &\langle\mathcal{T}\{\hat{\psi}^\dagger(x_1)\hat{\psi}^\dagger(x'_1)\hat{\psi}(x'_1)\hat{\psi}(x_1)\hat{\psi}(x)\hat{\psi}^\dagger(y)\}\rangle\\[10pt]&
        =(kq)^2\overparent{\hat{\psi}(x'_1)\hat{\psi}^\dagger(x'_1)}\overparent{\hat{\psi}(x_1)\hat{\psi}^\dagger(x_1)}\overparent{\hat{\psi}(x)\hat{\psi}^\dagger(y)}+(kq)^3\overparent{\hat{\psi}(x'_1)\hat{\psi}^\dagger(x_1)}\overparent{\hat{\psi}(x_1)\hat{\psi}^\dagger(x'_1)}\overparent{\hat{\psi}(x)\hat{\psi}^\dagger(y)}\\&
        +k(kq)^3\overparent{\hat{\psi}(x)\hat{\psi}^\dagger(x_1)}\overparent{\hat{\psi}(x_1)\hat{\psi}^\dagger(x'_1)}\overparent{\hat{\psi}(x'_1)\hat{\psi}^\dagger(y)}
        +(kq)^3\overparent{\hat{\psi}(x)\hat{\psi}^\dagger(x_1)}\overparent{\hat{\psi}(x_1)\hat{\psi}^\dagger(y)}\overparent{\hat{\psi}(x'_1)\hat{\psi}^\dagger(x'_1)}\\&
        +k(kq)^3\overparent{\hat{\psi}(x)\hat{\psi}^\dagger(x'_1)}\overparent{\hat{\psi}(x'_1)\hat{\psi}^\dagger(x_1)}\overparent{\hat{\psi}(x_1)\hat{\psi}^\dagger(y)}
        +k^2(kq)^3\overparent{\hat{\psi}(x)\hat{\psi}^\dagger(x'_1)}\overparent{\hat{\psi}(x'_1)\hat{\psi}^\dagger(y)}\overparent{\hat{\psi}(x_1)\hat{\psi}^\dagger(x_1)}.
    \end{aligned}
\end{equation}
That the Green's function resulting from the above relation can be written as follows:
\begin{equation}
    \begin{aligned}
     \langle\mathcal{T}\{\hat{\psi}^\dagger(x_1)&\hat{\psi}^\dagger(x'_1)\hat{\psi}(x'_1)\hat{\psi}(x_1)\hat{\psi}(x)\hat{\psi}^\dagger(y)\}\rangle\\[10pt]& 
     ={\color{white}
    \underbrace{{\color{black}(kq)^2 iG^{\mathcal{M}}(x'_1,x'_1)iG^{\mathcal{M}}(x_1,x_1)iG^{\mathcal{M}}(x,y)}}_{\textcolor{black}{(A)}}{\color{black}+}\underbrace{{\color{black}(kq)^3 iG^{\mathcal{M}}(x'_1,x_1)iG^{\mathcal{M}}(x_1,x'_1)iG^{\mathcal{M}}(x,y)}}_{\textcolor{black}{(B)}}}
     \\&
     ~{\color{white}{\color{black}+}\underbrace{{\color{black}k(kq)^3 iG^{\mathcal{M}}(x,x_1)iG^{\mathcal{M}}(x_1,x'_1)iG^{\mathcal{M}}(x'_1,y)}}_{\textcolor{black}{(C)}}{\color{black}+}\underbrace{{\color{black}(kq)^3 iG^{\mathcal{M}}(x,x_1)iG^{\mathcal{M}}(x_1,y)iG^{\mathcal{M}}(x'_1,x'_1)}}_{\textcolor{black}{(D)}}}
     \\&
     ~{\color{white}{\color{black}+}\underbrace{{\color{black}k(kq)^3 iG^{\mathcal{M}}(x,x'_1)iG^{\mathcal{M}}(x'_1,x_1)iG^{\mathcal{M}}(x_1,y)}}_{\textcolor{black}{(E)}}{\color{black}+}\underbrace{{\color{black}k^2(kq)^3 iG^{\mathcal{M}}(x,x'_1)iG^{\mathcal{M}}(x'_1,y)iG^{\mathcal{M}}(x_1,x_1)}}_{\textcolor{black}{(F)}}}.
     \label{FDaigram}
    \end{aligned}
\end{equation}
\newpage
    \begin{tikzpicture}
        \begin{feynman}
            \vertex (a) {$x$};
            \vertex [right=of a] (b);
            \vertex [right=of b] (c);
            \vertex [right=of c] (d) {$y$};
            \vertex [above right=of a] (e) ;
            \vertex [above right=of a] (ee) {$x_1~~~~$};
            \vertex [above left=of d] (f) ;
            \vertex [above left=of d] (ff) {$~~~~~x'_1$};
            \vertex [left=of e] (g);
            \vertex [right=of f] (h);

            \diagram{
            (a) -- [fermion] (d);
            (e) -- [boson] (f);
            (e) -- [fermion, half right] (g);
            (g) -- [half right] (e);
            (f) -- [fermion, half left] (h);
            (h) -- [half left] (f);
            };
        \end{feynman}
        \node [below] at (current bounding box.south) {(A)};
    \end{tikzpicture}~~~~~~
    \begin{tikzpicture}
        \begin{feynman}
            \vertex (a) {$x$};
            \vertex [right=of a] (b);
            \vertex [right=of b] (c);
            \vertex [right=of c] (d) {$y$};
            \vertex [above right=of a] (e) ;
            \vertex [above right=of a] (ee) {$x_1~~~~$};
            \vertex [above left=of d] (f) ;
            \vertex [above left=of d] (ff) {$~~~~~x'_1$};
            \vertex [left=of e] (g);
            \vertex [right=of f] (h);

            \diagram{
            (a) -- [fermion] (d);
            (e) -- [boson] (f);
            (e) -- [fermion, quarter right] (f);
            (f) -- [fermion, quarter right] (e);
            };
        \end{feynman}
        \node [below] at (current bounding box.south) {(B)};
    \end{tikzpicture}
    ~~~~~~
    \begin{tikzpicture}
        \begin{feynman}
            \vertex (a) {$x$};
            \vertex [right=of a] (b);
            \vertex [right=of b] (c);
            \vertex [right=of c] (d) {$y$};
            \vertex [above right=of a] (e) ;
            \vertex [above right=of a] (ee) {$x_1~~~~$};
            \vertex [above left=of d] (f) ;
            \vertex [above left=of d] (ff) {$~~~~~x'_1$};
            \vertex [left=of e] (g);
            \vertex [right=of f] (h);

            \diagram{
            (a) -- [fermion] (f);
            (e) -- [fermion] (d);
            (e) -- [boson] (f);
            (f) -- [fermion, quarter right] (e);
            };
        \end{feynman}
        \node [below] at (current bounding box.south) {(C)};
    \end{tikzpicture}
    \\
    
    \begin{tikzpicture}
        \begin{feynman}
            \vertex (a) {$x$};
            \vertex [right=of a] (b);
            \vertex [right=of b] (c);
            \vertex [right=of c] (d) {$y$};
            \vertex [above right=of a] (e) ;
            \vertex [above right=of a] (ee) {$x_1~~~~~$};
            \vertex [above left=of d] (f) ;
            \vertex [above left=of d] (ff) {$~~~~~x'_1$};
            \vertex [left=of e] (g);
            \vertex [right=of f] (h);

            \diagram{
            (a) -- [fermion] (e);
            (e) -- [fermion] (d);
            (e) -- [boson] (f);
            (f) -- [fermion, half left] (h);
            (h) -- [half left] (f);
            };
        \end{feynman}
        \node [below] at (current bounding box.south) {(D)};
    \end{tikzpicture}~~~~~~
    \begin{tikzpicture}
        \begin{feynman}
            \vertex (a) {$x$};
            \vertex [right=of a] (b);
            \vertex [right=of b] (c);
            \vertex [right=of c] (d) {$y$};
            \vertex [above right=of a] (e) ;
            \vertex [above right=of a] (ee) {$x_1~~~~~$};
            \vertex [above left=of d] (f) ;
            \vertex [above left=of d] (ff) {$~~~~~~x'_1$};
            \vertex [left=of e] (g);
            \vertex [right=of f] (h);

            \diagram{
            (a) -- [fermion] (e);
            (f) -- [fermion] (d);
            (e) -- [boson] (f);
            (e) -- [fermion, quarter right] (f);
            };
        \end{feynman}
        \node [below] at (current bounding box.south) {(E)};
    \end{tikzpicture}~~~~~~
    \begin{tikzpicture}
        \begin{feynman}
            \vertex (a) {$x$};
            \vertex [right=of a] (b);
            \vertex [right=of b] (c);
            \vertex [right=of c] (d) {$y$};
            \vertex [above right=of a] (e) ;
            \vertex [above right=of a] (ee) {$x_1~~~~$};
            \vertex [above left=of d] (f) ;
            \vertex [above left=of d] (ff) {$~~~~~~~x'_1$};
            \vertex [left=of e] (g);
            \vertex [right=of f] (h);

            \diagram{
            (a) -- [fermion] (f);
            (f) -- [fermion] (d);
            (e) -- [boson] (f);
            (e) -- [fermion, half right] (g);
            (g) -- [half right] (e);
            };
        \end{feynman}
        \node [below] at (current bounding box.south) {(F)};
    \end{tikzpicture}
\end{widetext}
\section{CHARGE-CHARGE RESPONSE FUNCTION}\label{LFanction}
The average potential energy is defined as follows
\begin{equation}
    \begin{aligned}
        \langle\hat{H}_\textbf{int}\rangle&
        =\frac{1}{2}\int d\textbf{x} d\textbf{x}'v(\textbf{x}-\textbf{x}')\langle\hat{\psi}^\dagger(\textbf{x})\hat{\psi}^\dagger(\textbf{x}')\hat{\psi}(\textbf{x}')\hat{\psi}(\textbf{x})\rangle\\&
        =\frac{k}{2}\int d\textbf{x} d\textbf{x}'v(\textbf{x}-\textbf{x}')\langle\hat{\psi}^\dagger(\textbf{x})\hat{\psi}^\dagger(\textbf{x}')\hat{\psi}(\textbf{x})\hat{\psi}(\textbf{x}')\rangle\\&
        =\frac{1}{2q}\int d\textbf{x} d\textbf{x}'v(\textbf{x}-\textbf{x}')
        \\ \times&\Big[\langle\hat{n}(\textbf{x})\hat{n}(\textbf{x}')\rangle-\frac{1}{V}\sum_{\textbf{p}}e^{i\textbf{p}(\textbf{x}-\textbf{x}')}q^{-n_{\textbf{p}}}
        \langle\hat{\psi}^\dagger(\textbf{x})\hat{\psi}(\textbf{x}')\rangle\Big]
        \label{Eq:P-Energy}
    \end{aligned}
\end{equation}
where in the second line we used Eq.~\eqref{algebra-mode-expantion-kqParticle}.
It is convenient to introduce the deviation operator 
\begin{equation}
    \tilde{n}(\textbf{x})\equiv \hat{n}(\textbf{x})-\langle\hat{n}(\textbf{x})\rangle,
\end{equation}
resulting to
\begin{equation}\label{Eq:interactionEnergy}
     \begin{aligned}
         \langle\hat{H}_{\text{int}}\rangle&=\frac{1}{2q}\int d\textbf{x} d\textbf{x}'v(\textbf{x}-\textbf{x}')
         \\&\times\Big[\langle\tilde{n}(\textbf{x})\tilde{n}(\textbf{x}')\rangle+\langle\hat{n}(\textbf{x})\rangle\langle\hat{n}(\textbf{x}')\rangle-\delta^{(kq)}(\textbf{x}-\textbf{x}')\Big].
     \end{aligned}
\end{equation}
where
\begin{equation}
\delta^{(kq)}(\textbf{x})\equiv \frac{1}{V}\sum_{\textbf{p}}e^{i\textbf{p}.(\textbf{x})}q^{-\bar{n}_{\textbf{p}}}\langle\hat{\psi}^\dagger(\tau\textbf{x})\hat{\psi}(\tau \textbf{0})\rangle.
\end{equation}
Therefore, we concentrate on the density correlation function $\langle\tilde{n}(x)\tilde{n}(x')\rangle$. We introduce a time-ordered correlation function, called the the polarization function 
\begin{equation}
iD_{kq}(x,x')=\langle\mathcal{T}[\tilde{n}^{}_H(x)\tilde{n}^{}_H(x')]\rangle,
\end{equation}
This relation is symmetric in its arguments $D_{kq}(x,x')=D_{kq}(x',x)$. 
The interaction energy~\eqref{Eq:interactionEnergy} can now be rewritten as 

\begin{equation}
    \begin{aligned}
        \langle\hat{H}_{\text{int}}\rangle=\frac{1}{2q}\int d\textbf{x} d\textbf{x}'v(\textbf{x}-\textbf{x}')\Big[iD_{kq}(\tau\textbf{x},\tau\textbf{x}')+n^2
    &\\
    -\delta^{(kq)}(\textbf{x}-\textbf{x}')\Big].&
    \end{aligned}
\end{equation}
Denoting the correlation function for a noninteracting system as $D^0_{kq}(x',x)$, one can separate the interaction energy into a first-order contribution and 
all the higher-order contributions: 
\begin{equation}
  \begin{aligned}
     \langle\hat{H}_{\text{int}}\rangle=&\frac{1}{2q}\int d\textbf{x} d\textbf{x}'v(\textbf{x}-\textbf{x}')\Big[iD^0_{kq}(\tau\textbf{x},\tau\textbf{x}')+n^2_{}
     \\&-\delta^{(kq)}(\textbf{x}-\textbf{x}')
     \\&+\frac{1}{2q}\int d\textbf{x} d\textbf{x}'v(\textbf{x}-\textbf{x}')\Big(i\delta D_{kq}(\tau\textbf{x},\tau\textbf{x}')\Big),
  \end{aligned}
\end{equation}
where $i\delta D_{kq}(\tau\textbf{x},\tau\textbf{x}')\equiv i D_{kq}(\tau\textbf{x},\tau\textbf{x}')-i D_{kq}^0(\tau\textbf{x},\tau\textbf{x}')$. One then can write
\begin{equation}\label{Hint}
    \begin{aligned}
        \langle\hat{H}_{\text{int}}\rangle&=\langle\hat{H}_{\text{int}}\rangle_0+\frac{1}{2q}\int d\textbf{x} d\textbf{x}'v(\textbf{x}-\textbf{x}')\Big(i\delta D_{kq}(\tau\textbf{x},\tau\textbf{x}')\Big),
    \end{aligned}
\end{equation}
where $\langle\hat{H}_{\text{int}}\rangle_0=\langle\Phi_0|\hat{H}_{\text{int}}|\Phi_0\rangle$. Introducing a variable coupling constant, $\lambda$, to the interaction Hamiltonian allows us to control the strength of the interaction. By setting $\lambda=0$ , we obtain a non-interacting system. Conversely, when $\lambda=1$, the system exhibits first-order interactions. The second term in Eq.~\eqref{Hint} can be expressed as follows:
\begin{equation}
    \begin{aligned}
        \langle\hat{H}_{\text{int}}\rangle&=\langle\hat{H}_{\text{int}}\rangle_0+ E_{corr},
    \end{aligned}
\end{equation}
where
\begin{equation}
    \begin{aligned}
        E_{corr}&=\frac{1}{2q}\int^1_0 \frac{d\lambda}{\lambda}\int d\textbf{x} d\textbf{x}'\lambda v(\textbf{x}-\textbf{x}')\Big(i\delta D_{kq}(\tau\textbf{x},\tau\textbf{x}')\Big).
    \end{aligned}
\end{equation}
In this part we calculate $\Pi_{kq}^0$. Using the definition of the Green's function, we can rewrite Eq.~\eqref{LindhardF} as follows:
\begin{equation}
    \begin{aligned}
        &\Pi_{k,q}^0(Q,\omega^{(j)}_n)=-\frac{kq}{\beta}\sum_{j',m}\\&\times\int \frac{d^3p}{(2\pi)^3} (\frac{q^{-\bar{n}^{}_p}_{}}{i\epsilon_m^{(j')}-f_{kq}(\zeta_p)})(\frac{q^{-\bar{n}^{}_{p+Q}}_{}}{i\epsilon_m^{(j')}+i\omega_n^{(j)}-f_{kq}(\zeta_{p+Q})}),
    \end{aligned}
\end{equation}
 Dividing the above equation into two separate parts, we obtain the following:
\begin{equation}\label{lindhard-function}
   \begin{aligned}
       &\Pi_{k,q}^0(Q,\omega^{(j)}_n)\\&=-\frac{kq}{\beta}\sum_{j',m}\int \frac{d^3p}{(2\pi)^3}\frac{q^{-\bar{n}^{}_p}q^{-\bar{n}^{}_{p+Q}}}{i\omega_n^{(j)}-f_{kq}(\zeta_{p+Q})+f_{kq}(\zeta_{p})}\\&
       \times\left(\frac{1}{i\epsilon_m^{(j')}-f_{kq}(\zeta_{p})}-\frac{1}{i\epsilon_m^{(j')}+i\omega_n^{(j)}-f_{kq}(\zeta_{p+Q})}\right).
   \end{aligned} 
\end{equation}
In the second part of the equation, we can modify the counter in the summation from $m\to m-n$ and $j'\to j$. This will cause the variable $\omega_n^{(j)}$ to vanish.
we have
\begin{equation}
    \begin{aligned}
        &\Pi_{k,q}^0(\omega^{(j)}_n,Q)
        \\&=-kq\int \frac{d^3p}{(2\pi)^3}\frac{q^{-(\bar{n}_{p+Q}^{}+\bar{n}_p^{})}\left[S^{}_{kq}(\zeta^{}_{p+Q})-S^{}_{kq}(\zeta^{}_{p})\right]}{i\omega_n^{(j)}-f_{kq}(\zeta^{}_{p+Q})+f_{kq}(\zeta^{}_{p})},
    \end{aligned}
\end{equation}
where
\begin{equation}
    S_{kq}(\zeta_{p})=\frac{1}{\beta}\sum_{j',m}\frac{f_{kq}(\zeta_{p})}{\epsilon_m^{(j')2}+f_{kq}(\zeta_{p})^2_{}},
\end{equation}
By summing over the indices $m$ and $j'$ for both q-boson and q-fermion Matsubara frequencies~\eqref{Eq:Matsubara}, we obtain the following relations:
\begin{equation}
\begin{aligned}
        S_B^{}(\zeta^{}_p)&=\sum_{m=-\infty}^{\infty}\sum_{j'}\frac{x}{\left(2m\pi-i(-1)^{j'}\ln{q}\right)^2+x^2}
        \\&=\frac{1}{2}\coth \frac{x-\log q}{2}+\frac{1}{2}\coth \frac{x+\log q}{2},
\end{aligned}
\end{equation}
\begin{equation}
    \begin{aligned}
        S_F^{}(\zeta^{}_p)&=\sum_{m=-\infty}^{\infty}\sum_{j'}\frac{x}{\left((2m+1)\pi+i(-1)^{j'}\ln{q}\right)^2+x^2}
        \\&=\frac{1}{2}\tanh \frac{x-\log q}{2}+\frac{1}{2}\tanh \frac{x+\log q}{2},
    \end{aligned}
\end{equation}
where $x=\beta f(\zeta_{p})$. Our goal is to express the aforementioned results in terms of the q-deformed distribution function~\eqref{ٍEq:distribution-function}. This will enable us to apply specific conditions for the distribution function. So we get to
\begin{equation}\label{S_{kq}-np1}
    \begin{aligned}
    &S_B(\zeta_p)
        =1+\frac{q}{\left(\frac{1-e^{A \bar{n}^{}_p} q^2}{q-e^{A\bar{n}^{}_p} q}\right)^{A'}-q}+\frac{1}{q
   \left(\frac{1-e^{A\bar{n}^{}_p} q^2}{q-e^{A\bar{n}_p} q}\right)^{A'}-1},
    \end{aligned}
\end{equation}
and
\begin{equation}\label{S_{kq}-np2}
    \begin{aligned}
        &S_F(\zeta_p)
        =1-\frac{q}{\left(\frac{e^{A\bar{n}^{}_p}-q^2}{q-e^{A\bar{n}^{}_p} q}\right)^{A''}+q}-\frac{1}{q
   \left(\frac{e^{A\bar{n}^{}_p}-q^2}{q-e^{A\bar{n}^{}_p} q}\right)^{A''}+1},
    \end{aligned}
\end{equation}
where 
\begin{equation}
    \begin{aligned}
        A&=q-\frac{1}{q}~~~,~~~A'=\frac{q^{-\bar{n}^{}_p}_{}+q^{1+\bar{n}^{}_p}_{}}{1+q},\\
        A&''=\frac{q^{-\bar{n}^{}_p}_{}(1+2q-q^{1+2\bar{n}^{}_p}_{})}{1+q}.
    \end{aligned}
\end{equation} 
also the expressions will be evaluated at their limiting conditions
\begin{align}
    &\lim_{q\to 1}S_B^{}(\zeta_p)=1+2\bar{n}^{}_p~,~~\text{For Bosons}~k=1,\\
    &\lim_{q\to 1}S_F^{}(\zeta_p)=1-2\bar{n}^{}_p~,~~\text{For Fermions}~k=-1.
\end{align}
\\
\subsection{Generalized Fermi Energy}
The total number of particles at zero temperature $(T=0)$ is equal to
\begin{equation}
    N=\sum_p \bar{n}_p,
\end{equation}
By approximating the sum as an integral, we obtain the following expression:
\begin{equation}
    N=V\int\frac{d^3p}{(2\pi)^3}\bar{n}_p,
\end{equation}
Note that, at zero temperature, the chemical potential $\mu$ is equivalent to the Fermi energy, $\mu(T=0)\equiv\epsilon^{(kq)}_F\equiv{p_F^{(kq)}}^2/2m$, and using Eq.~\eqref{density-particle} we have
\begin{equation}
    \begin{aligned}
        N=V\int_0^{p_F^{(kq)}}\frac{4\pi p^2 dp}{(2\pi)^3}\frac{2\ln{q}}{q-q^{-1}}=\frac{V}{2\pi^2}\frac{1}{3}{p_F^{(kq)}}^3\frac{2\ln{q}}{q-q^{-1}},
    \end{aligned}
\end{equation}
Given that the particle density $n$ is defined as $n\equiv N/V$, we can deduce from the previous equation that
\begin{equation}\label{Eq:kq-momentum}
    p_F^{(kq)}=\left(\frac{3\pi^2\bar{n}_p(q-q^{-1})}{\ln{q}}\right)^{1/3},
\end{equation}
We also need to calculate the total energy of the system
\begin{equation}
    E_{Total}=\sum_p\bar{n}_p\epsilon_p,
\end{equation}
In continuous space we have
\begin{equation}
    E_{Total}=\frac{V}{(2\pi)^3}\int_0^{p_F^{(kq)}} d^3p\frac{2\ln{q}}{q-q^-1}\frac{p^2}{2m},
\end{equation}
and
\begin{equation}
   \begin{aligned}
       E_{Total}&=\frac{V\ln{q}}{2m \pi^2(q-q^{-1})} \int_0^{p_F^{(kq)}}p^4 dp\\&=\frac{V\ln{q}}{10m \pi^2(q-q^{-1})}p^5\Bigg|^{p_F^{(kq)}}_0,
   \end{aligned}
\end{equation}
Using Eq.~\eqref{Eq:kq-momentum}, the total energy is written as follows
\begin{equation}
    E_{Total}=\frac{V\ln{q}~{p_F^{(kq)}}^3}{5\pi^2(q-q^{-1})}\frac{{p_F^{(kq)}}^2}{2m}=\frac{3}{5}N\epsilon^{(kq)}_F.
\end{equation}
\\
\subsection{Calculation Of The Polarization Function}
We present the calculation of the expression denoted by Eq.~\eqref{Eq.Li} in this section
\begin{equation}
   \begin{aligned}
        &R^{(q)}_{1}
        \\&=\frac{q^{-(2\bar{n}(0)-1)}S^{F}_{q}(0)}{(2\pi)^3}\int_0^{p_F^{(q)}} \frac{\text{d}^3p}{\frac{\hbar^2}{2m}C_{q}^F(0)\left(Q^2+2p.Q\right)-i\omega^{(j)}_n}.
   \end{aligned}
\end{equation}
Employing the transformation $d^3p = p^2 \sin{\theta} \, dp \, d\theta \, d\phi$, it is possible to re-express the denominator, which we write like this $C_{q}^F(0)\frac{\hbar^2Q^2}{2m}+C_{q}^F(0)\frac{\hbar^2p.Q}{m}-i\omega^{(j)}_n=C_{q}^F(0)\frac{\hbar^2Q}{m}(Z^{}_1p_F^{(q)}+p\cos{\theta})$ that $Z^{}_1=\frac{Q}{2p_F^{(q)}}-\frac{m(i\omega_n^{(j)})}{\hbar^2C_q^F(0)p_F^{(q)}Q}$,
\begin{equation}
\begin{aligned}
R^{(q)}_{1}
&=\frac{q^{-(2\bar{n}(0)-1)}S^{F}_{q}(0)}{(2\pi)^3}
\\&\times\int_0^{2\pi}\int_0^{p_F^{(q)}}\int_0^{\pi}\frac{p^2\sin{\theta}~dp~d\theta~d\phi}{C_q^F(0)\frac{\hbar^2Q}{m}(Z_1p_F^{(q)}+p\cos{\theta})},
\end{aligned}
\end{equation}
and
\begin{equation}
    \begin{aligned}
        &R^{(q)}_{1}\\&=\frac{q^{-(2\bar{n}(0)-1)}S^{}_{q}(0)m}{\hbar^2(2\pi)^2C_q^F(0)Q}\int_0^{p_F^{(q)}}p^2 dp\int_0^{\pi}\frac{\sin{\theta}~d\theta}{Z_1p_F^{(q)}+p\cos{\theta}}
        \\
        &=\frac{q^{-(2\bar{n}(0)-1)}S^{}_{q}(0)m}{\hbar^2(2\pi)^2C_q^F(0)Q}\int_0^{p_F^{(q)}}p^2\left(\frac{1}{p}\ln\frac{Z_1p_F^{(q)}+p}{Z_1p_F^{(q)}-p}\right)dp,
    \end{aligned}
\end{equation}
using the following indefinite integrate
\begin{equation}
    \int x \ln{\frac{a+x}{a-x}}dx=ax+\frac{1}{2}(x^2-a^2)\ln{\frac{a+x}{a-x}},
\end{equation}
we obtain
\begin{equation}
    \begin{aligned}
        &\int_0^{p_F^{(q)}}p\left(\ln\frac{Z^{}_1p_F^{(q)}+p}{Z^{}_1p_F^{(q)}-p}\right)dp
        \\&={p_F^{(q)}}^2\left(Z^{}_1+\frac{1-Z_1^2}{2}\ln{\frac{Z^{}_1+1}{Z^{}_1-1}}\right),
    \end{aligned}
\end{equation}
as a result
\begin{equation}
    \begin{aligned}
        R^{(q)}_{1}&=\frac{q^{-(\bar{n}(0)-1)}S^{}_{q}(0)m{p_F^{(q)}}^2}{\hbar^2(2\pi)^2C_q^F(0)Q}\\&\times\left[Z^{}_1+\frac{1-Z_1^2}{2}\ln{\frac{Z^{}_1+1}{Z^{}_1-1}}\right].
    \end{aligned}
\end{equation}
\end{document}